\documentclass[12pt]{article}
\usepackage[dvips]{graphicx}
\usepackage{amsmath}
\usepackage{array}
\usepackage{cite}
\usepackage{epsfig}
\usepackage[english]{babel}
\usepackage[latin1]{inputenc}
\usepackage[T1]{fontenc}
\usepackage{amsfonts}
\usepackage{float}
\usepackage{sidecap}
\usepackage{caption}
\usepackage{multirow}
\usepackage{lscape}

\begin{document}
\begin{titlepage}
\begin{flushright}
DAMTP-2010-4\\
IC/2010/005
\end{flushright}
\vspace{1.1cm}
\begin{center}
{\bf \Large D-branes at Toric Singularities:\\[0.2cm] {\large Model Building, Yukawa Couplings and Flavour Physics}}\\[1cm]
Sven Krippendorf$^1$, Matthew J. Dolan$^1$, Anshuman Maharana$^1$, Fernando Quevedo$^{1,2}$\\[0.4cm]
$^1$ {\it DAMTP, University of Cambridge,}\\
 {\it Wilberforce Road, CB3 0WA, Cambridge, UK.}\\
$^2$ {\it ICTP, Strada Costiera 11, Trieste  34151, Italy.}\\[1cm]
\date{\today}
\end{center}
\begin{abstract} We discuss general properties of D-brane model building at toric singularities. Using dimer techniques to obtain the gauge theory from the structure of  the singularity, we extract results on the matter sector and superpotential of the corresponding gauge theory.  We show that the number of families in toric phases is always less than or equal to three, with a
unique exception being the zeroth Hirzebruch surface. 
With the physical input of three generations we find that the lightest family of quarks is massless and the masses of the other two can be hierarchically separated. We compute the CKM matrix for explicit models in this setting
and find the singularities possess sufficient structure to allow for realistic mixing between generations and CP violation. 
\end{abstract}

\end{titlepage}
\newpage
\tableofcontents
\newpage

\section{Introduction}

Explaining the structure of the Standard Model of particle physics is one of the biggest outstanding problems in string phenomenology. The success story of the Standard Model is uncontested
but an ultraviolet completion, as in principle provided by string theory, is an important challenge. 

There are many different proposals for embedding the Standard Model and its extensions into various classes of string models. 
Besides the general features of the Standard Model, such as the gauge symmetries and spectrum, there are also many observables that should be calculable from the corresponding string model and confronted with experimental results. In particular, realistic models must explain the hierarchy of masses of the different generations of quarks and leptons and the information encoded in the CKM matrix.

Efforts in string model building go back to heterotic string theory and include more recently constructions in brane engineering within type II flux compactifications. An attractive feature of type IIB model building is the presence of successful scenarios for moduli stabilisation~\cite{KKLT, lvs}.
The LARGE volume scenario (LVS)~\cite{lvs} is particularly attractive: the LARGE volume ensures that perturbative corrections are well under control and does not require fine-tuning of the flux superpotential. 
The hierarchy problem can be solved by gravity/moduli mediated supersymmetry breaking \cite{0505076,0605141,0610129,0906.3297}.
Depending on the overall volume of the bulk geometry one can lower the string scale to values smaller than or similar to  the GUT scale  $M_{GUT}\sim10^{16}\, {\rm GeV}$, making them attractive from the perspective of gauge unification.

The existence of a large volume implies that the Standard Model has to be located at a local region within the full Calabi-Yau space \cite{0005067}.  It is natural to think of those models as supersymmetric models, with supersymmetry breaking dealt with from the bulk \cite{0810.5660,0906.3297, 0912.2950}. We hence will, from now on, talk about supersymmetric model building and expect supersymmetry to be broken by bulk/moduli effects.

This viewpoint is applicable to all model building efforts in this category:  intersecting brane models~\cite{0007024,0011132,0502005, 0702094}, F-theory models~\cite{0802.3391,0802.2969,0806.0102,0808.2969,0901.4941,0904.3932,0908.1784,0911.3008,1001.0577} and branes at singularities \cite{0005067,0105042,0312051,0508089,0703047,0810.5660,0910.3616}. 
Several models with realistic matter content and potentially  couplings have been developed in these scenarios. However, obtaining the correct flavour physics remains challenging to date. For example in the widely studied F-theory models, the flavour structure is not complete at tree-level \cite{0811.2157,0811.2417,0904.3101,0904.1584,0910.0477,0912.0853} and one has to introduce non-commutative fluxes \cite{0910.2413} or appropriate instanton effects \cite{0910.5496} to achieve for Yukawa matrices with rank greater than one.

In this article we concentrate on the class of models corresponding to D-branes at singularities which represents one of the most promising avenues to obtain the Standard Model from string theory:
\begin{itemize}
\item  The models are truly local in the sense that many features can be addressed in a way that is essentially independent of the moduli stabilisation problem.
\item  Even though the distances involved are in a regime smaller than the string scale, string theory is well  under control.
\item
 In the case of IIB string theory, the models can be determined from local F-theory models in the limit of vanishing cycles, yet their properties are much more constrained and in some sense represent a
minimal class of F-theory models. In particular there is much less freedom to include the matter sector and couplings, making realistic model-building more rigid. If a realistic property is obtained, its presence can be argued to be more robust than in local F-theory models where there is much freedom to engineer the models.  For instance simple GUT models are not possible to obtain with non-vanishing quark masses. However,  several other generalisations of the Standard Model are easy to obtain. Contrary to simple group GUT models, Yukawa couplings giving rise to fermion masses are not forbidden by anomalous $U(1)$ symmetries and the structure of these couplings is elegantly determined in terms of the corresponding dimer diagrams~\cite{0503149,0504110j,0505211,0511063,0511287,0706.1660,0803.4474}.
\item
 In principle, if the standard model is on D3 branes at a singularity, gauge coupling unification is automatic since, to leading order, all the gauge couplings of a product gauge group are given by the expectation value of the dilaton field, a property shared by  heterotic models but not by intersecting brane models. The question of unification reduces to whether the matter sector is such that the measured low energy values of the gauge couplings can be obtained after renormalisation group running. Notice that a simple GUT group is not needed to achieve unification.
\item
To leading order a vanishing value of the blow-up mode is a natural extremum of the effective action and then considering the model at a singularity is a
promising ansatz once moduli stabilisation is taken into account.
\item
 Over the past few years very powerful techniques have been developed~\cite{0503149,0504110j,0505211,0511063,0511287,0706.1660,0803.4474} to describe these models in terms of quiver and dimer diagrams that provide the relevant information concerning the spectrum and couplings of the corresponding gauge theory. These techniques to study branes at general singular points have been used to address several aspects of the AdS/CFT correspondence as well as local supersymmetry breaking and properties of M2-branes, but, except for the simplest cases of orbifold singularities and some del Pezzo surfaces, their phenomenological aspects for string model building have not been explored.
\item The gauge theories that can be obtained at the singularities are highly restricted. Both the matter content and superpotential are completely determined. One may only vary the ranks of the gauge groups and vevs of fields (corresponding to resolution of cycles).

\end{itemize}

\subsection{Summary of Results}

In this paper we focus on local model building with branes at singularities and describe general constraints on flavour physics within a large class of these models, the so-called toric singularities for which the matter content and superpotential can be obtained systematically. 
We start with a detailed introduction to the relevant aspects of toric singularities in terms of quiver and dimer diagrams. In particular we describe how to determine the matter spectrum and perturbative superpotential for D3 branes  at the singularity,  including also D3-D7 states. The most efficient way to  determine the gauge theory from the toric singularity is through an algorithm~\cite{gulotta} which we describe in detail since it shall be the primary tool used in obtaining our results. Our main results can be summarised as follows:

\begin{description}
\item{\bf Three families bound.}
We find as a general result that the number of families in toric models is bounded by $N_f=3$. The only exception is the well known case of the zeroth Hirzebruch surface which has four families, but has a Seiberg dual phase with two families.\footnote{As the example of the zeroth Hirzebruch surface suggests the origin of the 3-families is not due to the 3-complex dimensions of the bulk geometry.} The bound $N_f\leq 3$ was found before in $Z_N$ singularities, with $Z_3$ saturating the bound \cite{0005067}. It is remarkable that this result extends to the much larger  class of toric singularities. Given the arbitrariness in the number of families in most string constructions, it is intriguing that the physical value plays an important role in this class of models. In principle, non-toric phases obtained by Higgsing could generate models with more families.

\item{\bf Hierarchy of masses.} 
The explicit knowledge of the  superpotential allows us to compute the mass matrix for the quarks. When a physically realistic choice of quarks is made, we find that there is always one vanishing mass eigenvalue in a toric singularity. Generically it is possible to find hierarchical masses for the other two eigenstates (except for the zeroth del Pezzo singularity $dP_0$). For three families we find then the masses $(M,m,0)$ with $M>>m$. This result  was found for $dP_1$ in \cite{0810.5660} and it is remarkable that extends to the general toric case.

\item{\bf CKM matrix.} 
 We have also computed the CKM matrix for two classes of models at toric del Pezzo singularities: the first is where both the up and down quarks arise from D3-D3 states, and the second has one type of quarks as D3-D3 states and the other type as D3-D7 states.
We can construct the correct CKM matrix in both type of models by appropriate values in the ratios of the Higgs vevs. We illustrate our findings with concrete examples based on SM-like and left-right symmetric models at del Pezzo singularities. We show that in the first class of models the $dP_1$ singularity allows for the correct flavour mixing. In the second class of models we find the correct mixings for the $dP_2$ and $dP_3$ singularities. In our analysis we neglect perturbative corrections to kinetic terms or non-perturbative corrections to the superpotential which in principle could influence the small mixing angles in the CKM-matrix.

\item{\bf CP violation.}
Given the structure of the Yukawa matrices, the amount of CP violation can be easily computed in terms of the Jarlskog invariant $J$ \cite{Jarlskog:1985cw,Jarlskog:1985ht}. With the hierarchical structure in the CKM matrix, the magnitude of the Jarlskog invariant is automatically in the desired range \cite{0605217}. In our examples we express the complex phase in the CKM matrix in terms of the Higgs fields.

\end{description}

\subsection{Branes at singularities with a broad brush}
\label{sec:branebrush}

Placing D3 branes at singular points in the bulk geometry can reduce the amount of supersymmetry and generate chiral matter, providing a rich arena for particle physics model building.
Models have been constructed on orbifold singularities such as $\mathbb{C}^3/\mathbb{Z}^3$~\cite{0005067}, as well as toric and non-toric singularities~\cite{0508089,0703047}.
In this section we provide an introduction to recent methods in extracting the gauge theories which arise from branes probing toric singularities (much more detailed reviews include \cite{0803.4474, 0706.1660}).

We focus on {\it conical singularities.} This means that we are interested in a Calabi-Yau threefold $Y$ whose metric can be locally written in a cone-like form
\begin{equation}
ds^2=dr^2+r^2 g_{ij}\, dx^i dx^j\, ,
\end{equation}
where $0<r<\infty$ and $g_{ij}$ is the metric on a five dimensional Sasaki-Einstein manifold $X.$\footnote{As a Calabi-Yau manifold, $Y$ is K\"ahler and Ricci flat which implies that $X$ has to be a Sasaki-Einstein space \cite{9807080}.} The point $r=0$ desribes the tip of the cone, where the D3 branes are placed. This geometry is non-compact, however one should think of this infinite cone as being embedded as a local region within a full Calabi-Yau compactification (for a concrete example see \cite{0005067,0610007}).

Generally, an explicit metric for $X$ is not known. However, the gauge theory is only determined by the topology of the collapsing hypersurface $X,$ as can be seen in the following way: the matter content and superpotential of the D3 branes correspond to the matter content and couplings D5 branes encounter when wrapping the two-cycles in the non-singular surface $X$. This is the large volume perspective \cite{9906242,9906200,0110028,0212021}. The gauge theory on the D5 branes is determined as in the intersecting brane picture. D5 branes can wrap distinct 2-cycles, corresponding to different gauge groups. The intersection of those 2-cycles gives rise to chiral matter and triple intersections correspond to superpotential terms.

Focusing on toric Calabi-Yau manifolds allows  the powerful techniques of toric geometry to be used to extract the gauge theory on the branes at the singularity \cite{0003085,0505211} (for a review of toric geometry see for example \cite{9711013}). In this context, the six dimensional cone can be represented as a $T^3$ fibration over a convex rational polyhedral cone in $\mathbb{R}^3,$ which is parametrised by the normal vectors $v_i$ of each facet. At the facets of the polyhedral cone the $T^3$ fibre degenerates to $T^2,$ at the edges to $S^1,$ and at the tip $T^3$ degenerates completely. Imposing the Calabi-Yau condition on $Y$ implies that by an $SL(3, \mathbb{Z})$ transformation of the torus, all normal vectors can be transformed to $v_i=(1,w_i).$ Being rational vectors the basis $w_i$ can then be drawn in the integer lattice of $\mathbb{R}^2,$ forming a convex polygon. This is the {\it toric diagram} or Newton polytope.
\begin{center}
\includegraphics[width=0.45\textwidth]{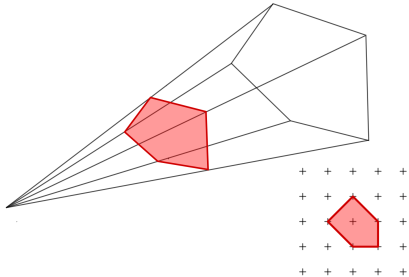}
\captionof{figure}{\footnotesize{A five-faceted polyhedral cone in $\mathbb{R}^3.$ The normal vectors determine the toric diagram. At the facets of the polyhedral cone the $T^3$ fibre degenerates to $T^2,$ at the edges to $S^1,$ and at the tip $T^3$ degenerates completely.}}
\end{center}
The polygon encodes all the topological information relevant for our purposes. To extract particle physics models from a singularity, the analysis always starts with the toric diagram of $Y$. For example, the number of gauge groups is given by twice the area of the toric diagram.

In Section 5 when exploring flavour physics we focus on models which are based on toric del Pezzo surfaces.  Recall that the del Pezzo surfaces $dP_n$ are defined as blow-ups at $n$ points ($0\leq n\leq 8$) of the compact projective space $\mathbb{P}^2$, the first four of which admit a toric description. The associated two dimensional base spaces can be obtained by modifying the toric base of $\mathbb{P}^2$, which is a triangle.  For the first toric del Pezzo surfaces the blowing up corresponds to replacing the vertices of the triangle with line segments. This is shown for the first del Pezzo surface $dP_1$ in Figure~\ref{fig:blowup}. In this case the Calabi-Yau 3-folds $Y$ are complex cones over the complex two surfaces $dP_n$.

\begin{center}
  \includegraphics[width=0.45\textwidth]{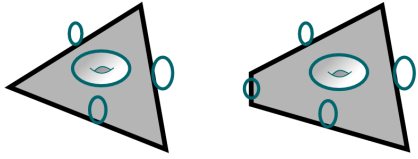}
\captionof{figure}{\footnotesize{{\bf Left:} $\mathbb{P}_2$ can be described as a $T^2$ fibration over a triangle. At the edges of the triangle the torus $T^2$ degenerates to a circle $S^1$ and at a vertex the fibre degenerates to a point. {\bf Right:} Blowing up a vertex in $\mathbb{P}_2$ and replacing it with a line generates the first del Pezzo surface.  }\label{fig:blowup}}
\end{center}

One of the best known examples of a gauge theory from a toric singularity is that residing on D3 branes probing the conifold~\cite{9807080}.
In this example the five dimensional Sasaki-Einstein manifold is $T^{(1,1)},$ and the associated gauge theory is the following ${\cal N}=1$ superconformal quiver gauge theory:
\begin{center}
 \begin{tabular}{c c}
 \begin{tabular}{c}$W=\epsilon_{ij}\epsilon_{kl}A^{i}_{12}B^{k}_{21}A^{j}_{12}B^{l}_{21}$ \end{tabular}& \begin{tabular}{c}\includegraphics[width=0.35\textwidth]{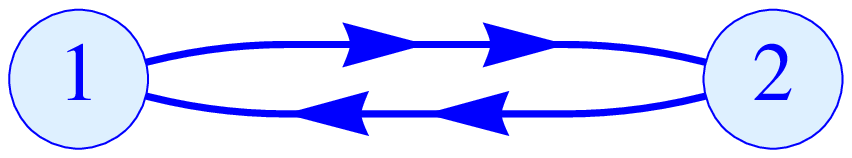}\end{tabular}
\end{tabular}
\captionof{figure}{\footnotesize{The superpotential and quiver gauge theory of the conifold.\label{conifold}}}
\end{center}
The ranks of the two gauge groups are equal, the fields $A_{12}^i$ transform as $(N_1,\bar{N}_2)$ and $B_{21}^i$ as $(N_2,\bar{N}_1)$, where $N_i$ refers to gauge group $i$.

From the point of view of model building, a theory with equal ranks is not desirable. However,
by adding {\it fractional branes}, the ranks of the gauge groups change. In the example of the conifold, adding $M$ fractional branes changes the gauge group  $SU(N)\times SU(N)\to SU(N+M)\times SU(N).$ The matter content and superpotential remain unchanged but the theory is no longer conformal \cite{9911096}. The addition of fractional branes allows us to choose the rank of the gauge groups freely and hence to construct realistic gauge group content. In addition, the loss of conformality makes realistic model building possible.

\subsubsection*{Quiver gauge theories}
It can be shown that gauge theories associated to toric singularities are always quiver gauge theories~\cite{0503149,0504110j}. 
 Quiver gauge theories are gauge theories whose matter content can be described by directed graphs. Nodes in the graph correspond to gauge groups and arrows between two nodes to bi-fundamental matter. Multiple arrows correspond to multiple copies of that type of matter, i.e. families.
The language of quivers is also useful in the context of model building. For instance, Higgsing of fields can be implemented by collapsing arrows between nodes. The cancellation of non-abelian gauge anomalies can be checked by verifying whether the number of incoming arrows and the number of outgoing arrows are equal, where each arrow is weighted with the size of the gauge group from which it originates or ends.

The superpotential cannot be read off from the quiver, only all gauge invariant operators, which correspond to cycles in the quiver diagram.\footnote{How to obtain the number of gauge invariant operators is discussed in Appendix \ref{sec:operators}.} Obtaining the superpotential leads us to a discussion of dimer techniques.

\subsubsection*{Dimer diagrams}
Dimer diagrams encode the complete gauge theory and provide the relation to the toric diagram~\cite{0503149,0504110j,0505211,0511063,0511287,0706.1660,0803.4474}.
A dimer (also known as a brane tiling) is a graph on a torus $T^2,$ drawn as a parallelogram in $\mathbb{R}^2$ with opposite edges identified. 
Each distinct face of the dimer corresponds to a gauge group.  Edges between faces correspond to bi-fundamental matter charged under the associated gauge groups.

For an edge which divides gauge group faces $A$ and $B,$ the transformation properties (i.e. $(A, \bar{B})$ or $(B, \bar{A})$) of the bi-fundamental matter are determined by {\it zigzag paths.}
Zigzag paths are directed paths in the dimer which have the property that they turn maximally at each node, and they cross once the edges along which they run~\cite{0511063,0511287}.
The intersection of two zigzag paths along an edge corresponds to bi-fundamental matter whose gauge group transformation properties are determined by the orientation of the crossing (cf. Figure~\ref{fig:zigzagmatter}). The number of common edges between two gauge group faces gives the multiplicity of the bi-fundamental matter charged under these two gauge groups. 
\begin{center}
\includegraphics[width=0.3\textwidth]{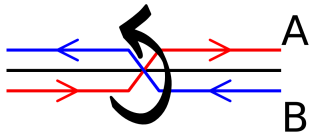}
\captionof{figure}{\footnotesize{This crossing of zigzag paths corresponds to bi-fundamental matter transforming as $(B,\bar{A}).$ }\label{fig:zigzagmatter}}
\end{center}
Zigzag paths form closed paths on the torus. This property is crucial in providing a connection with the toric diagram. In particular, the winding numbers of the zigzag paths are the inverse slopes of the edges of the toric diagram. 

Vertices in the dimer correspond to terms in the superpotential. All vertices have an intrinsic orientation provided by the circulation of zigzag paths around them. Neighbouring vertices always have opposite orientation, therefore dimers are bipartite graphs.  The bi-fundamental matter going into a vertex forms a superpotential term with the sign of the term corresponding to the sense of orientation, shown in Figure~\ref{fig:zigzagsup} below.
\begin{center}
\begin{tabular}{c c}
\begin{tabular}{c}\includegraphics[width=0.3\textwidth]{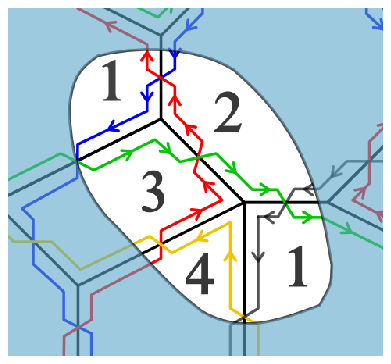}\end{tabular} & 
\begin{tabular}{c} {\bf Superpotential for these two vertices:}\\
$W=X_{13}X_{32}X_{21}-X_{14}X_{43}X_{32}X_{21}$
\end{tabular}\\
\end{tabular}
\captionof{figure}{\footnotesize{A part of a dimer showing zigzag paths around two nodes, and the corresponding superpotential terms.}\label{fig:zigzagsup}}
\end{center}
With this information in hand, one can obtain both the quiver diagram and superpotential from the dimer. This is explicitly illustrated for the cases of $dP_0$ and $dP_1$ in Figure~\ref{fig:dp0dp1}, which shows the toric diagram, dual web diagram, dimer, zigzag paths and quiver diagram for each of these theories.
\newpage
\setlength{\textheight}{25.1cm}
\begin{landscape}

\begin{tabular}{| c | c || c | c |}
\hline
\begin{tabular}{c}
Factor $A$ $(1,-2):$\\
\includegraphics[width=0.25\textwidth]{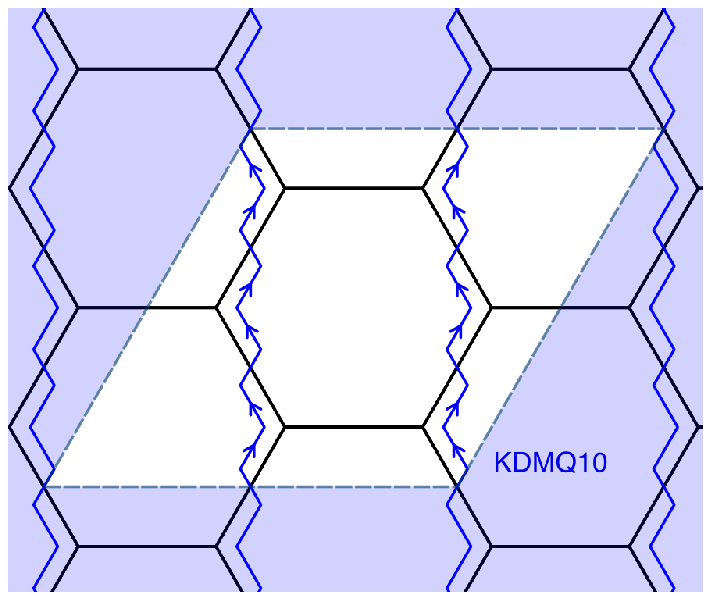}\\
Factor $B$ $(1,1):$\\
\includegraphics[width=0.25\textwidth]{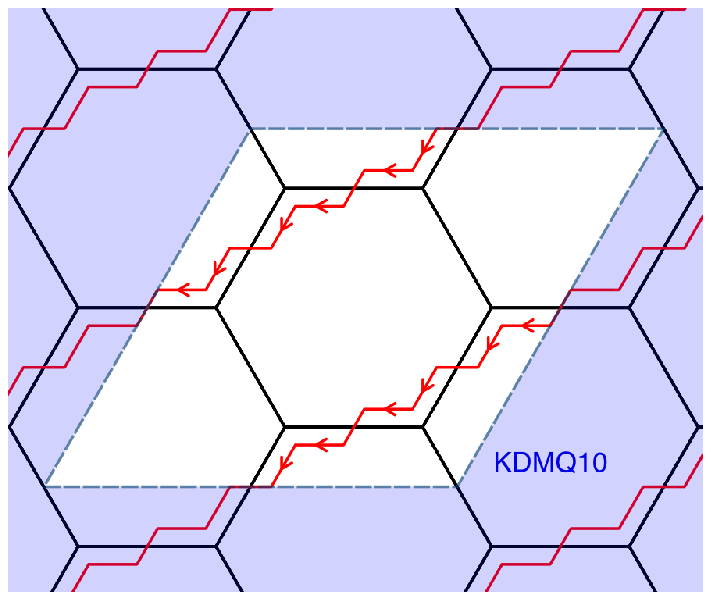}\\
Factor $C$ $(-2,1):$\\
\includegraphics[width=0.25\textwidth]{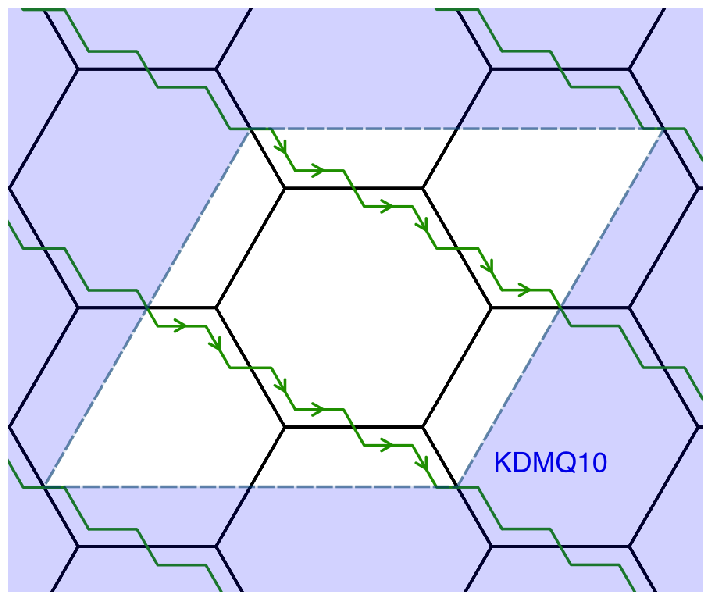}\\
\end{tabular}
&
\begin{tabular}{c}
{\bf The Toric Diagram and $(p,q)$-web}\\
\includegraphics[width=0.25\textwidth]{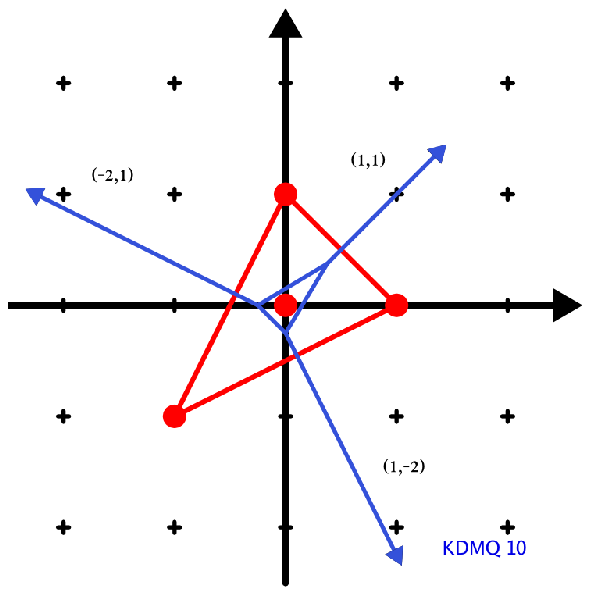}\\[0.5cm]
\hline
{\bf Complete Dimer:}\\
\includegraphics[width=0.35\textwidth]{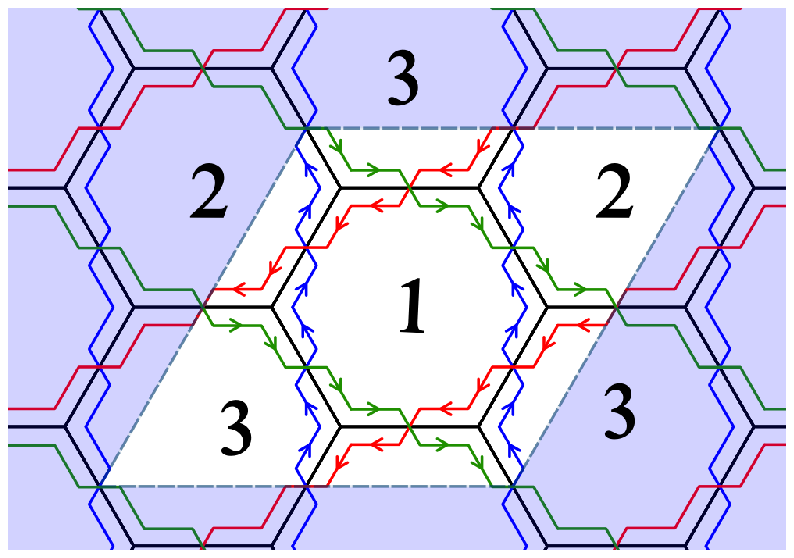}\\[0.5cm]
\hline
{\bf Quiver diagram:}\\
\includegraphics[width=0.25\textwidth]{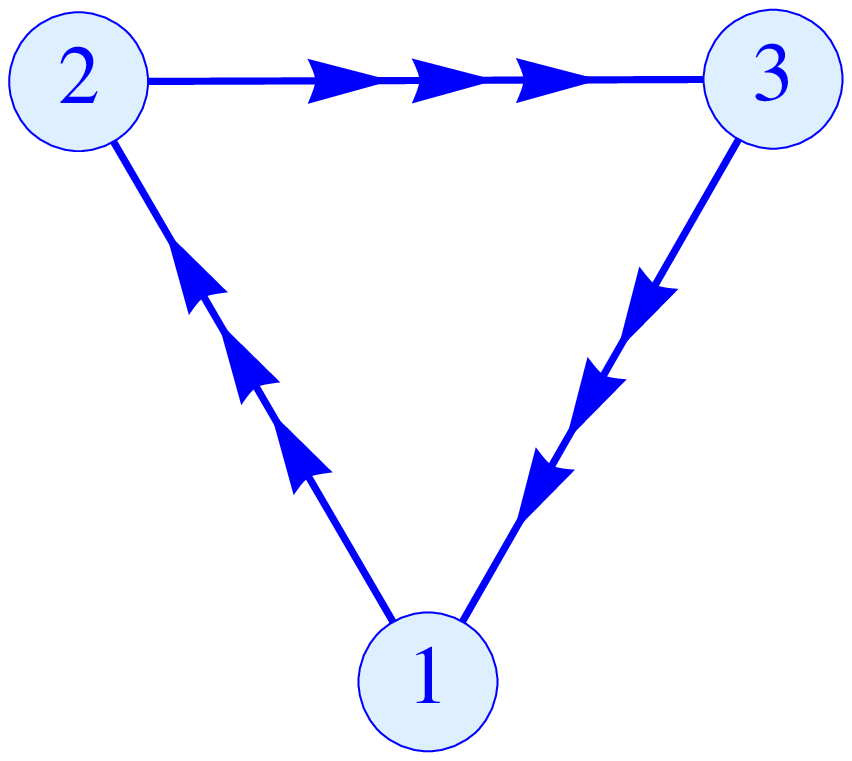}\\
\end{tabular}
&
\begin{tabular}{c}
Factor $A$ $(0,-1):$\\
\includegraphics[width=0.24\textwidth]{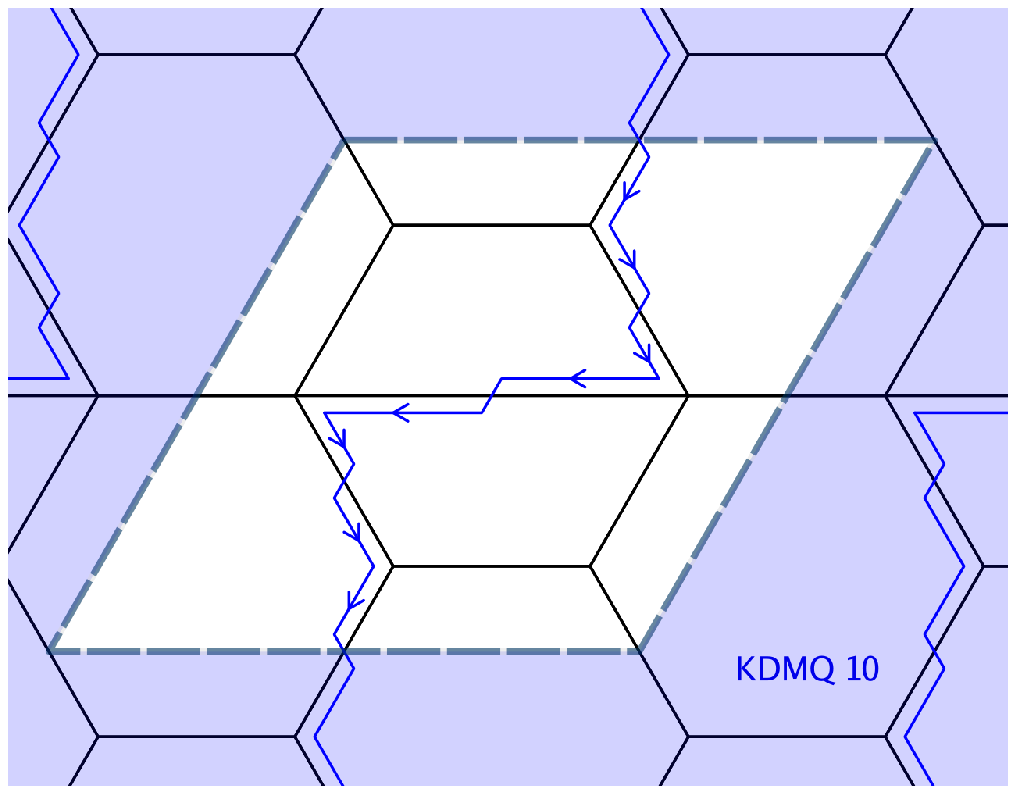}\\
Factor $B$ $(1,-1):$\\
\includegraphics[width=0.24\textwidth]{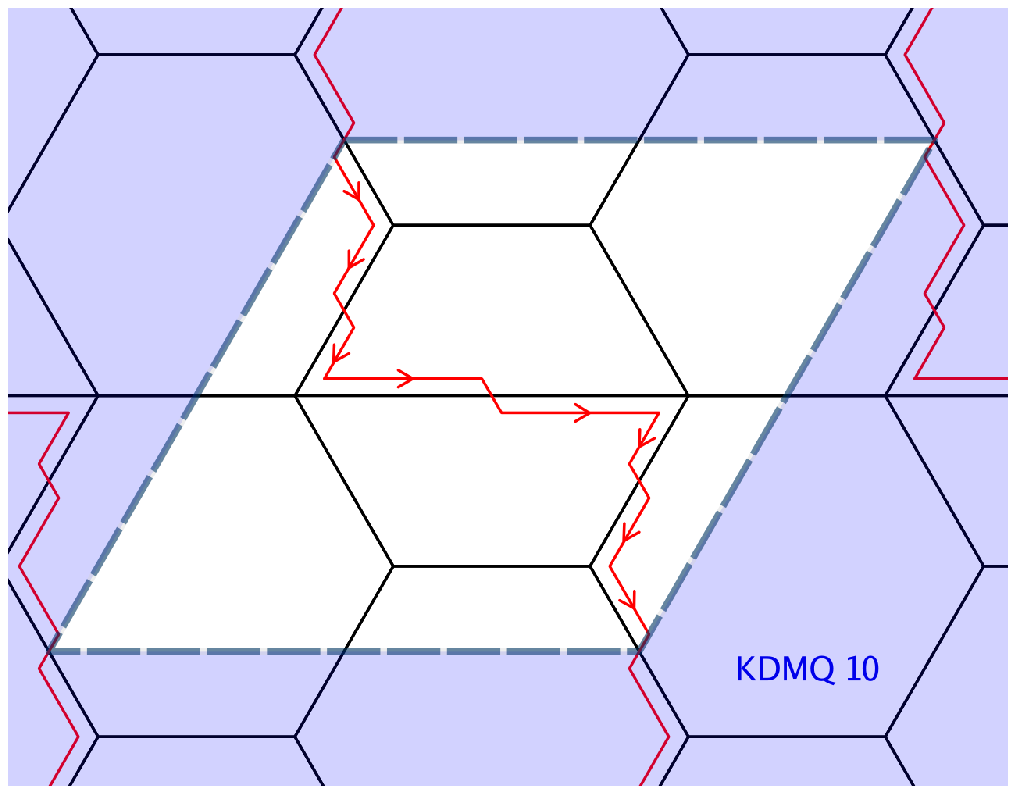}\\
Factor $C$ $(-2,1):$\\
\includegraphics[width=0.24\textwidth]{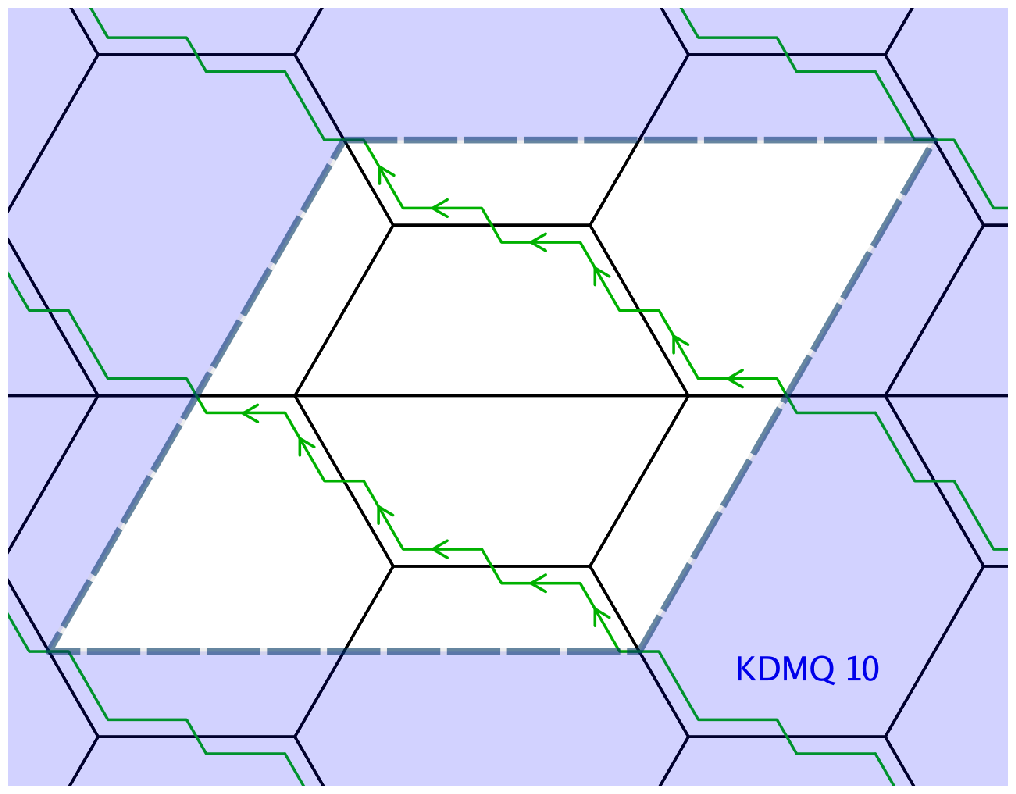}\\
Factor $D$ $(1,1):$\\
\includegraphics[width=0.24\textwidth]{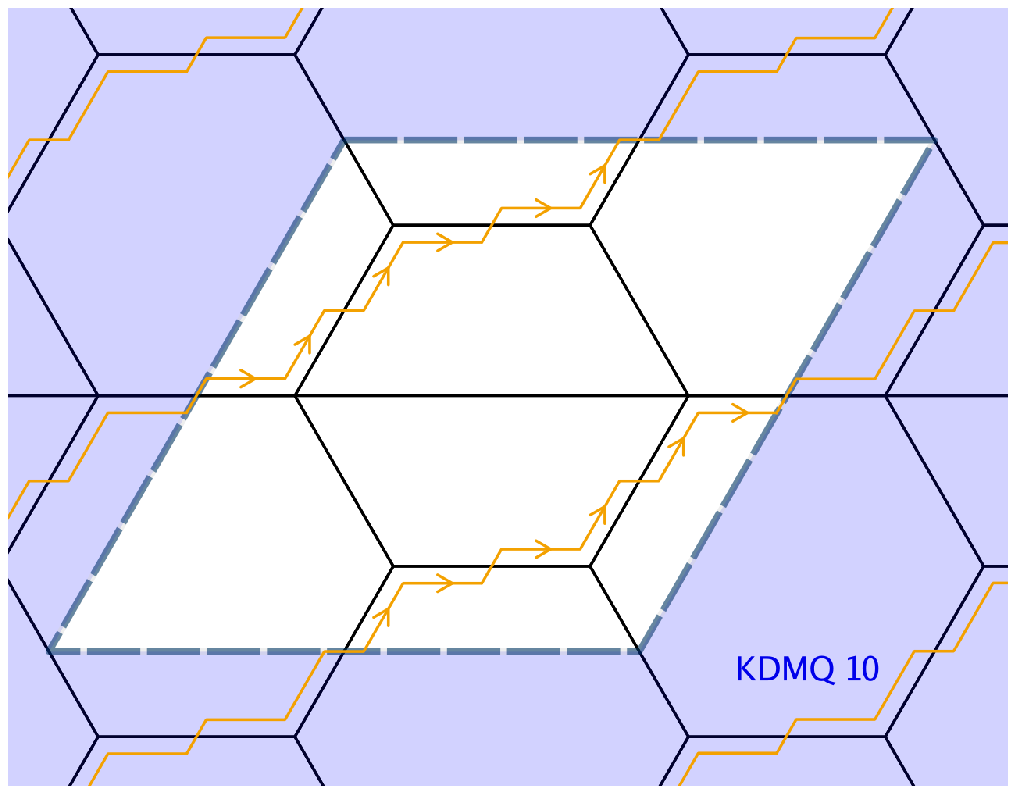}\\
\end{tabular}
&
\begin{tabular}{c}
{\bf The Toric Diagram and $(p,q)$-web}\\
\includegraphics[width=0.24\textwidth]{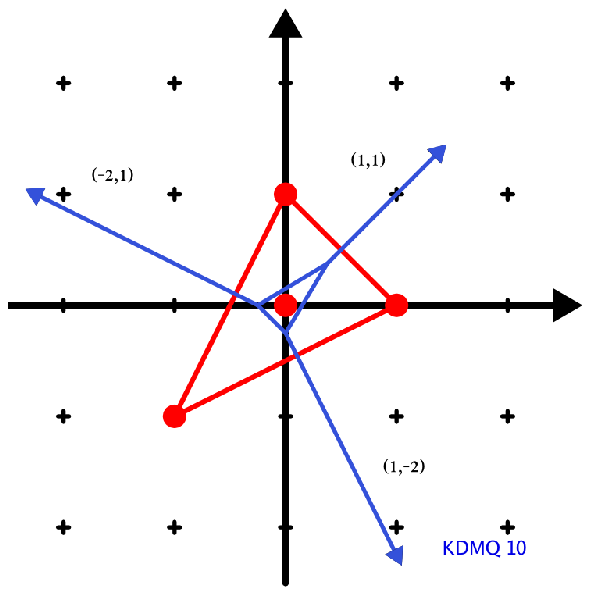}\\[0.5cm]
\hline
{\bf Complete Dimer:}\\
\includegraphics[width=0.35\textwidth]{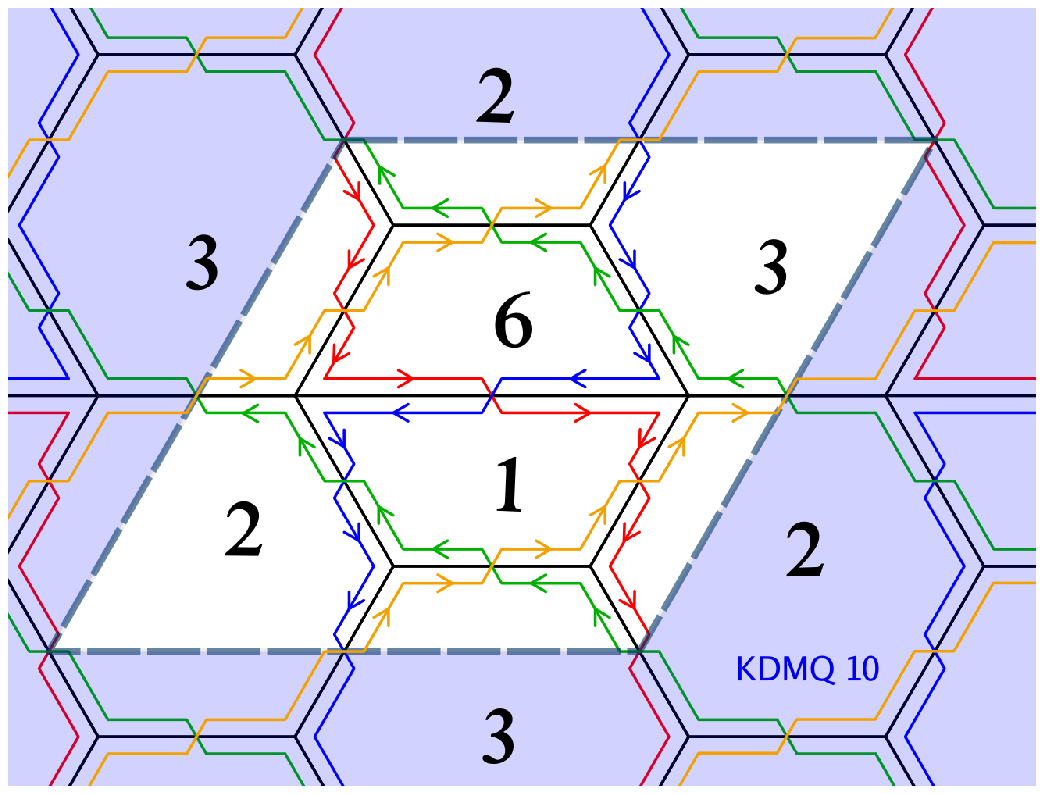}\\[0.5cm]
\hline
{\bf Quiver diagram:}\\
\includegraphics[width=0.25\textwidth]{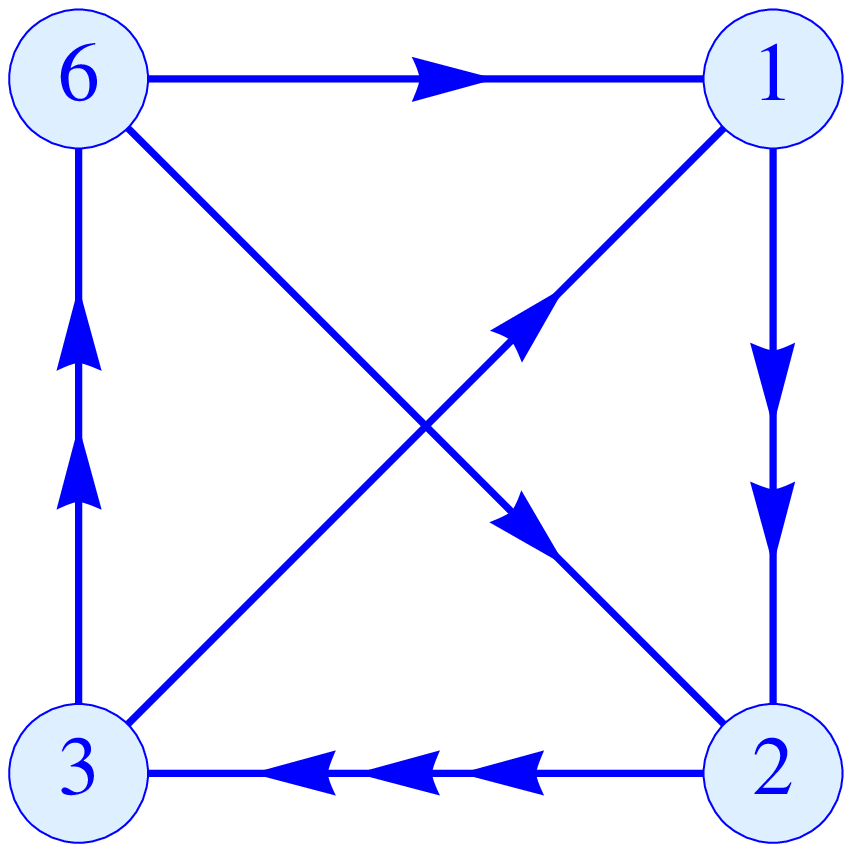}\\
\end{tabular}
\\
\hline
\label{fig:dp0dp1}
\end{tabular}
\end{landscape}
\setlength{\topmargin}{-0.3cm}
\setlength{\textheight}{23.2cm}

However, it should be noted that there can be many different gauge theories which correspond to a given singularity, all of which are connected by Seiberg duality \cite{0109053,0109063}. The class of theories where the ranks of gauge groups are equal are known as the toric phases of the singularity. There can be more than one toric phase associated to any given singularity; we will come back to this feature in the context of constraining the number of families in Section 3.

\section{Algorithmic view on gauge theories of toric singularities}
\label{sec:gulotta}

The first algorithmic way to construct the gauge of a toric singularity was presented in~\cite{0003085}, and the dimer interpretation was introduced in~\cite{0503149}. Recently, Gulotta~\cite{gulotta} presented an extremely efficient inverse algorithm to construct the dimer directly which allows to construct the gauge theory and provides a visual perspective on many properties of the gauge theory. In this section we provide a review of this algorithm, as we shall make heavy use of it in the following sections. For proof and further details we refer the reader to the original paper~\cite{gulotta}.

The starting point for Gulotta's algorithm is the toric diagram of the singularity of interest. When rotated by 90 degrees, the slopes of the edges of the rotated toric diagram are equal to the slopes of the dual web diagram. One then embeds the rotated toric diagram into the minimal rectangular toric diagram into which it fits. This rectangular toric diagram corresponds geometrically to an orbifold of the conifold $\mathcal{C}$. This process is illustrated for the case of the third del Pezzo surface $dP_3$ in Figure~\ref{oldinverse}. The left hand side of this figure shows the toric diagram for $dP_3$, and the right hand side shows the rotated toric diagram embedded into a minimal rectangular grid.
The philosophy of Gulotta's algorithm is to obtain the dimer and gauge theory of interest by partial resolution of this orbifold of the conifold.
\begin{center}
\begin{tabular}{c c}
\includegraphics[width=0.3\textwidth]{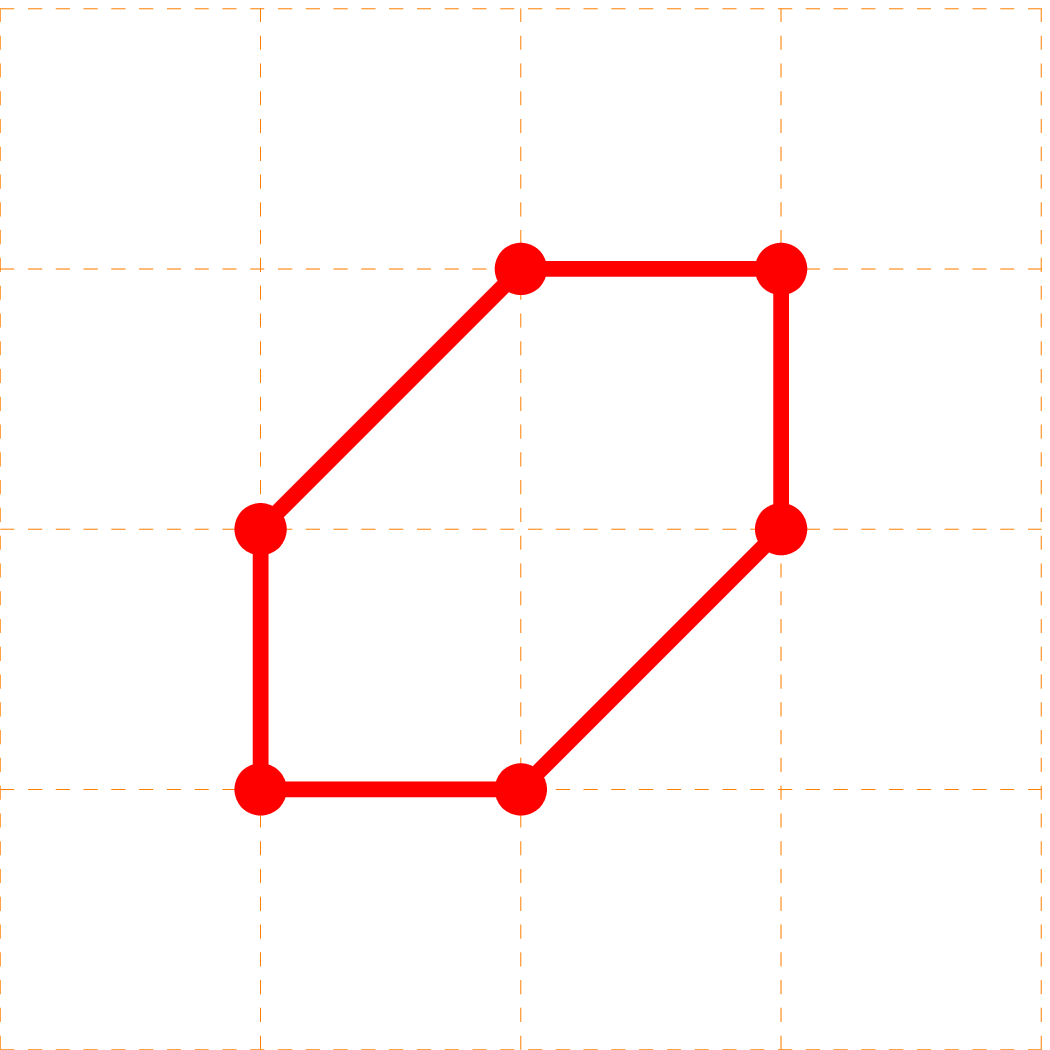}
& \includegraphics[width=0.3\textwidth,angle=90]{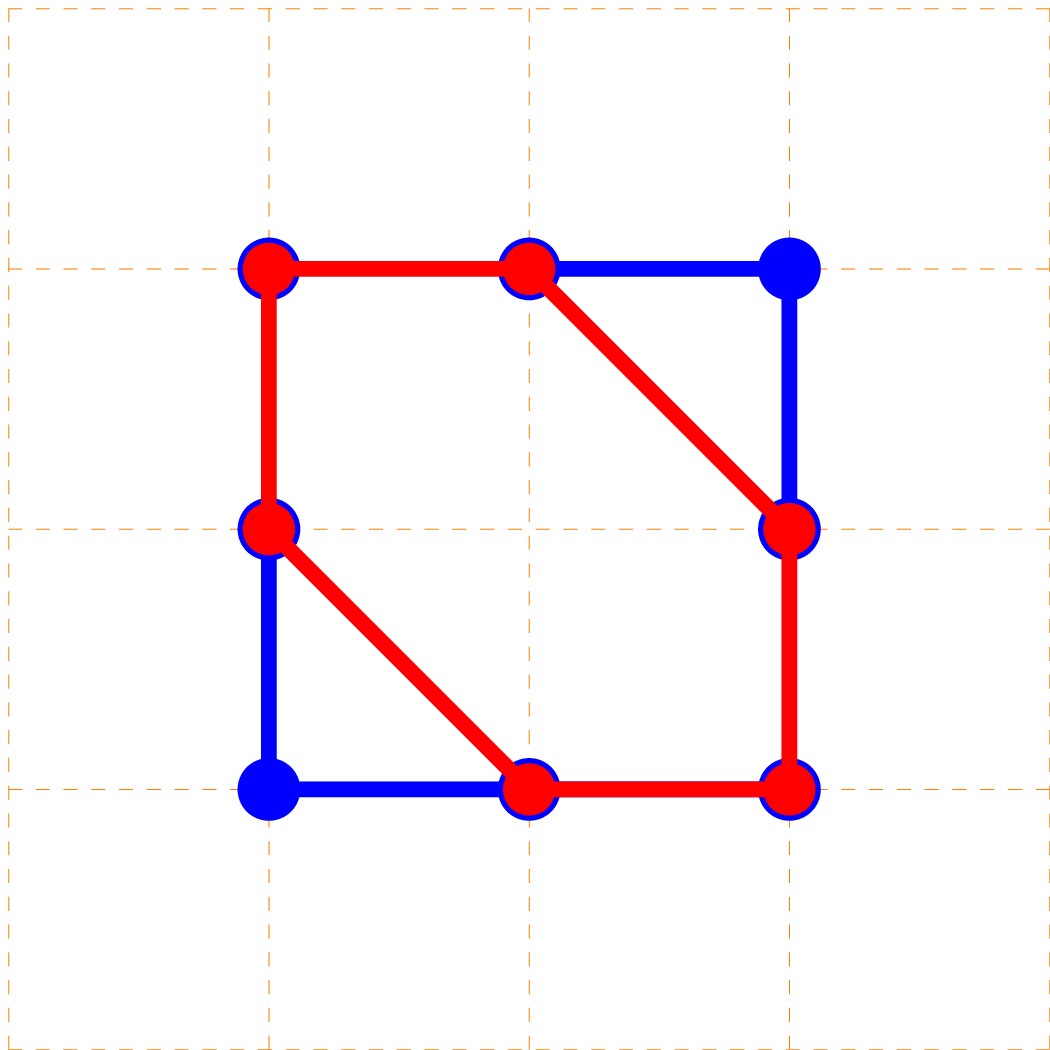}\\
\end{tabular}
\captionof{figure}{\footnotesize{ {\bf LHS}: The toric diagram for the third del Pezzo surface $dP_3$. {\bf RHS}: The toric diagram rotated by 90 degrees and embedded into its minimal rectangular diagram.}\label{oldinverse}}
\end{center}
\newpage
Since the rectangular toric diagram plays a central role in the algorithm, we would like to describe the associated dimer and gauge theory in some detail.
An $n\times m$ grid corresponds to the $\mathbb{Z}_{n}\times \mathbb{Z}_{m}$ orbifold of the conifold  ${\cal C}/(\mathbb{Z}_{n}\times \mathbb{Z}_{m})$.
The dimer associated to the $n\times m$ rectangular diagram consists of $n$ $(1,0)$ and $(-1,0)$ horizontal paths, and $m$ $(0,1)$ and  $(0,-1)$ vertical zigzag paths, where the zigzag path $(a,b)$ wraps the torus $a$ times horizontally (left to right) and $b$ times vertically (bottom to top).
The zigzag paths are drawn in alternating fashion such that no paths with the same winding numbers are next to each other, e.g.~alternating between $(1,0)$ and $(-1,0)$ for the horizontal paths.
 In this dimer there are two types of faces: one corresponding to gauge groups and the other to superpotential terms. 
They are distinguished by the fact that the zigzag paths surround superpotential faces in a clockwise or anti-clockwise fashion, whereas for the gauge group faces the zigzag paths clash. 
The intersection of two zigzag paths corresponds to bi-fundamental matter charged under the two gauge groups at the intersection, as in Figure~\ref{fig:zigzagmatter}.
Finally, the different orientations in the superpotential faces correspond to different signs in the superpotential terms.
Figure~\ref{gridexample} shows the toric diagram, dimer and quiver associated with a $2\times 3$ rectangular grid to illustrate these points.
\begin{center}
\begin{tabular}{c c c}
\begin{tabular}{c}\includegraphics[width=0.2\textwidth]{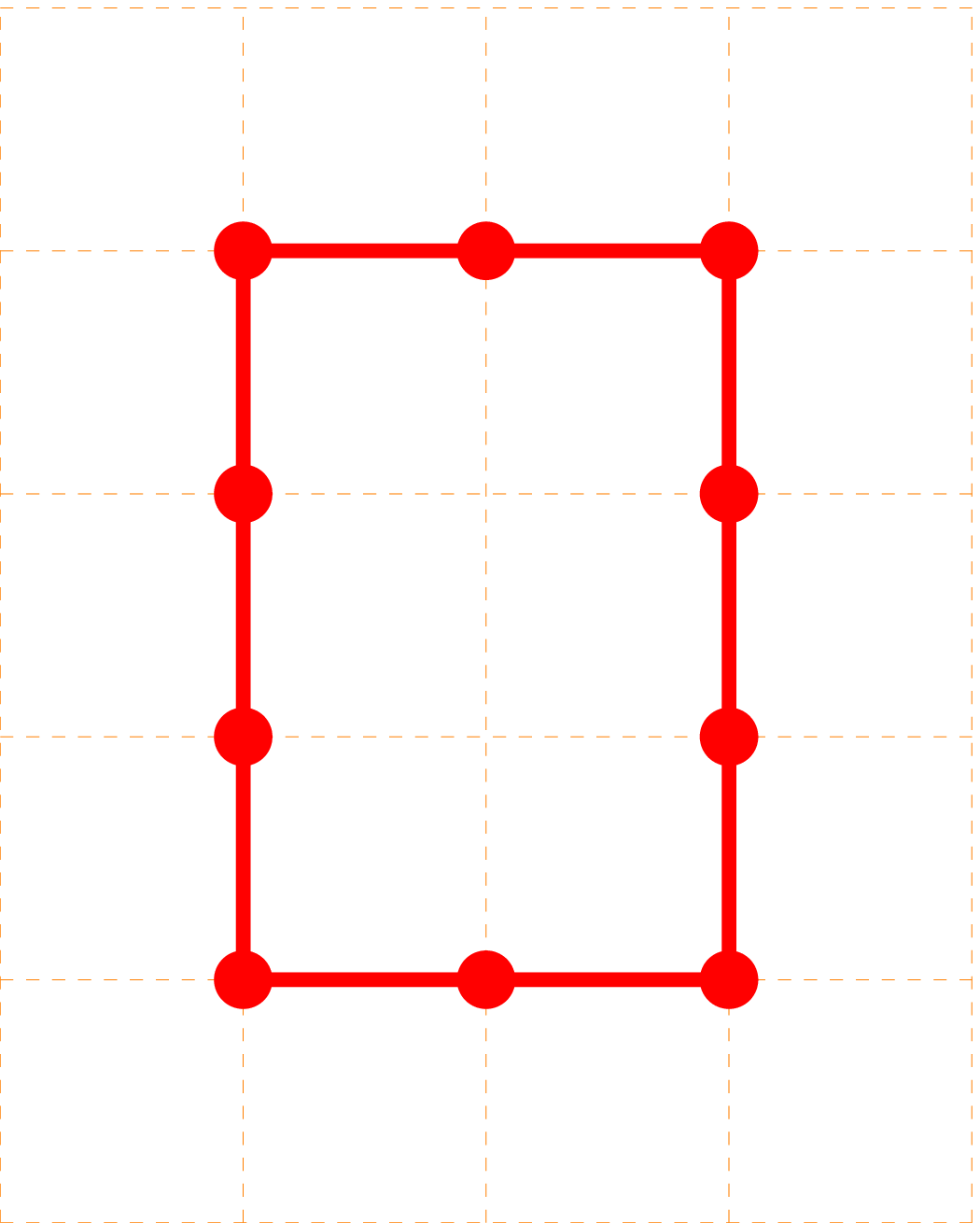} \end{tabular}&\begin{tabular}{c}
\includegraphics[width=0.38\textwidth]{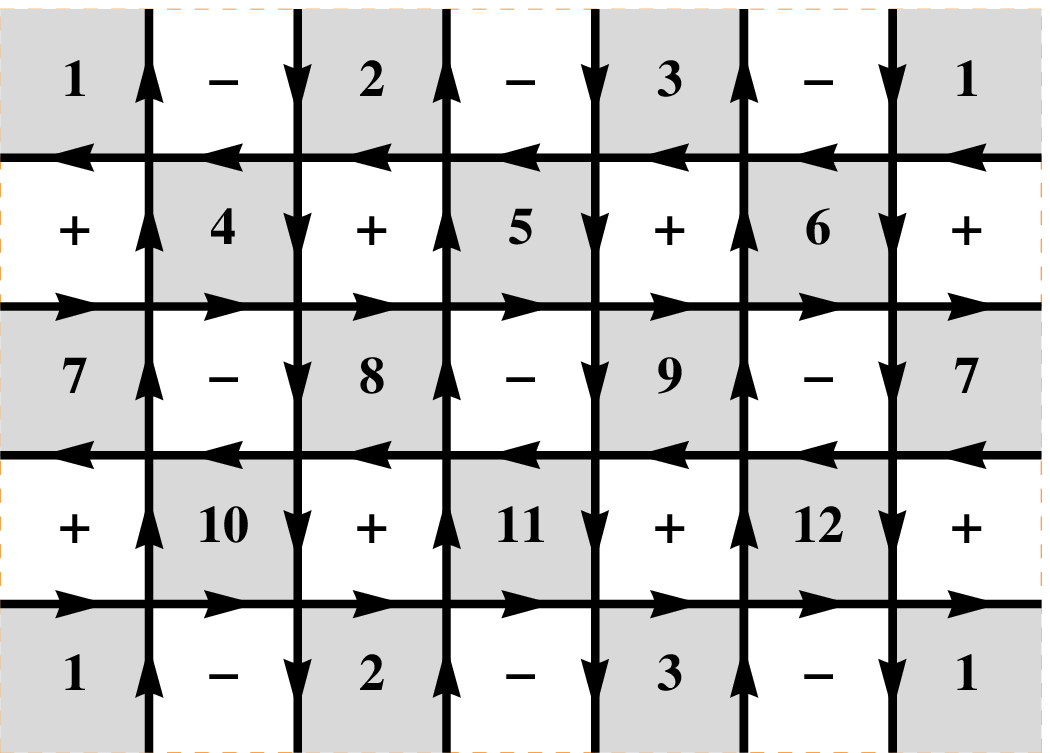} \end{tabular}& \begin{tabular}{c}\includegraphics[width=0.33\textwidth]{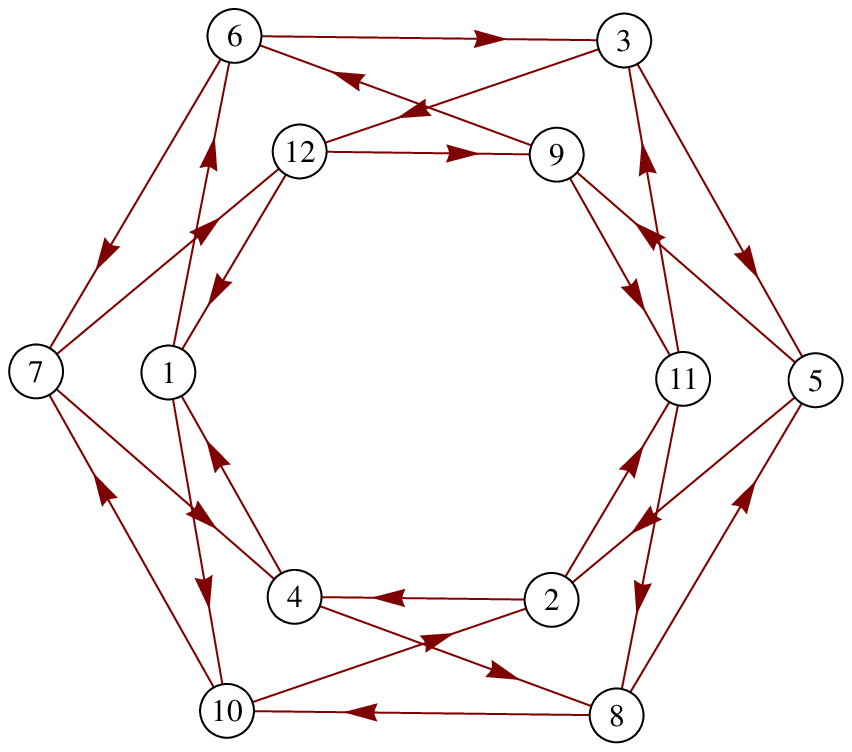}\end{tabular}\\
\end{tabular}
\captionof{figure}{\footnotesize{{\bf Left:} Rectangular toric diagram. {\bf Middle:} The associated chess-board dimer, gauge groups are labelled by numbers, superpotential terms correspond to faces that can be surrounded by cycles (the orientation determines the sign of the superpotential term). {\bf Right:} The associated quiver.}\label{gridexample}}
\end{center}
Starting from this rectangular grid, Gulotta's algorithm prescribes how to transform the dimer in a way which corresponds to collapsing cycles in the toric singularity, and cutting the toric diagram to the desired shape.
In this framework, collapsing cycles corresponds to merging zigzag paths in an appropriate fashion.
Cutting the toric diagram with a line of a given slope corresponds to making a zigzag path of precisely this slope. For example if we cut off the top left corner of the $2\times 3$ rectangular toric diagram above with a line of slope $1$, this corresponds to merging a $(0,1)$ and $(1,0)$ path in the dimer to give a $(1,1)$ zigzag path. We give precise details on how to merge zigzag paths in the next section, and now describe the algorithm to obtain the desired toric diagram by repeated cuttings.

\begin{enumerate}
\item We start with making all paths of slope $+1.$ If the edge removed from the toric diagram is an $(\pm n,\pm n)$ vector on the lattice, then $n$ $(1,1)$ or $(-1,-1)$ zigzag paths should be created.

\item We then cut the toric diagram further with lines of lower and higher slopes, e.g.  $1/2$ or $2.$ This corresponds to creating zigzag paths of higher winding numbers such as $(1,2).$ Paths of higher winding number are created by combining already existing paths. The order in which these paths are made is  given by the Farey tree, which is shown in Figure~\ref{fig:farey}.

\end{enumerate}
\begin{center}
\includegraphics[width=0.3\textwidth]{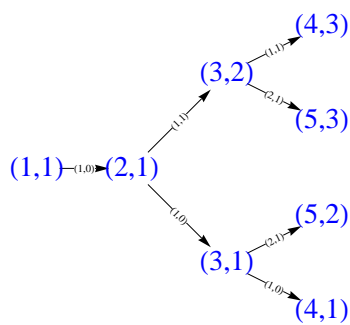}
\captionof{figure}{\footnotesize{The Farey tree tells the order in which zigzag paths are made. The part of the tree displayed shows the paths with slope $<1,$ reversing the winding numbers gives the paths with slope $>1.$}\label{fig:farey}}
\end{center}
\begin{enumerate}
  \setcounter{enumi}{2}
\item After cutting the toric diagram with all positive slopes, we cut the edges with negative slopes in the same fashion.
\end{enumerate}
This procedure is continued until the desired toric diagram and dimer are obtained.
We now return to the issue of merging zigzag paths.

\subsection{Merging zigzag paths}
\label{sec:gulotta2}

 Let us start with the example of creating a $(1,1)$ path from the $2\times 2$ grid. Since merging paths is a local operation, one can think of this grid as being embedded into any larger grid without any loss of generality.
\begin{center}
\begin{tabular}{c c}
\includegraphics[width=0.3\textwidth]{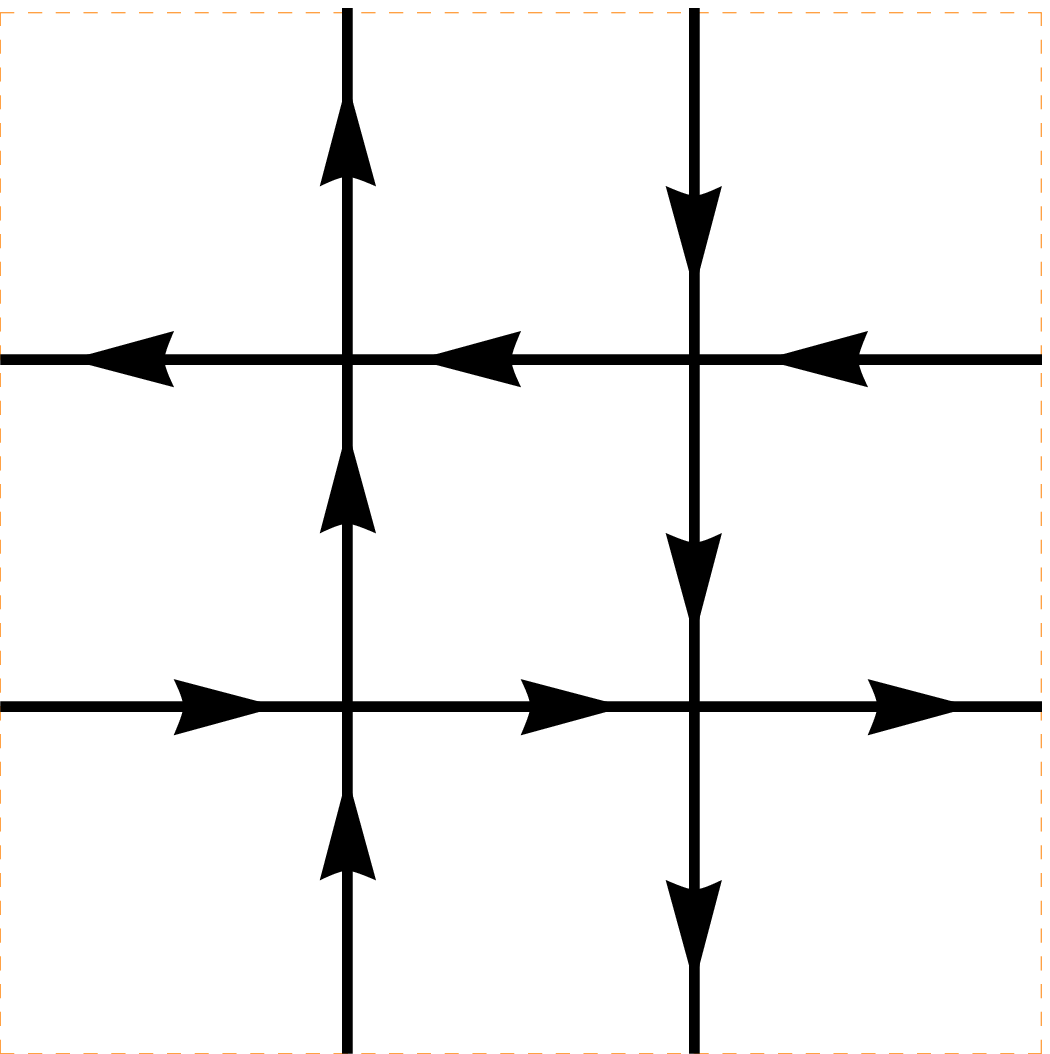} & \includegraphics[width=0.3\textwidth]{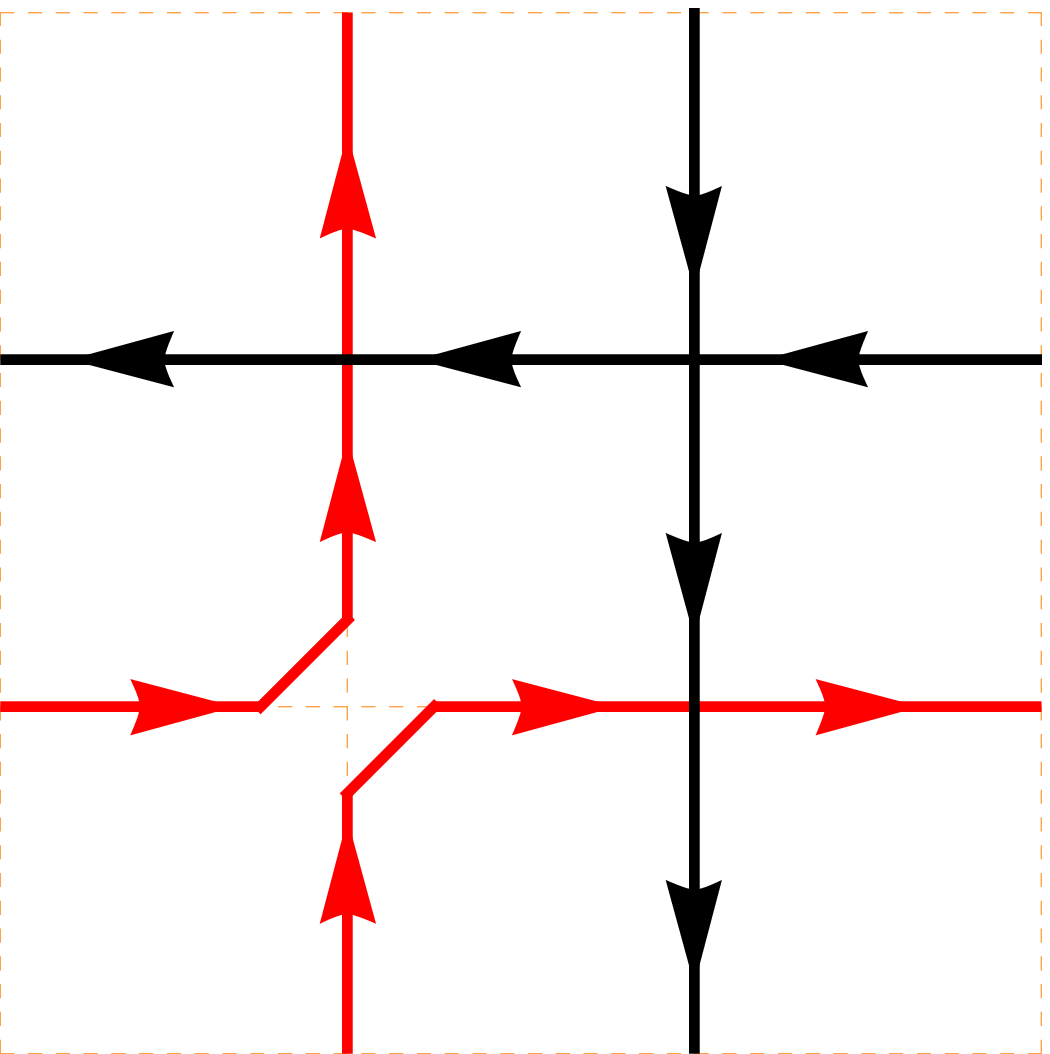}\\
\end{tabular}
\captionof{figure}{\footnotesize{Merging a $(1,0)$ and a $(0,1)$ zigzag path creates a $(1,1)$ path.
}\label{fig:11path}}
\end{center}
Combining the $(1,0)$ and $(0,1)$ paths gives the desired $(1,1)$ path, as shown in Figure~\ref{fig:11path}. In a similar manner one can create paths with winding number $(1,-1),$ $(-1,1)$ and $(-1,-1).$
Paths with higher winding numbers can be obtained by combining previously created paths with further paths of winding number $(0,1)$ or $(1,0)$. 

Given two zigzag paths a fundamental topological quantity is their oriented intersection number. For two paths of winding numbers $(m_a,n_a)$ and $(m_b,n_b)$ this is $I_{ab} = m_a n_b - n_a m_b$. On the other hand the total number of intersections is not topological, and can change by local deformations of the paths. Note that the number of intersections has to be at least as large as the number of oriented intersections. If the number of unsigned intersections is greater than the oriented intersection number, we say that the two paths have additional crossings. Gulotta's algorithm requires that while merging paths we do not generate any additional crossings. Thus while merging paths one always performs local operations so as to get rid of any additional crossings.

As an example, we show in Figure~\ref{fig:addcrossing} additional crossings that can arise while merging $(1,0)$, $(0,1)$, $(-1,0)$ and $(0,-1)$ paths to form a $(1,1)$ and a $(-1,-1)$ path. 
\begin{center}
\includegraphics[width=0.3\textwidth]{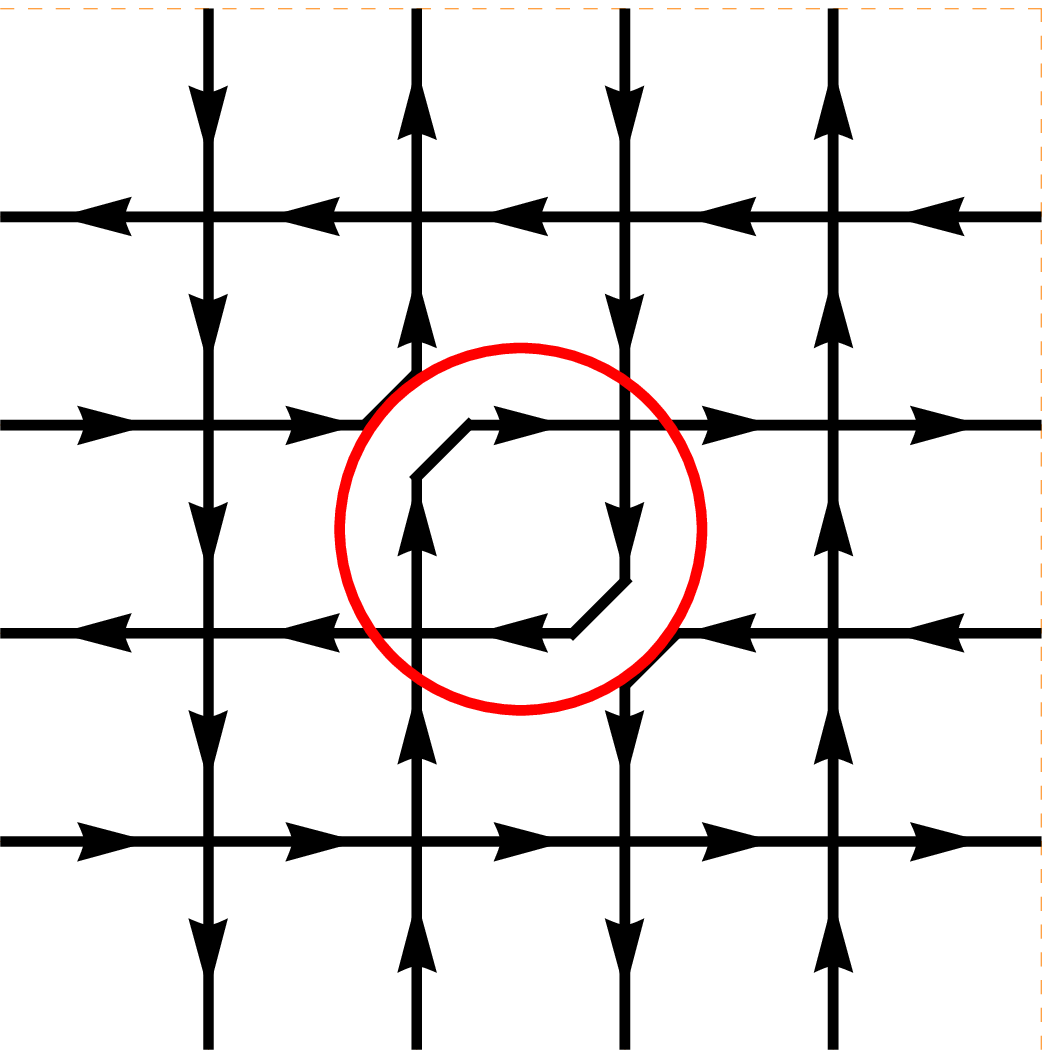}
\captionof{figure}{\footnotesize{The additional crossing corresponding to a mass term when a $(1,1)$ and a $(-1,-1)$ path are created after each other.}\label{fig:addcrossing}}
\end{center}
In this case, to avoid additional crossings one constructs the $(1,1)$ and $(-1,-1)$ paths simultaneously as shown in Figure~\ref{fig:addcross2}.
\begin{center}
\begin{tabular}{c c}
 \includegraphics[width=0.3\textwidth]{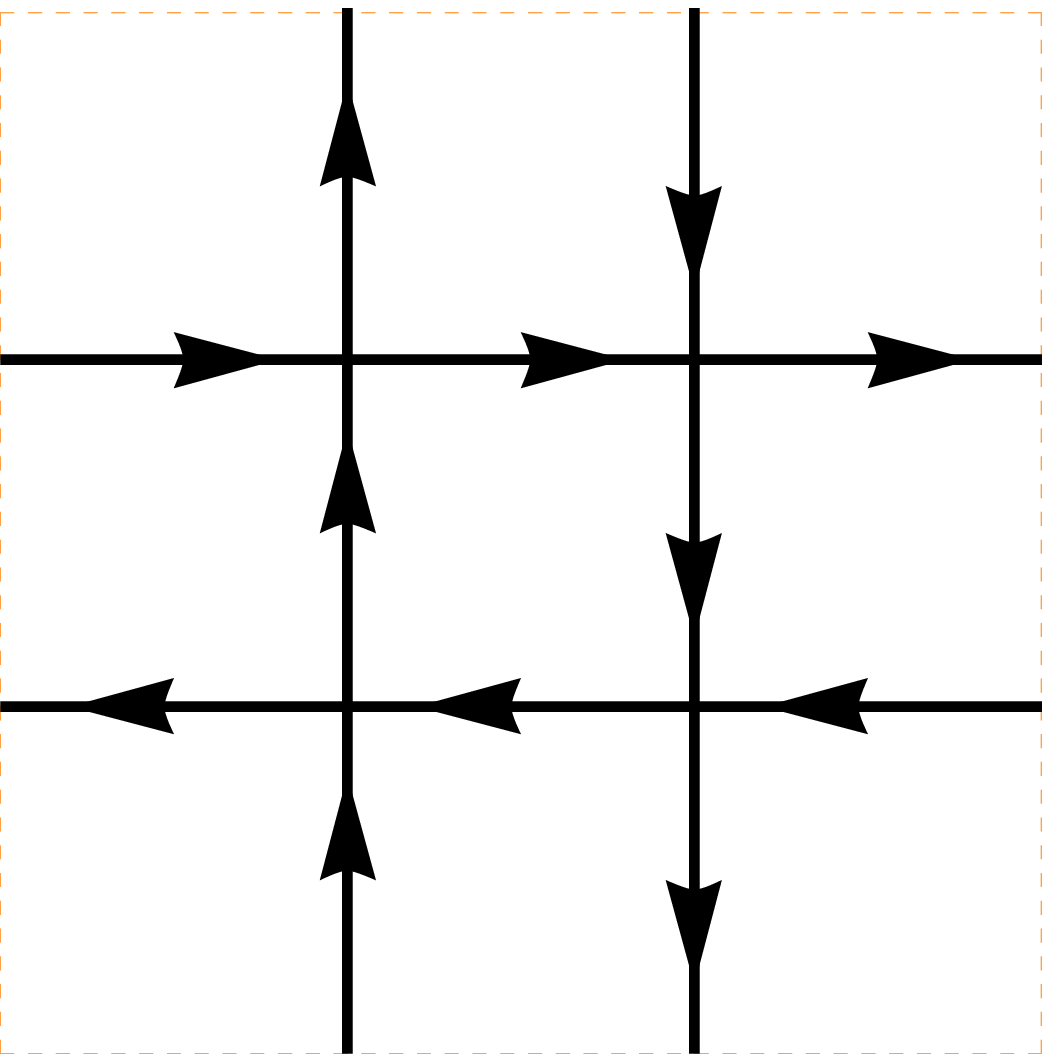}  & \includegraphics[width=0.3\textwidth]{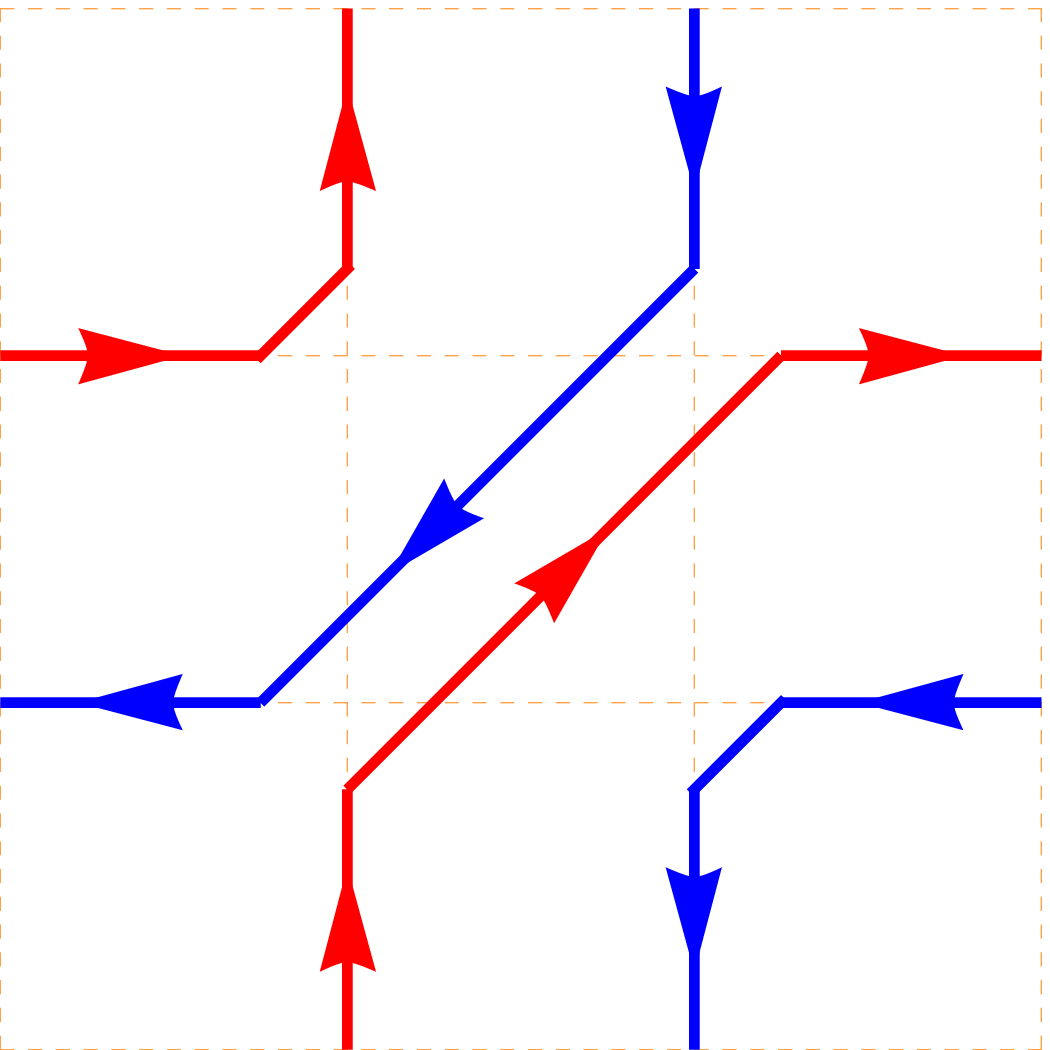} \\
\end{tabular}
\captionof{figure}{\footnotesize{Creating a pair of $(1,1)$ and $(-1,-1)$ paths simultaneously to avoid additional crossings.}\label{fig:addcross2}}
\end{center}
This crossing has the following interpretation. 
Recall that the dimer represents the massless spectrum of the theory. The circled region in Figure~\ref{fig:addcrossing} gives a mass term in the superpotential. Integrating out the massive fields gives the dimer shown on the right hand side of Figure~\ref{fig:addcross2}.

However, not all additional crossings have the interpretation as mass terms in the superpotential.
The presence of such additional crossings can render the dimer inconsistent.
For example, in the case of multiple paths of one type where $I_{ab}=0$, it is essential to avoid additional crossings since otherwise the number of gauge groups does not correspond to twice the area of the toric diagram. 
Whether or not a dimer with additional crossings is consistent needs to be decided on a case by case basis.\footnote{Examples of consistent dimers with additional crossings are the second toric phase of the zeroth Hirzebruch surface $\mathbb{F}_0$ and phases 1 and 3 of the third del Pezzo surface $dP_3$, which are discussed in Section 3.3.} For these reasons, Gulotta's algorithm requires the removal of all additional crossings, and provides three basic operations for merging paths which guarantee the absence of such crossings. 

\begin{itemize}
\item {\bf Operation I} involves making $n$ $(\pm 1,\pm 1)$ paths.  One first creates the paths and then moves them apart from each other to avoid additional crossings. Some further discussion on removing additional crossings in Operation I can be found in Appendix~\ref{sec:moredimer}.
\item {\bf Operation II} creates $n$ pairs of $(1,1)$ and $(-1,-1)$ paths. We saw in Figure~\ref{fig:addcross2} how to make one pair of such paths without additional crossings. 
Multiple pairs can be achieved by creating one pair and then placing the other pairs parallel to the one of the already created paths.
An example of this is shown in Figure~\ref{fig:op2} where the dimer of the toric diagram is obtained by creating two pairs of $(1,1)$ and $(-1,-1)$ paths.
\end{itemize}
\begin{center}
\begin{tabular}{c c c c}
\includegraphics[width=0.2\textwidth]{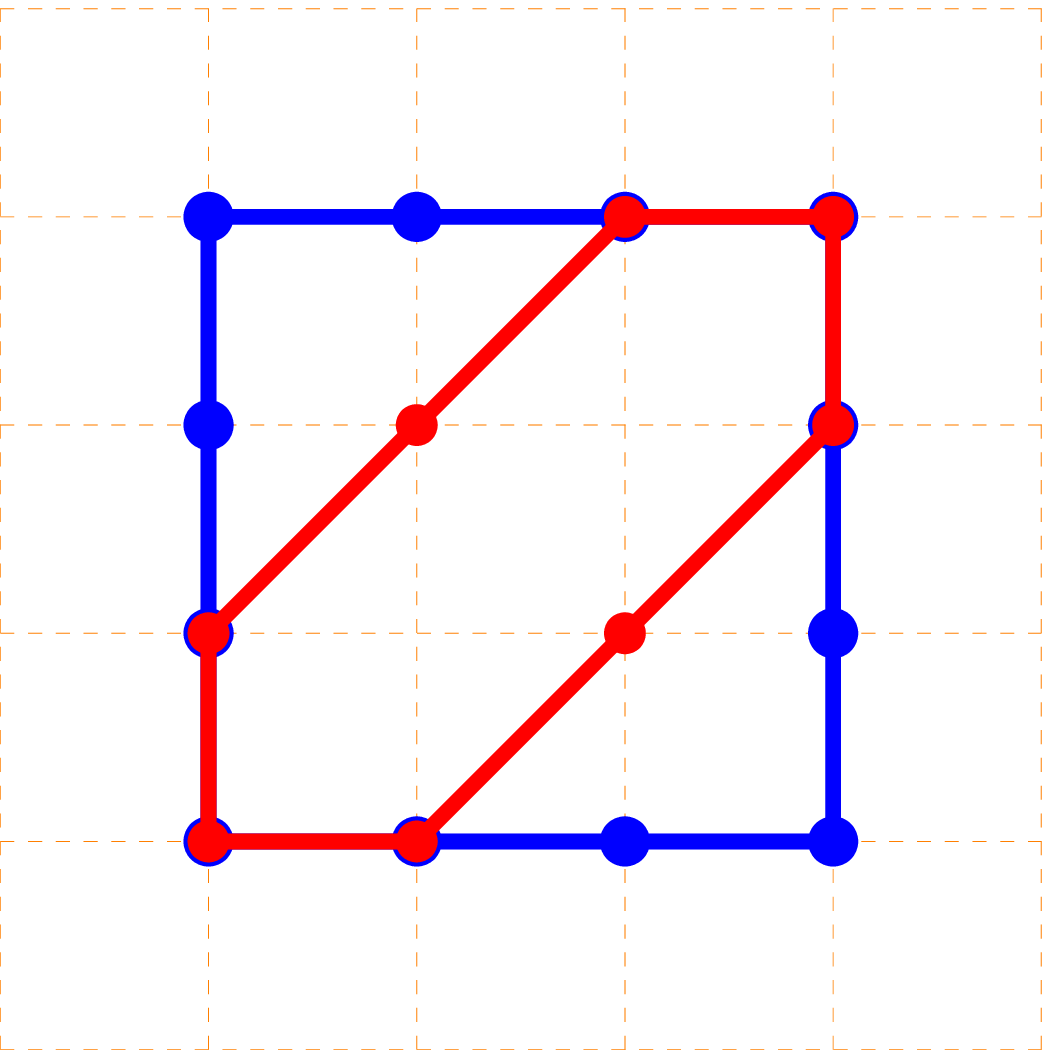} & \includegraphics[width=0.2\textwidth]{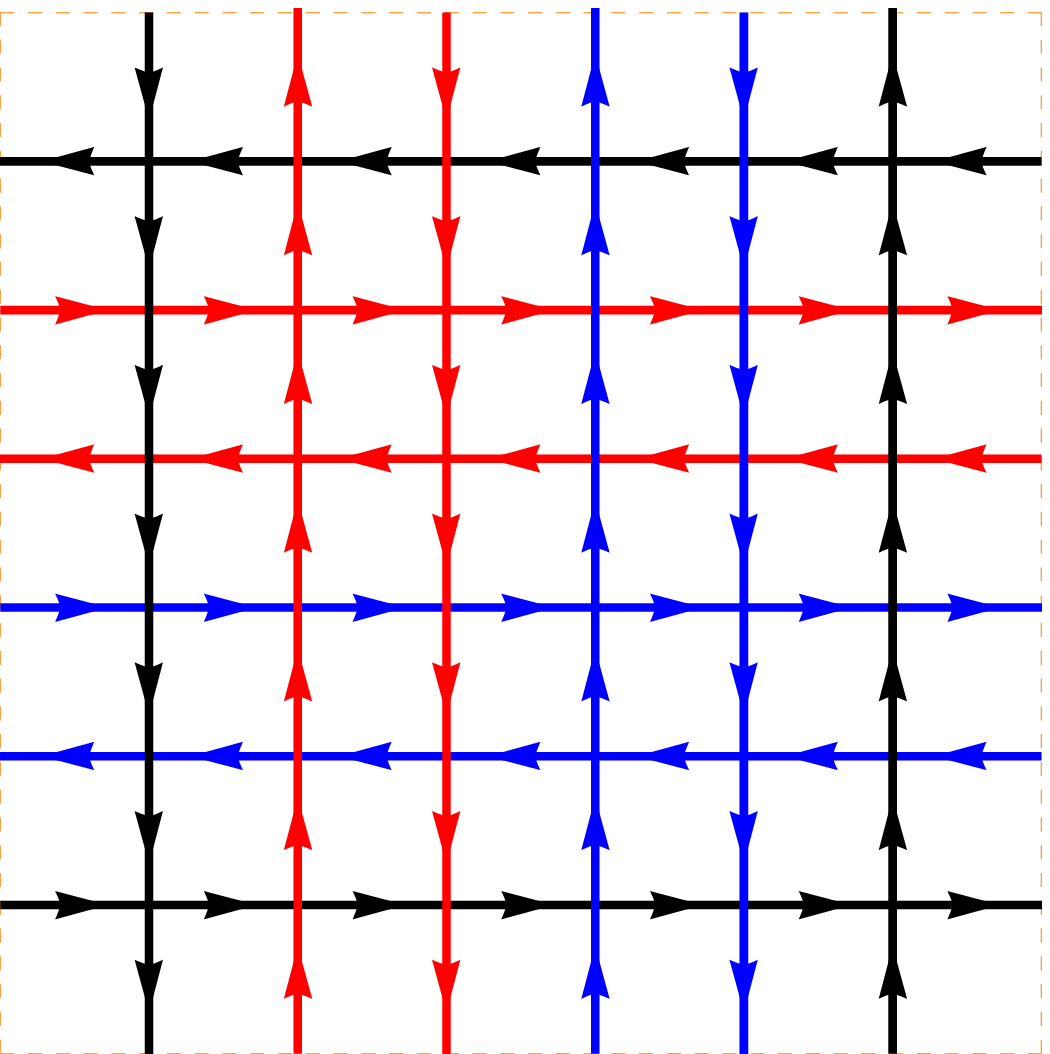}  & \includegraphics[width=0.2\textwidth]{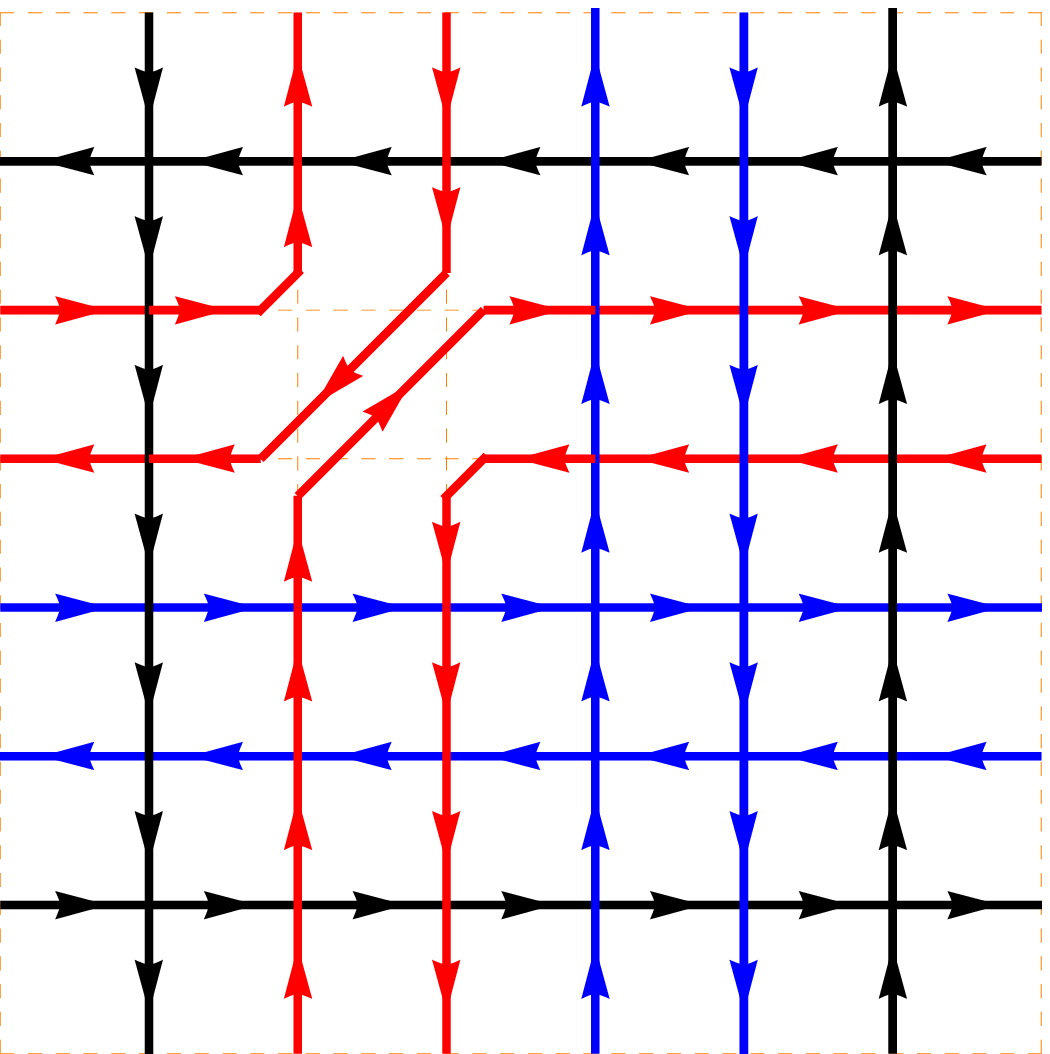} &\includegraphics[width=0.2\textwidth]{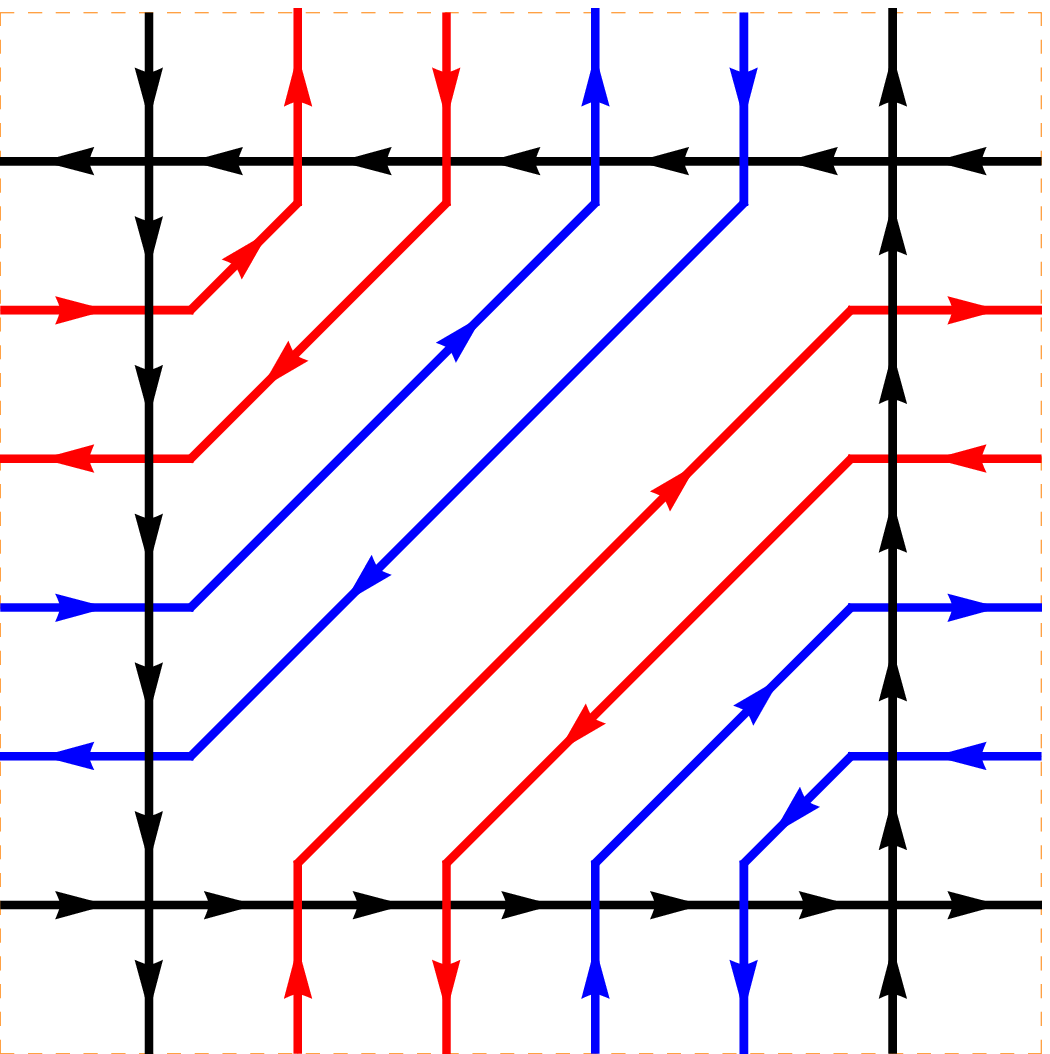} \\
\end{tabular}
\captionof{figure}{\footnotesize{Example of creating two pairs of $(1,1)$ and $(-1,-1)$ paths via Operation 2. 
The red pair is drawn first. Then we draw the second pair (shown in blue) parallel to the $(-1,-1)$ path of the first pair.}\label{fig:op2}}
\end{center}
\begin{itemize}
 \item  {\bf Operation III} involves creating  different numbers of  $(1,1)$ and  $(-1,-1)$ paths, $n$ and $m$, say. Suppose  $n-m>0$, then  $n-m$ $(1,1)$ paths are created with Operation I and the remaining pairs of $(1,1)$ and $(-1,-1)$ paths are then drawn as in Operation II parallel to one of the $(1,1)$ paths. If $n-m<0$, one creates $m-n$ $(-1,-1)$ paths and then  proceeds as before. An example of Operation III is shown in Figure~\ref{fig:op3}.
\end{itemize}
\begin{center}
\includegraphics[width=0.8\textwidth]{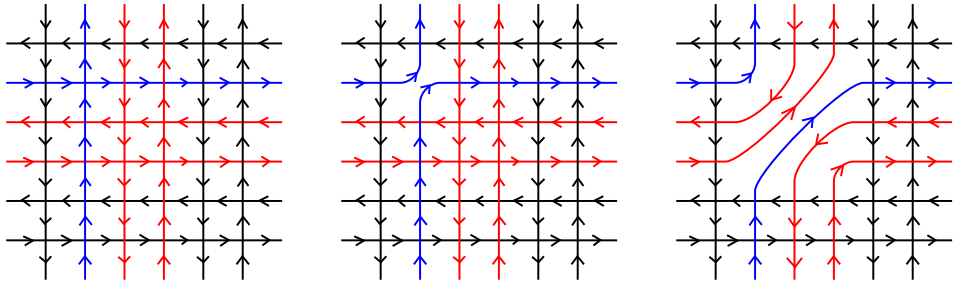}
\captionof{figure}{\footnotesize{Creating two $(1,1)$ paths and one $(-1,-1)$ path by Operation III. We firstly create one $(1,1)$ path (blue) and then create a pair of $(1,1)$ and $(-1,-1)$ paths parallel to the blue $(1,1)$ path.}\label{fig:op3}}
\end{center}

The one-to-one relationship between winding numbers of zigzag paths and slopes in the web-diagram is essential~\cite{0511063}. Violating this relationship will lead to a dimer that has no interpretation in terms of the geometry. This has an important consequence if one is trying to construct dimers by merging zigzag paths in a fashion which does not adhere to the prescriptions of Gulotta's algorithm. While merging zigzag paths one must not create paths which correspond to moving backwards in or out of the Farey tree. One implication of this is that zigzag paths are never self-intersecting.  For example, combining a $(2,1)$ and a $(-1,0)$ path corresponds to moving backwards in the Farey tree, while combining a $(2,1)$ and $(0,1)$ path moves out of the Farey tree. These features have been discussed in the context of $\mathbb{F}_0$ in \cite{0511063}.

\subsection{Application to toric del Pezzo surfaces}
\label{sec:toricapp}

To illustrate the operations of Gulotta's algorithm we discuss the familiar examples of the first three del Pezzo surfaces.
Here we are not strict about the order in which the paths are created, however it is easily checked that the results do not change, when following the algorithm strictly. 

We can embed the toric diagrams of the first four del Pezzo surfaces into the $2\times 2$ square, as shown in Figure~\ref{oldinverse} for $dP_3$.  
The dimer associated to this square is the chess-board with two paths of each type: $(0,1),$ $(0,-1),$ $(1,0),$ and $(-1,0).$ The dimer for $dP_3$ is obtained by creating a pair of $(1,-1)$ and $(-1,1)$ paths, where we use Operation II. This process is shown in Figure~\ref{fig:dp3dimer}. Removing additional crossings, as done in Operation II,  decreases the number of terms in the superpotential. 

\begin{center}
\begin{tabular}{c c c}
\begin{tabular}{c} \includegraphics[width=0.28\textwidth]{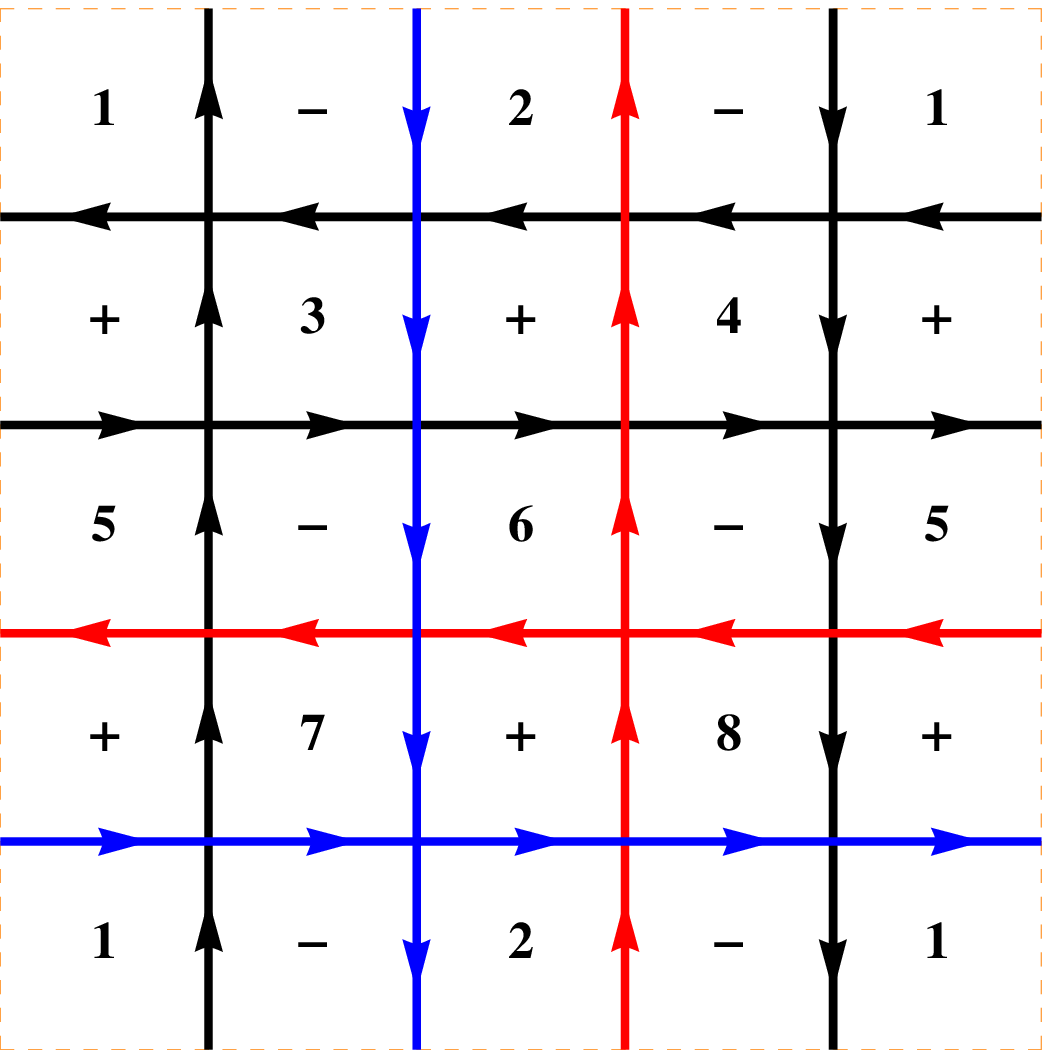}\end{tabular} &\begin{tabular}{c} \includegraphics[width=0.23\textwidth]{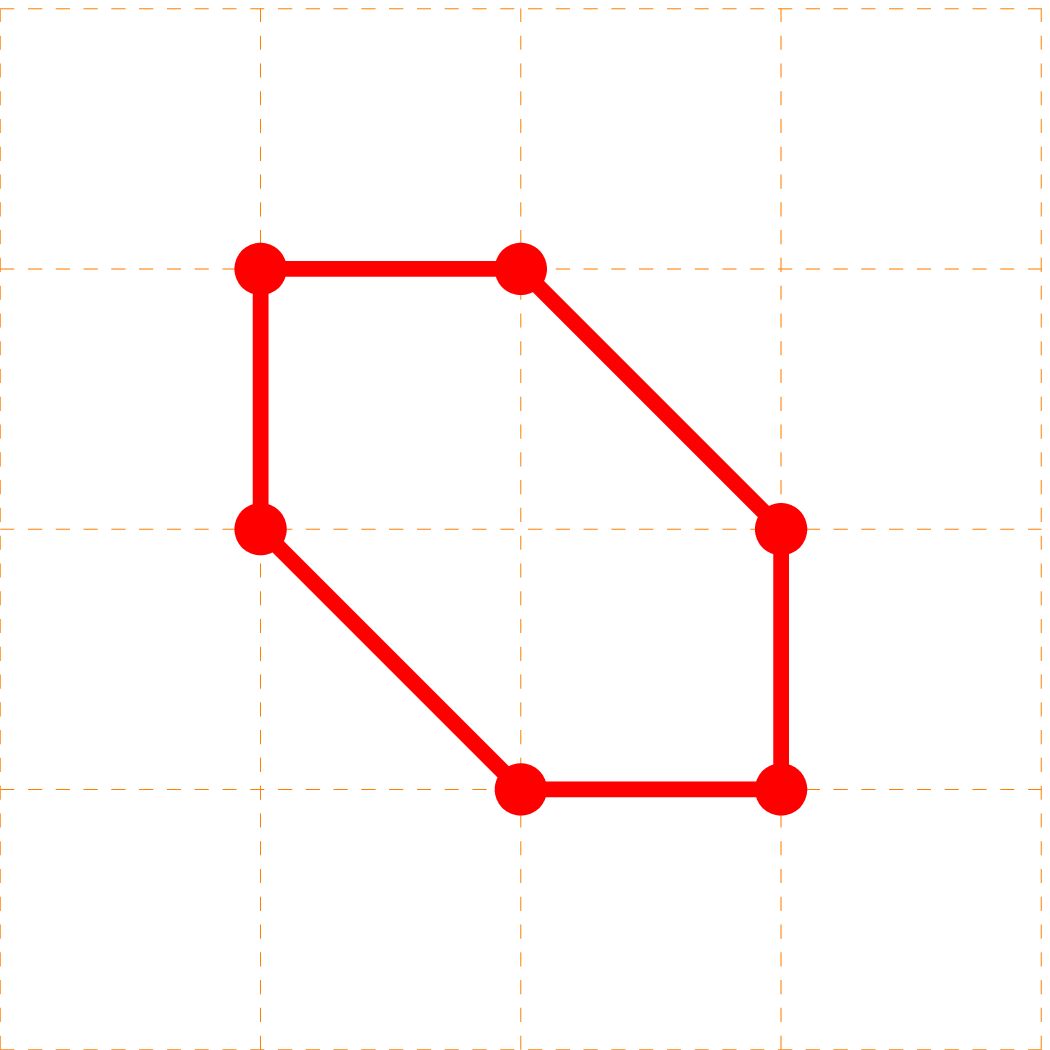}\end{tabular}& \begin{tabular}{c} \includegraphics[width=0.28\textwidth]{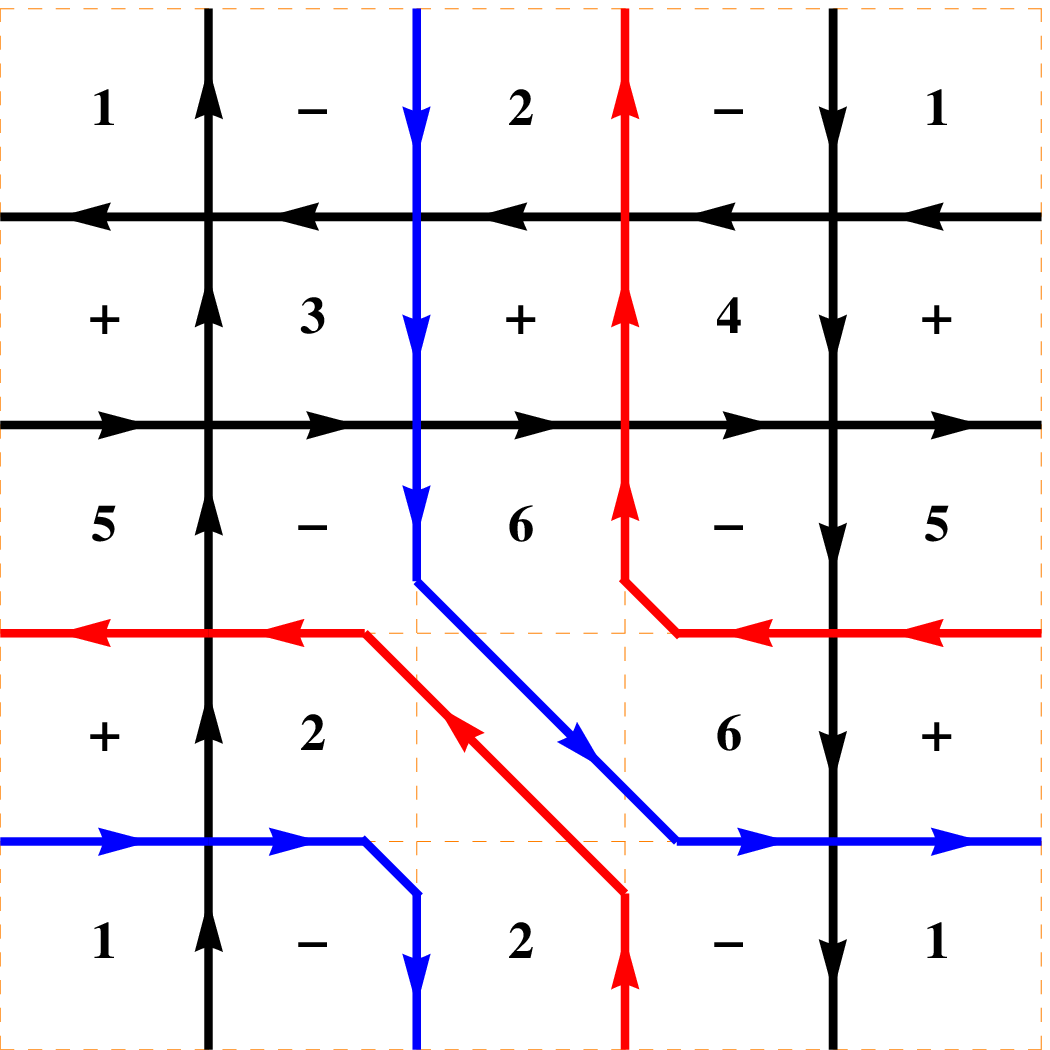}\end{tabular}
\end{tabular}
\captionof{figure}{\footnotesize{\textit{Left}: The conifold dimer with the paths to be merged shown in blue and red. \textit{Middle}: The toric diagram for $dP_3$. \textit{Right}: The $dP_3$ dimer with the $(1,-1)$ path in blue and the $(-1,1)$ path in red.}\label{fig:dp3dimer}}
\end{center}
The superpotential can be read off to be:
\begin{eqnarray}
\nonumber W_{dP_{3}}&=&-X_{12} Y_{31} Z_{23}-X_{45} Y_{64} Z_{56}+X_{45} Y_{31} Z_{14} \rho _{53}+X_{12} Y_{25} Z_{56} \Phi _{61}\\ \nonumber & &+X_{36} Y_{64} Z_{23} \Psi _{42}-X_{36}Y_{25} Z_{14} \rho _{53} \Phi _{61} \Psi _{42}\\ \vspace{14pt}
&=&\left(
\begin{array}{c}
 X_{45} \\
 Y_{25} \\
 Z_{23}
\end{array}
\right)\left(
\begin{array}{ccc}
 0 & Z_{14} \rho _{53} & -Y_{64} \\
 -Z_{14} \rho _{53} \Phi _{61} \Psi _{42} & 0 & X_{12} \Phi _{61} \\
 Y_{64} \Psi _{42} & -X_{12} & 0
\end{array}
\right)\left(
\begin{array}{c}
 X_{36} \\
 Y_{31} \\
 Z_{56}
\end{array}
\right).
\end{eqnarray}
Note that every field appears in exactly two terms in the superpotential, once with a positive sign and once with a negative sign. In fact, this is a general feature of toric singularities, and follows from the bipartiteness of the dimer~\cite{0504110j}.
We also write the superpotential in a matrix form, which we will use in Sections 4 and 5. The matrix will play the role of a Yukawa matrix, and the fields in the column vectors will be quarks.

In order to obtain $dP_2$ we join the remaining $(1,0)$ and $(0,1)$ path from $dP_3$, as shown in Figure~\ref{fig:dp2}.
\begin{center}
\begin{tabular}{c c c}
\begin{tabular}{c}\includegraphics[width=0.28\textwidth]{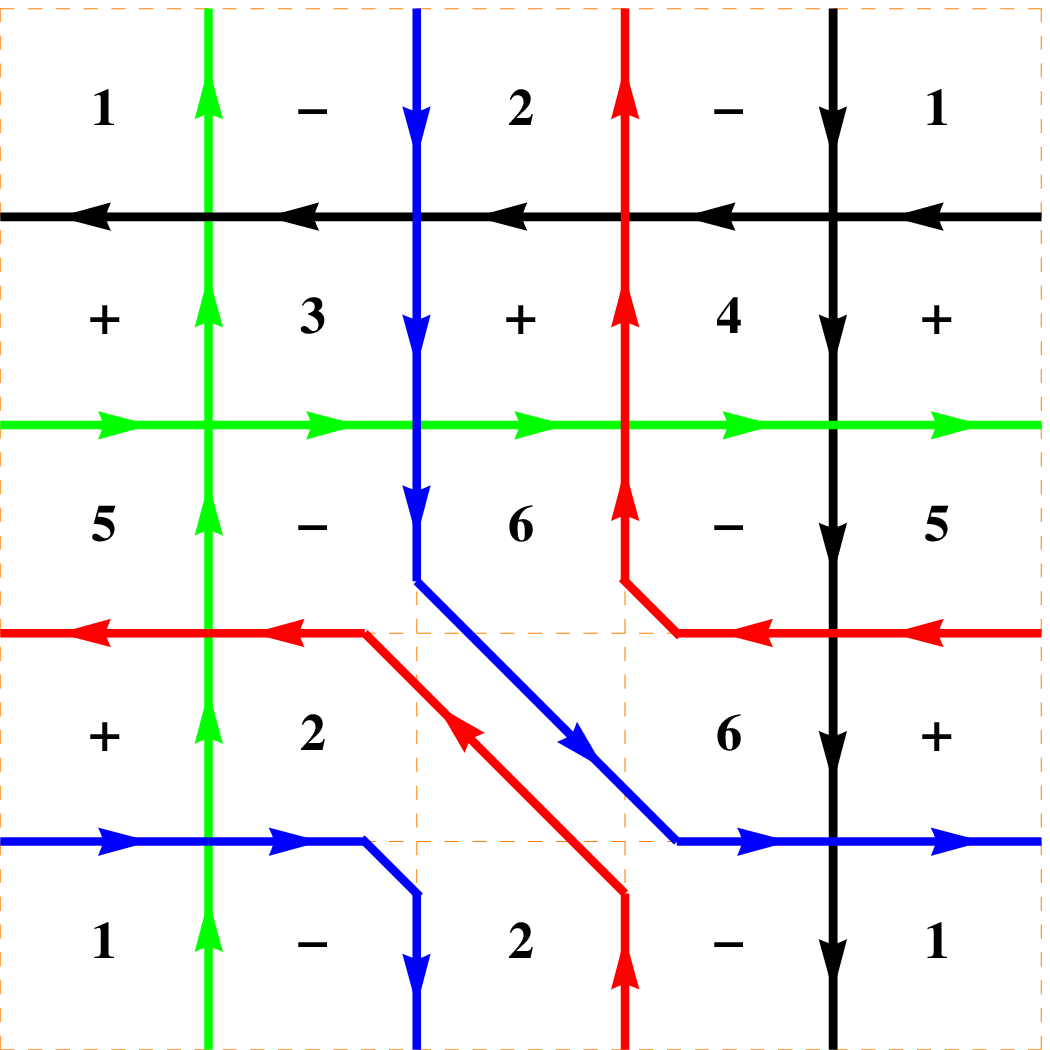}\end{tabular} & \begin{tabular}{c} \includegraphics[width=0.23\textwidth]{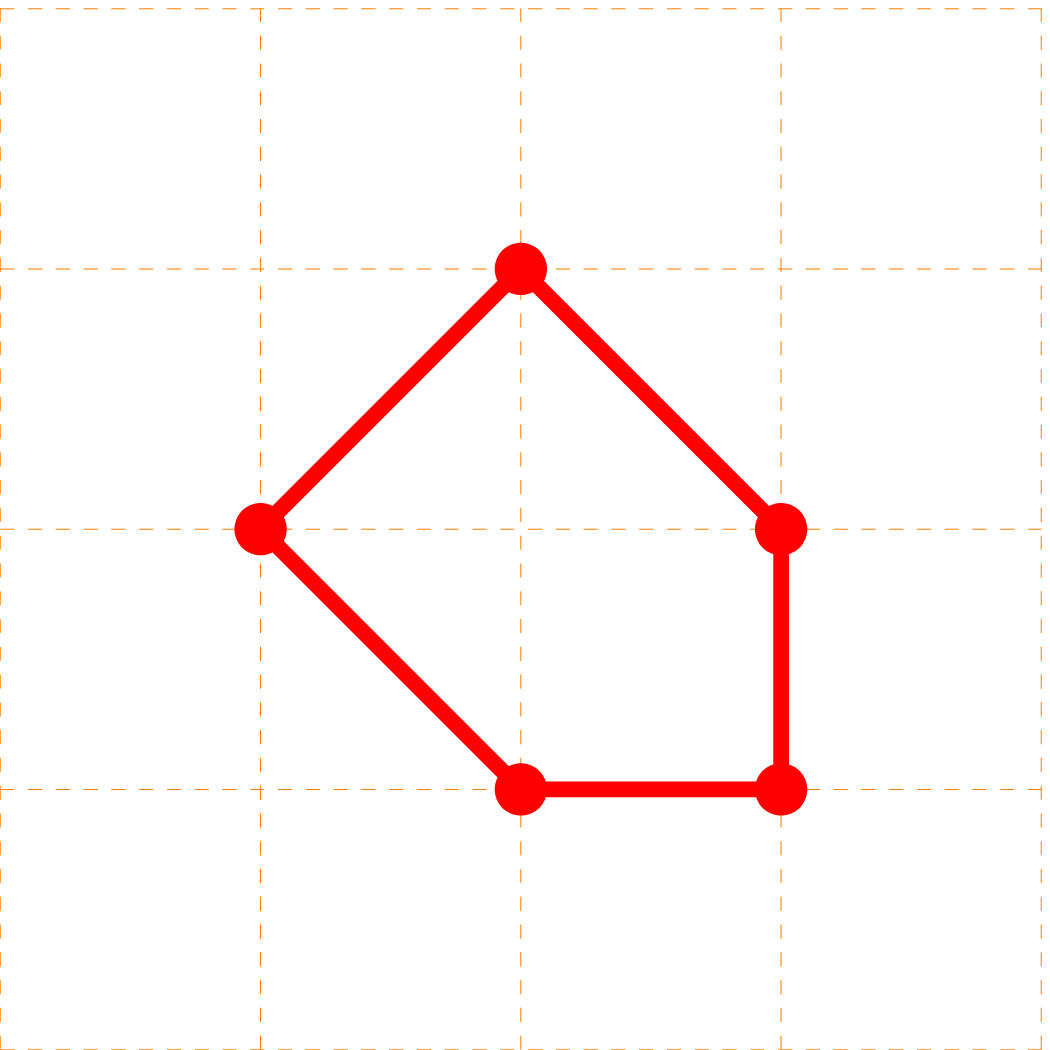}\end{tabular} & \begin{tabular}{c}\includegraphics[width=0.28\textwidth]{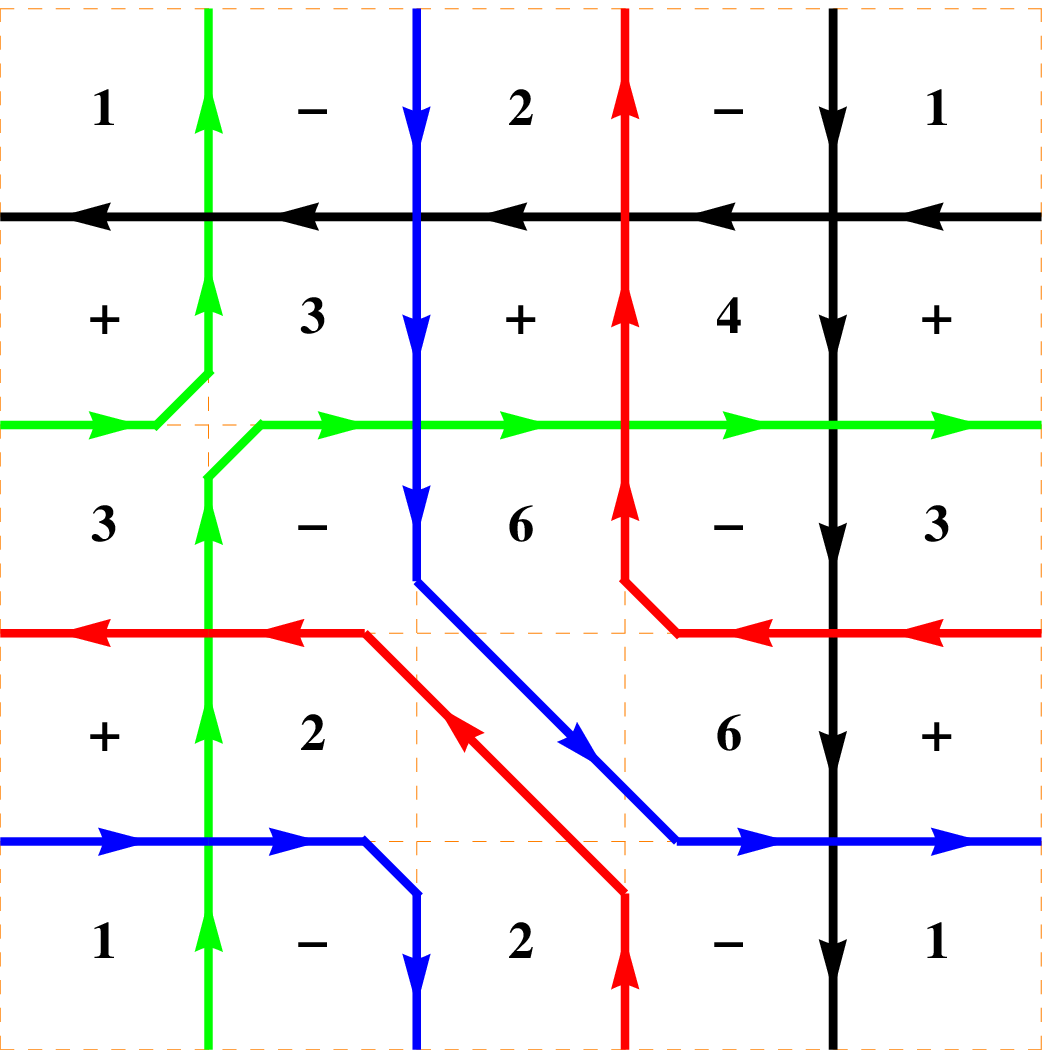}\end{tabular}
\end{tabular}
\captionof{figure}{\footnotesize{{\it Left:} The $dP_3$ dimer with the paths we will merge highlighted in green. {\it Middle:} The toric diagram for $dP_2$. {\it Right:} The $dP_2$ dimer with  the newly created (1,1) path in green.}\label{fig:dp2}}
\end{center}
Note that the $dP_2$ superpotential can be obtained from that of $dP_3$ by vevving and integrating out the field $\rho_{53}$ which was a bi-fundamental under gauge groups 5 and 3. These groups were combined in the merging of the zigzag paths. This is a general feature when cycles are collapsed.

\begin{eqnarray}
\nonumber W_{dP_2}&=&X_{43} Y_{31} Z_{14}-X_{12} Y_{31} Z_{23}-X_{43} Y_{64} Z_{36}+X_{12} Y_{23} Z_{36} \Phi _{61}\\ \nonumber && +X_{36}
Y_{64} Z_{23} \Psi _{42}-X_{36} Y_{23} Z_{14} \Phi _{61} \Psi _{42}\\
&=&\left(
\begin{array}{c}
 X_{43} \\
 Y_{23} \\
 Z_{23}
\end{array}
\right)\left(
\begin{array}{ccc}
 0 & Z_{14} & -Y_{64} \\
 -Z_{14} \Phi _{61} \Psi _{42} & 0 & X_{12} \Phi _{61} \\
 Y_{64} \Psi _{42} & -X_{12} & 0
\end{array}
\right)\left(
\begin{array}{c}
 X_{36} \\
 Y_{31} \\
 Z_{36}
\end{array}
\right).
\end{eqnarray}
$dP_1$ is obtained by collapsing another cycle, this involves generating a $(-2,1)$ path. We note that even when paths of higher winding number are created, the structure of the dimer only changes locally.
\begin{center}
\begin{tabular}{c c c}
\begin{tabular}{c}\includegraphics[width=0.28\textwidth]{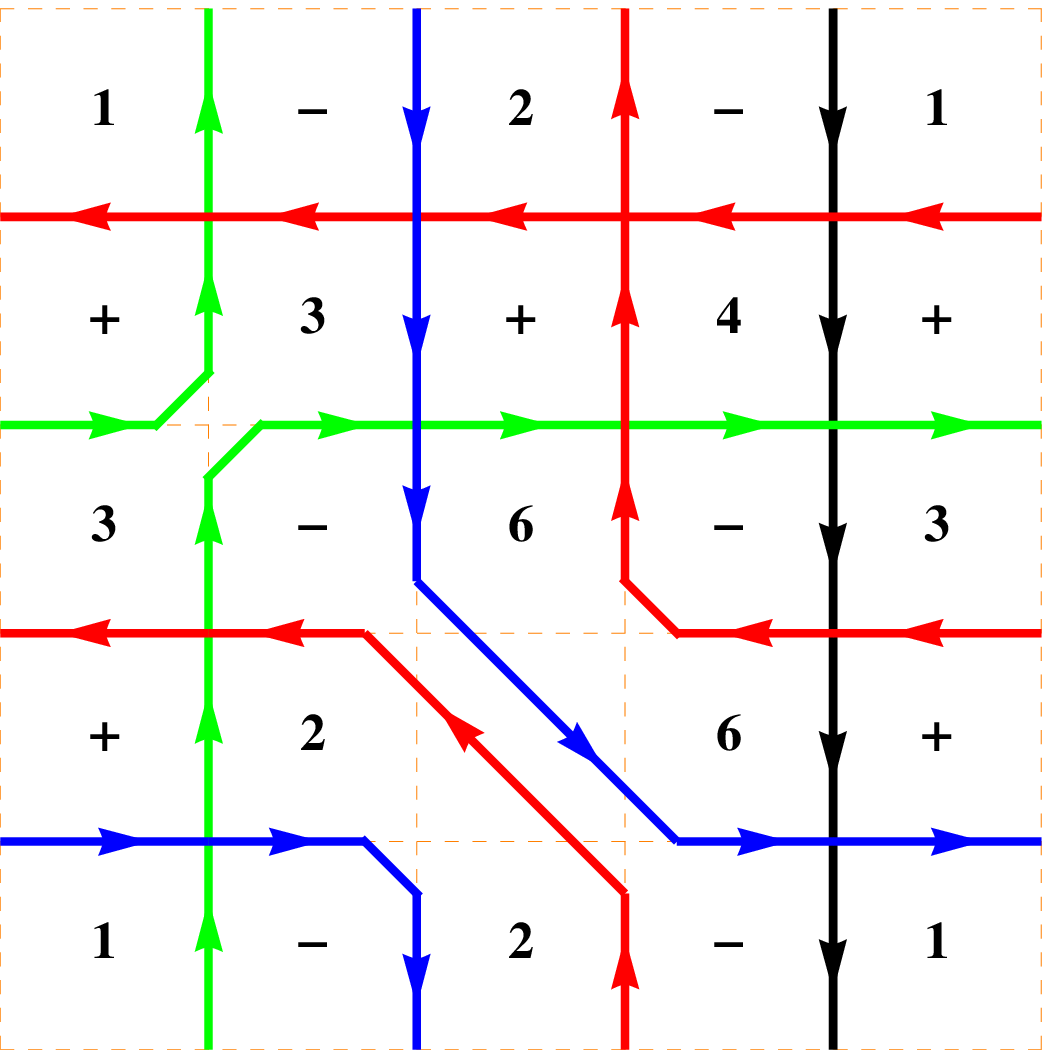} \end{tabular}&\begin{tabular}{c} \includegraphics[width=0.23\textwidth]{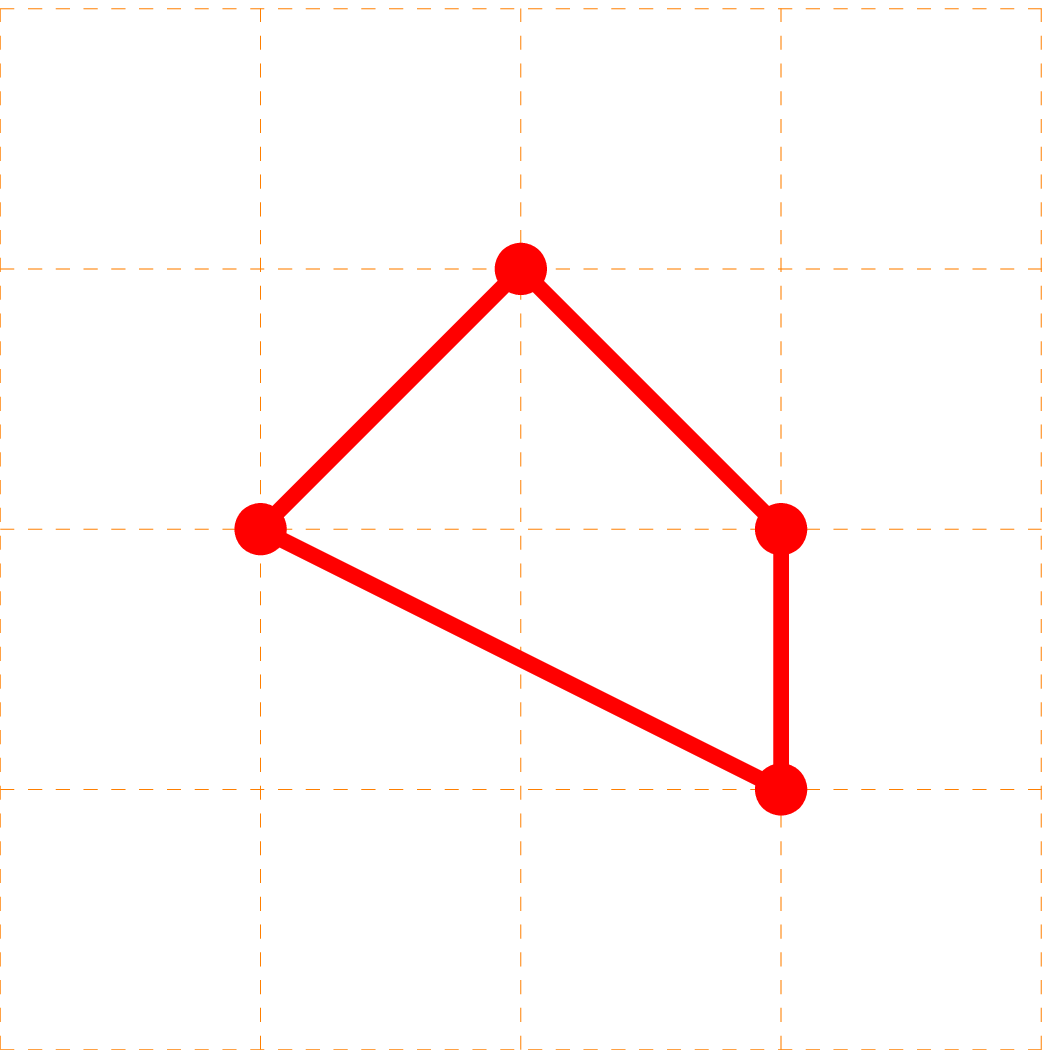} \end{tabular}& \begin{tabular}{c}\includegraphics[width=0.28\textwidth]{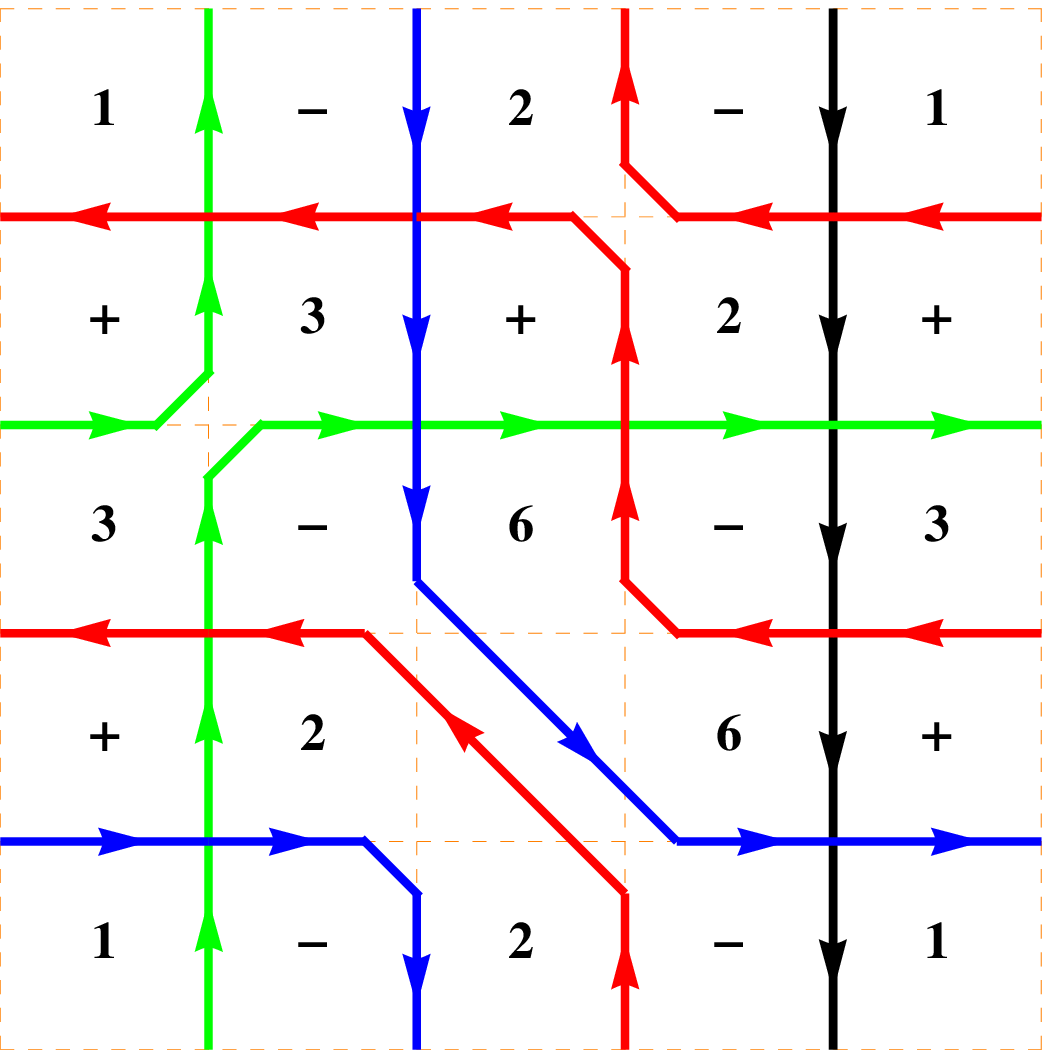}\end{tabular}
\end{tabular}
\captionof{figure}{\footnotesize{{\it Left:} The $dP_2$ dimer with a $(-1,0)$ path highlighted in red. {\it Middle:} The toric diagram for $dP_1$. {\it Right:} The $dP_1$ dimer with a $(-2,1)$ zigzag path (red).}\label{fig:dp1dimer}}
\end{center}
\begin{eqnarray}
\nonumber W_{dP_1}&=&X_{23} Y_{31} Z_{12}-X_{12} Y_{31} Z_{23}+X_{36} Y_{62} Z_{23}-X_{23} Y_{62}
Z_{36}\\ \nonumber && -X_{36} Y_{23} Z_{12} \Phi _{61}+X_{12} Y_{23} Z_{36} \Phi _{61}\\
&=&\left(
\begin{array}{c}
 X_{23} \\
 Y_{23} \\
 Z_{23}
\end{array}
\right)\left(
\begin{array}{ccc}
 0 & Z_{12} & -Y_{62} \\
 -Z_{12} \Phi _{61} & 0 & X_{12} \Phi _{61} \\
 Y_{62} & -X_{12} & 0
\end{array}
\right)\left(
\begin{array}{c}
 X_{36} \\
 Y_{31} \\
 Z_{36}
\end{array}
\right).
\label{dp1super}
\end{eqnarray}
Generating a $(1,-2)$ path out of the existing $(1,-1)$ and $(0,-1)$ path leads us to the dimer of $dP_0$.
\begin{center}
\begin{tabular}{c c c}
\begin{tabular}{c}\includegraphics[width=0.28\textwidth]{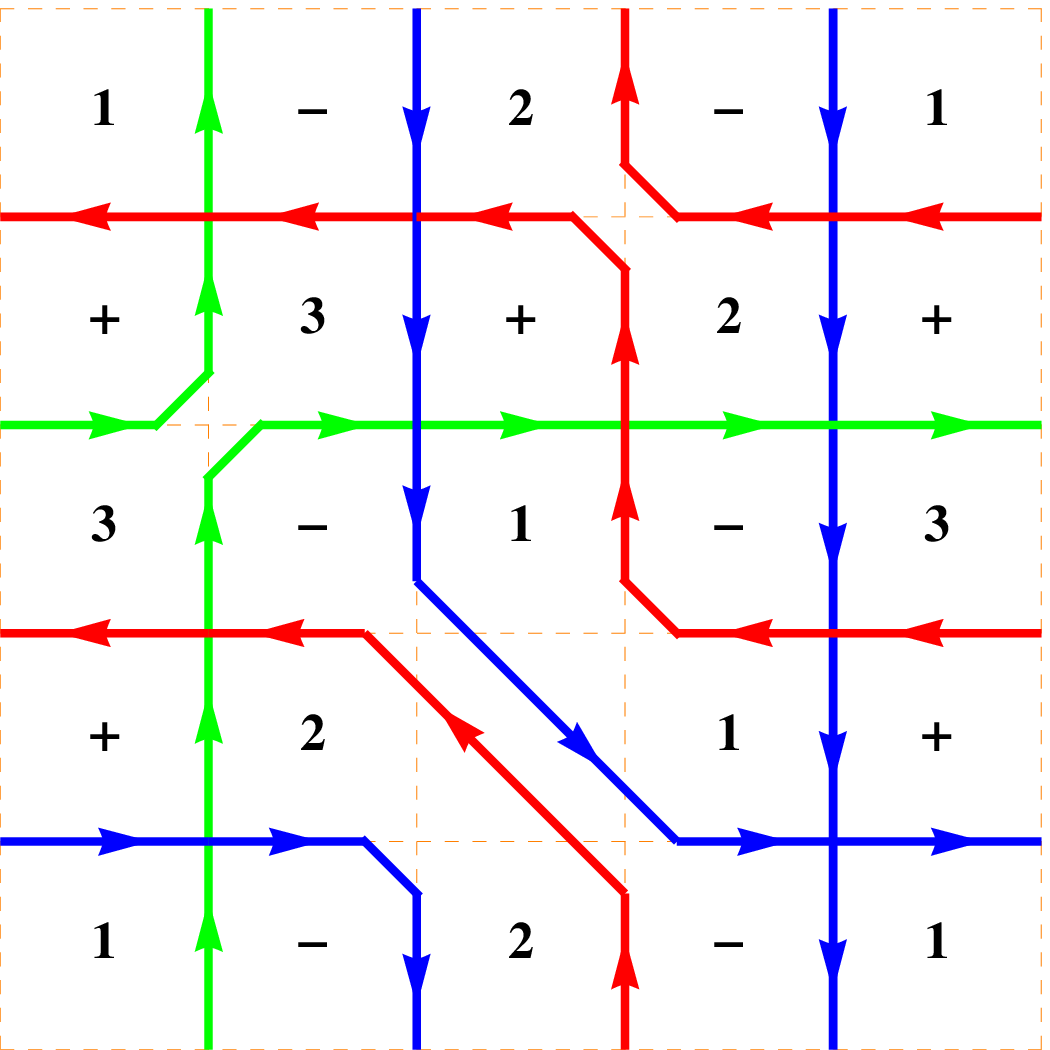} \end{tabular}&\begin{tabular}{c}\includegraphics[width=0.23\textwidth]{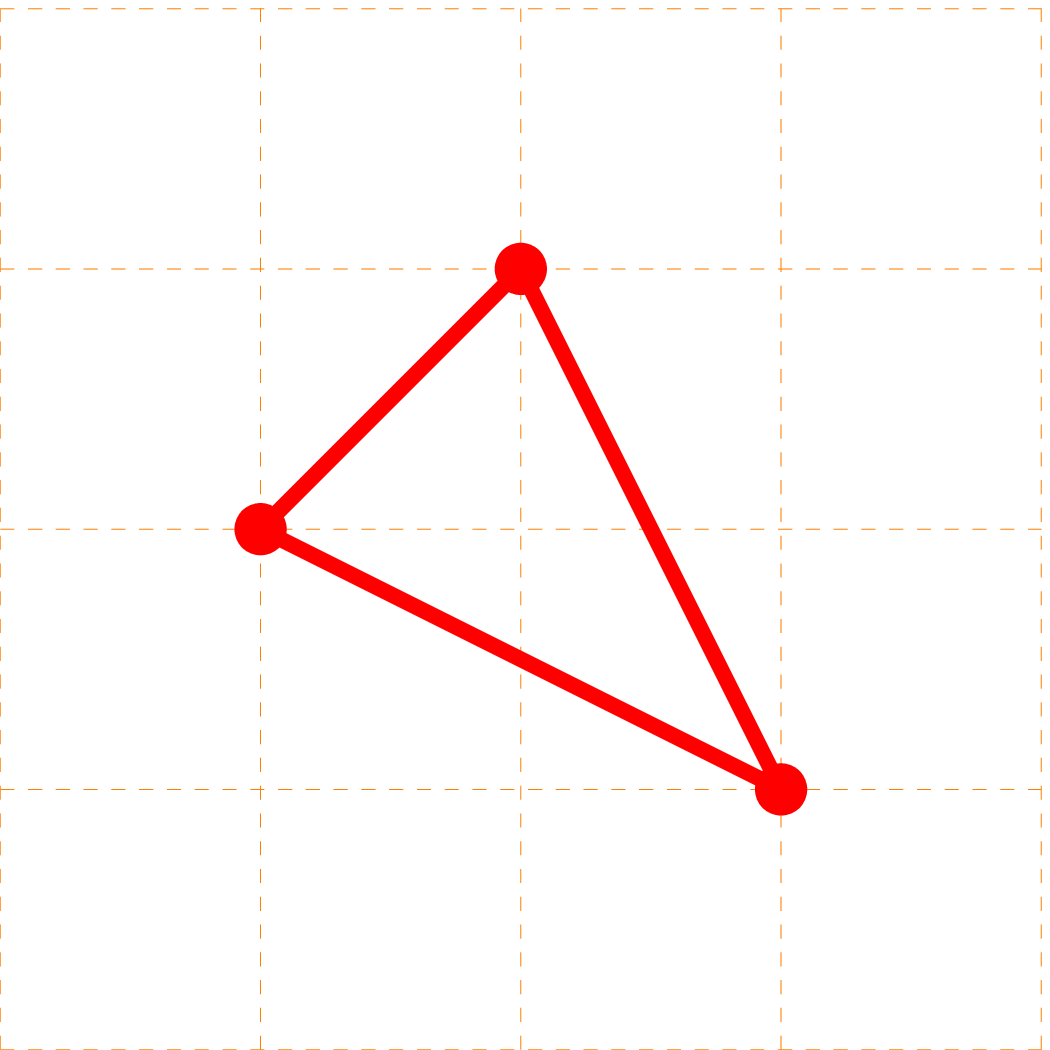} \end{tabular}& \begin{tabular}{c}\includegraphics[width=0.28\textwidth]{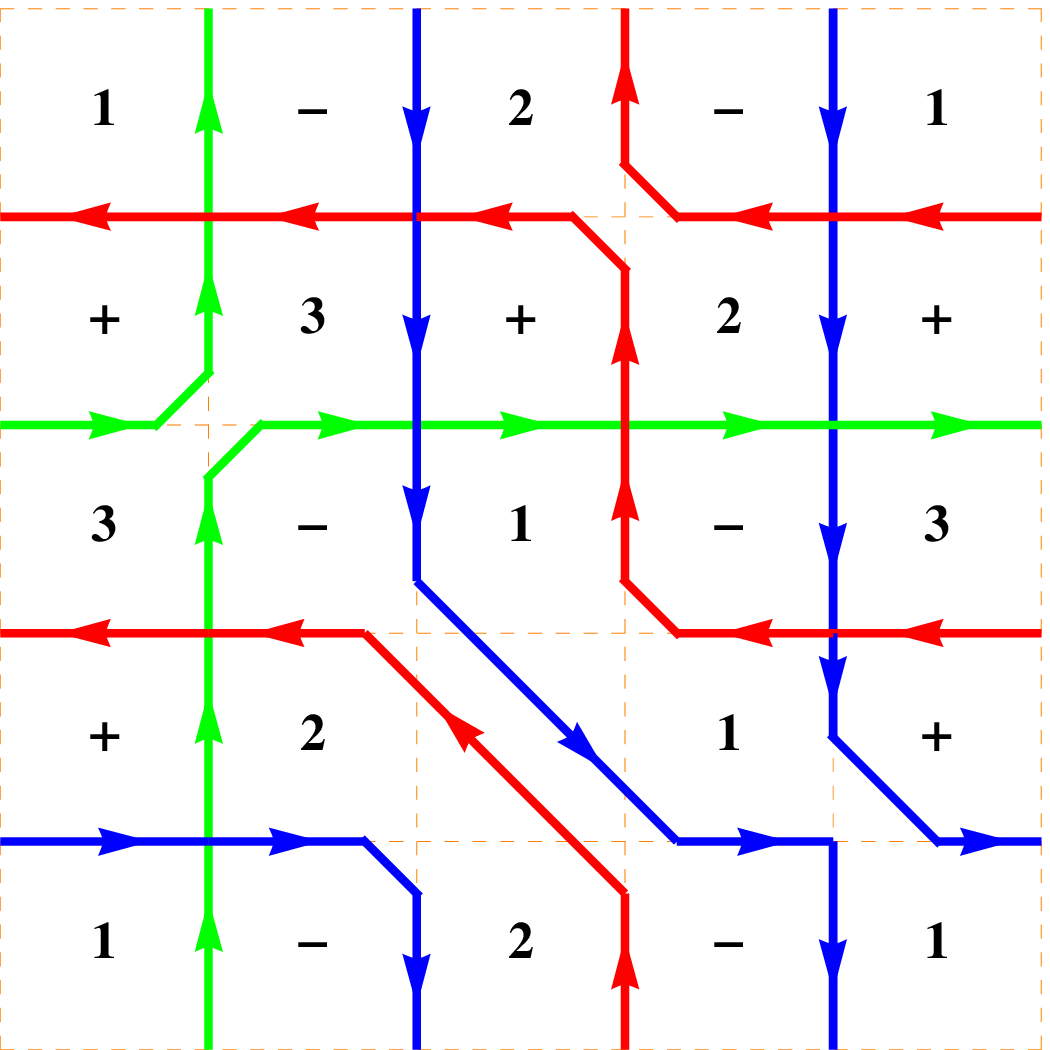}\end{tabular}
\end{tabular}
\captionof{figure}{\footnotesize{{\it Left:} The $dP_1$ dimer. {\it Middle:} The toric diagram for $dP_0$. {\it Right:} The $dP_0$ dimer with the $(1,-2)$ path shown in blue.}}
\end{center}
The superpotential for $dP_0$ exhibits an  $SU(3)$ flavour symmetry under the rotation of families. 
\begin{eqnarray}
\nonumber W_{dP_0}&=&-X_{31} Y_{23} Z_{12}+X_{23} Y_{31} Z_{12}+X_{31} Y_{12} Z_{23}\\ \nonumber &&-X_{12} Y_{31}
Z_{23}-X_{23} Y_{12} Z_{31}+X_{12} Y_{23} Z_{31}\\
&=&\left(
\begin{array}{c}
 X_{23} \\
 Y_{23} \\
 Z_{23}
\end{array}
\right)
\left(
\begin{array}{ccc}
 0 & Z_{12} & -Y_{12} \\
 -Z_{12} & 0 & X_{12} \\
 Y_{12} & -X_{12} & 0
\end{array}
\right)
\left(
\begin{array}{c}
 X_{31} \\
 Y_{31} \\
 Z_{31}
\end{array}
\right).
\end{eqnarray}

\subsubsection*{Connecting Gulotta's dimer with the traditional dimer}

While the zigzag path diagram introduced by Gulotta contains all the information encoded in dimers, their visual appearence is quite different. Here we discuss how one can reassemble dimers from dimers constructed with Gulotta's algorithm. This process is shown in Figure~\ref{fig:zigzagdimers}. Beginning with the zigzag path diagram corresponding to the conifold $\mathcal{C}$, we label the gauge groups and superpotential terms. We then draw lines between the centres of the superpotential faces. The zigzag paths can then be bent, to make the structure resemble the dimer diagrams of~\cite{0504110j}.

\begin{center}
\includegraphics[width=0.5\textwidth]{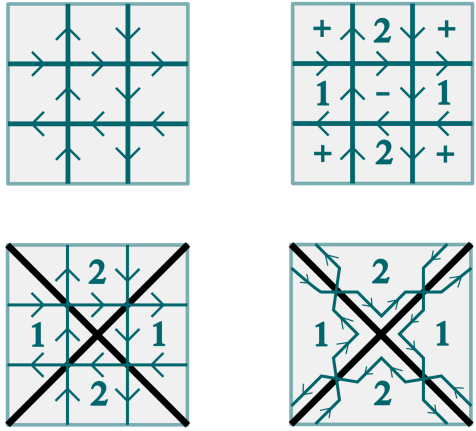}
\captionof{figure}{\footnotesize{The connection between zigzag path diagrams and dimers.}\label{fig:zigzagdimers}}
\end{center}

\subsection{D7 branes}
\label{sec:d7branes}

In local models D7 branes enter by wrapping non-compact holomorphic four cycles which pass through the singular locus. In a given singularity the number of different four-cycles is given by the number of external points in the toric diagram. In a compact model the D7 branes would wrap cycles which would extend into the bulk region of the Calabi-Yau and have non-vanishing gauge coupling. The presence of the D7 branes leads to three new kinds of string states, 37, 73 and 77 strings which we discuss in turn. 

For every bi-fundamental 33 field there exists a pair of 37 and 73 states which interact with the 33 state via a 33-37-73 coupling. Thus, when families of 33 states exist, there can be multiple 37 and 73 states with which they interact. Dimer techniques for the computation of the 37 and 73 spectrum and superpotential have been developed in~\cite{0604136}.

If there are multiple D7 branes there will be D7-D7' strings leading to 77' matter fields which interact with 37 fields via a 37-77'-7'3 coupling in the superpotential. Unfortunately, since the 77' fields are supported on noncompact cycles dimer techniques are not sensitive to the full details of their spectrum and interactions. 

For general geometries, the 33-37-73 couplings are diagonal in flavour space. If the singularity possesses isometries there is no longer a unique correspondence between one pair of 37 and 73 states and 33 states. For instance, in the case of the $\mathbb{C}^3/\mathbb{Z}^3$ orbifold~\cite{0005067} the orbifold action is $z_j \to z_j e^{\frac{i\pi}{3}}$. This implies that any equation $\sum_j \beta_j z_j =0$ is invariant under the orbifold action,  leading to a freedom in the choice of holomorphic four cycles that D7 branes wrap. This leads to a more general form of the superpotential, in which a given 33 state might couple to an arbitrary 37 state with the strength of the coupling being determined by $\beta_j$. This freedom arises from the $SU(3)$ global symmetry associated with being able to perform $SU(3)$ transformations on the coordinates $z_i$. Moving away from orbifolds, these arguments can be generalised~\cite{0604136} to show that the presence of a continuous global symmetry implies off-diagonal interactions between 37 and 33 states.\footnote{We thank A. Uranga for pointing this out.} In $dP_0$ the singularity has an $SU(3)$ isometry, in $dP_{1}$ there is an $SU(2)$ isometry and for the higher del Pezzo surfaces there are no non-abelian isometries.

In the context of models which involve 37 states~\cite{0005067, 0810.5660} this implies that at the $dP_0$ singularity the Yukawa matrix for the down generation is a generic $3\times 3$ complex matrix, while for $dP_1$ one has mixing between two of the three generations. For the higher del Pezzo surfaces the Yukawa matrix is diagonal. These facts shall play an important role in our discussion of flavour physics in Section~\ref{sec:ckm}.

\section{Families from toric singularities}

Families are matter in gauge theories with the same quantum numbers, which in quiver gauge theories corresponds to the number of arrows between two gauge group nodes.
In this section we explore the bounds on the number of families which exist in gauge theories probing toric singularities. 
We first discuss families in the context of Gulotta's algorithm and find that there is an upper bound of three families. After that, we include the possibility of addtional crossings and see that there is a unique exception to this bound, namely the zeroth Hirzebruch surface, one phase of which has four families.
Finally we discuss the effects of Seiberg duality and how the bound may be circumvented.

\subsection{Families \`a la Gulotta}

In this section we consider the effects which the operations of Gulotta's algorithm have on the number of matter fields and their couplings. We will show that no more than six matter fields exist charged under any given gauge group. Since the toricity condition implies that three of these are fundamental and three anti-fundamental, there can be no more than three families.

The starting point for obtaining the gauge theory associated with a given toric singularity is the smallest rectangular diagram in which the toric diagram can be embedded, corresponding geometrically to an orbifold of the conifold. In such a 
theory there is at most one arrow connecting any two gauge groups; every second square in the zigzag path diagram corresponds to a different gauge group, and there is not more than one field between any two gauge groups. An example is shown in Figure \ref{gridexample} which shows a $2\times 3$ rectangular toric diagram, its zigzag diagram and its quiver. Our starting point will therefore always be a quiver with only one family.\footnote{This does not apply to the conifold itself, which has two families.} 

While the effects of Gulotta's algorithm are 
local in the dimer for the gauge group faces and couplings, they are global from the point of view of zigzag paths. However, the changes in the global nature of the zigzag paths only affect the gauge groups and couplings in the region where zigzag paths are being merged.
Thus, while we present our results based on a small part of a larger dimer, the  embedding this into a larger dimer will have no effects on our results. 
We now proceed systematically through the effects of the operations of Gulotta's algorithm.

\begin{itemize}
\item \textbf{Operation I} creates $n$ (1,1) paths. Consider the creation of a single $(1,1)$ path, which  corresponds to joining two gauge groups and changing two quartic couplings to cubic couplings, as shown in Figure~\ref{fig:op1families}.
\end{itemize}
\begin{center}
\begin{tabular}{c c}
\includegraphics[width=0.38\textwidth]{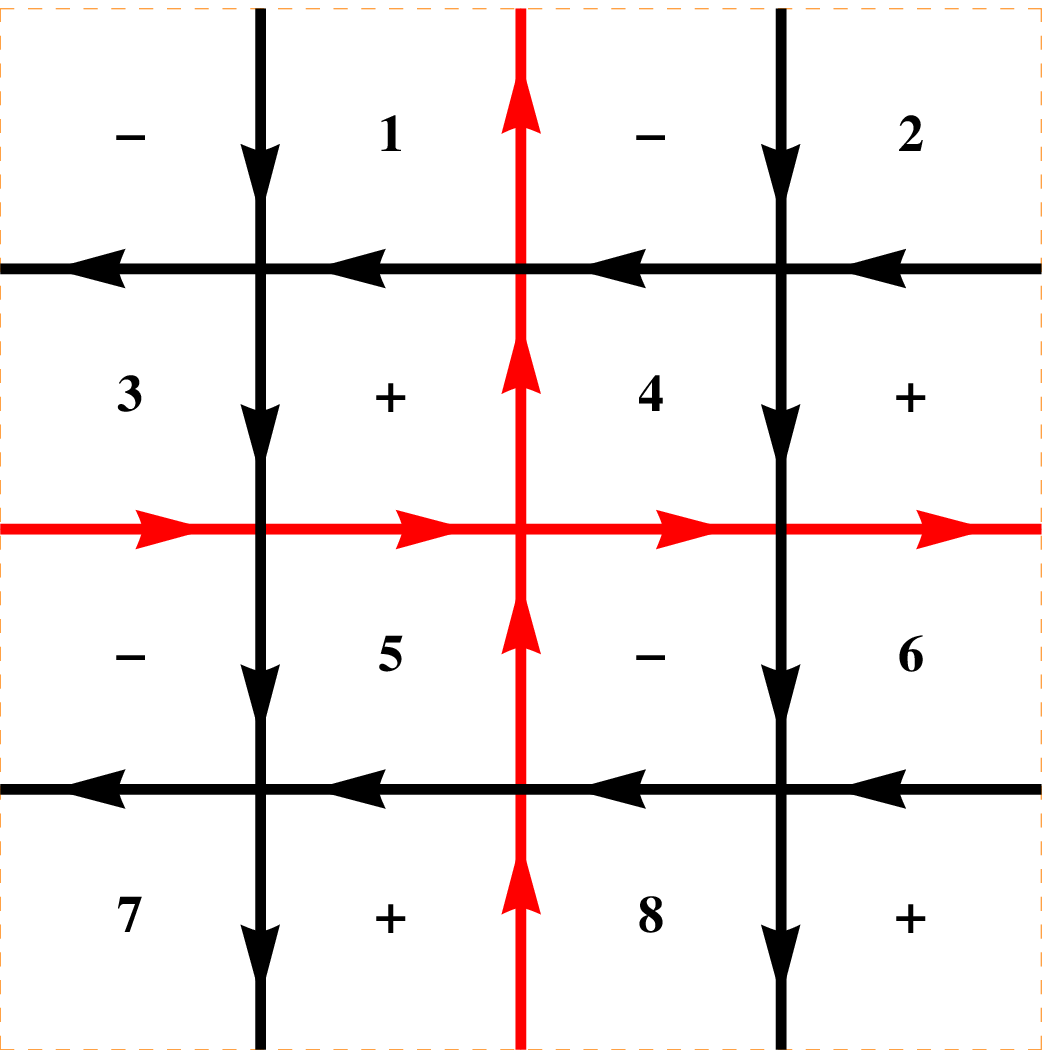} & \includegraphics[width=0.38\textwidth]{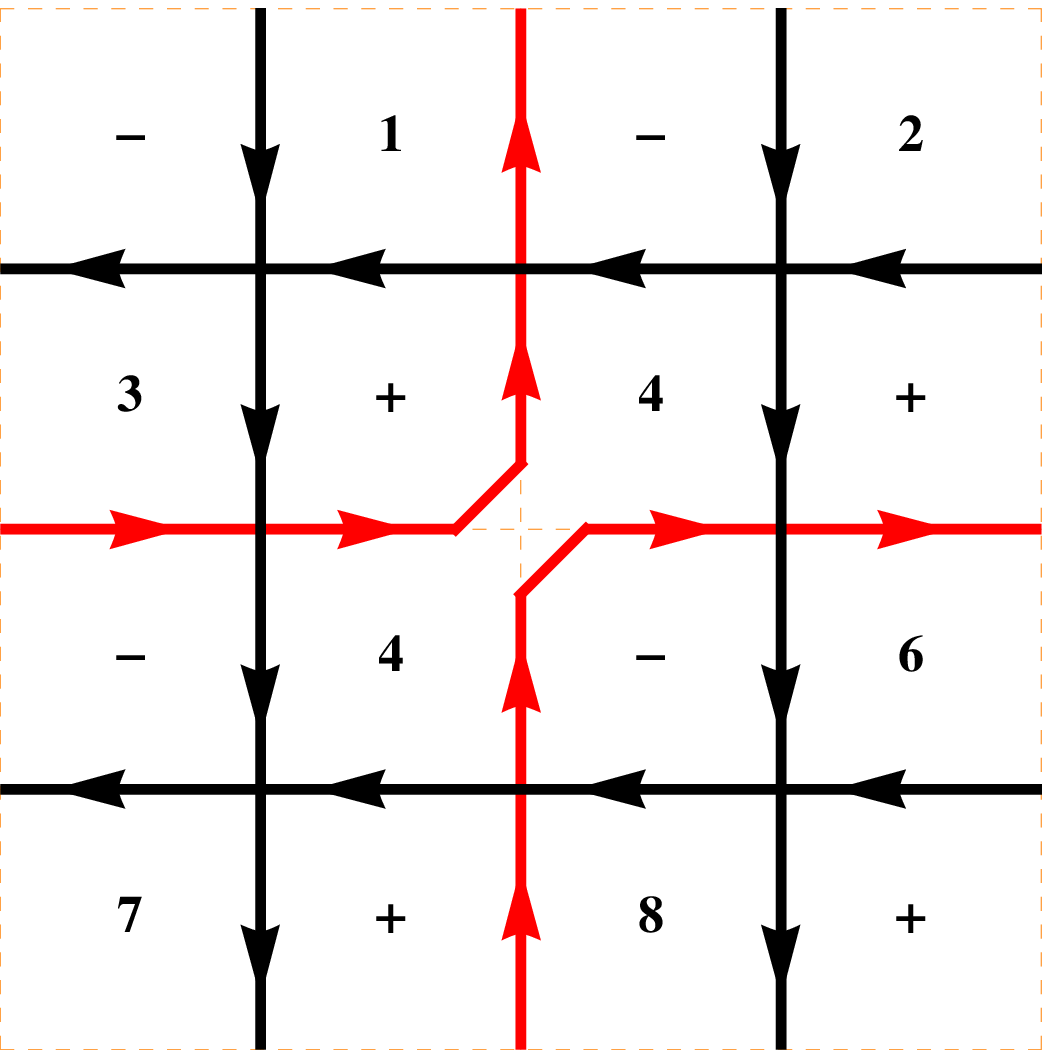}\\
\end{tabular}
\captionof{figure}{\footnotesize{The effect of joining a (1,0) and (0,1) path (Operation I) on a part of a larger dimer. Gauge groups 4 and 5 are joined, and two quartic couplings become two cubic couplings.}\label{fig:op1families}} 
\end{center}
\begin{itemize}
\item[] When creating multiple $(1,1)$ paths additional crossings are inevitable, and have to be removed. While this leads to global changes in the structure of the zigzag diagram, locally the situation always resembles the figure above since removing the additional crossings creates a situation locally the same as in the above figure.  This is illustrated in Figure~\ref{fig:appaddcross} in Appendix \ref{sec:moredimer}.
From Figure~\ref{fig:op1families} it is clear that there are 6 matter fields charged under gauge group 4. The toric condition implies that three of these transform in the fundamental representation of gauge group 4, and the other three as anti-fundamentals. There can therefore be a maximum of three families arising from  this operation.

\item \textbf{Operation II} creates $n$ pairs of oppositely winding paths. As an example in Figure~\ref{fig:op2families} we present the case of a pair of $(1,-1)$ and $(-1,1)$ paths.
By this procedure two pairs of gauge groups are joined, 6 with 12 and 4 with 11, and five quartic couplings become two cubic and one sextic coupling.
\end{itemize}
\begin{center}
\begin{tabular}{c c}
\includegraphics[width=0.38\textwidth]{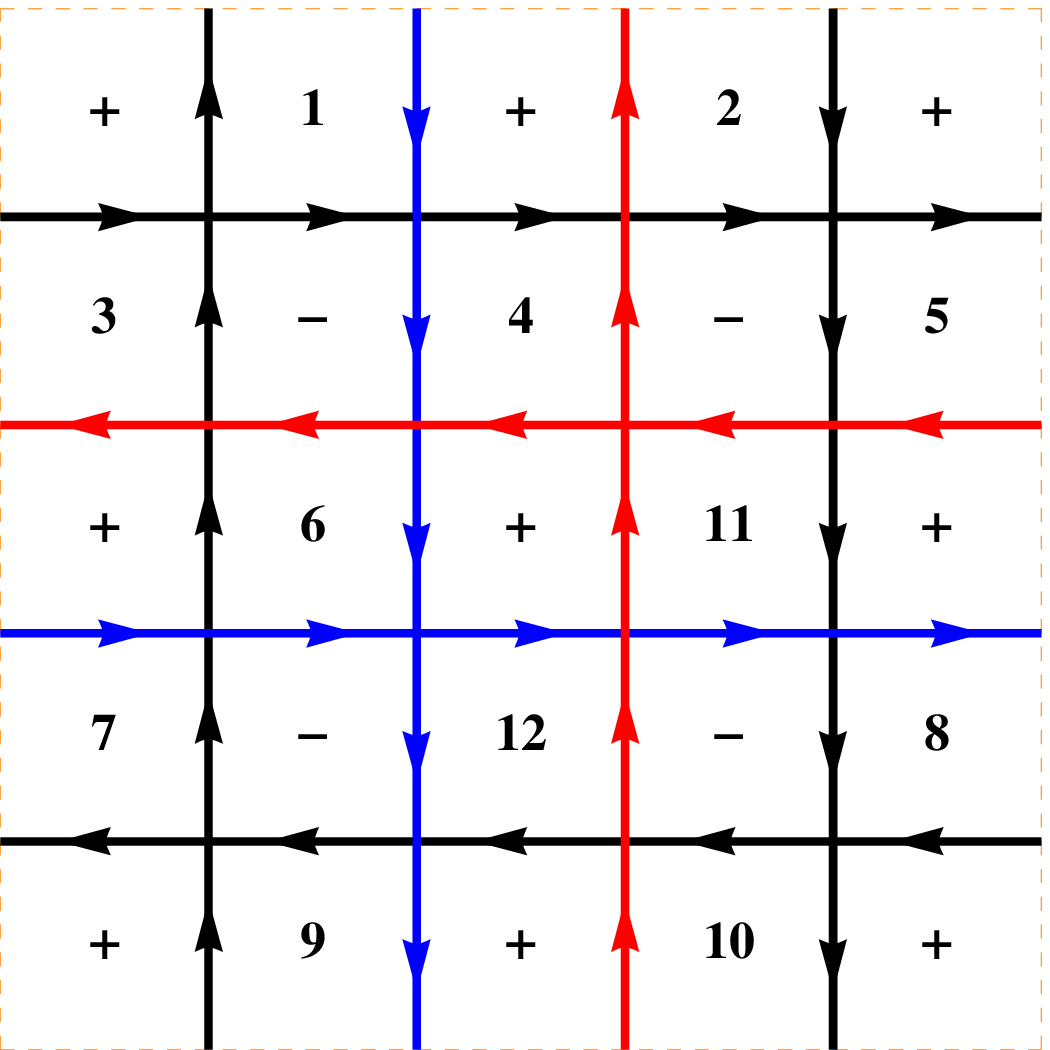} & \includegraphics[width=0.38\textwidth]{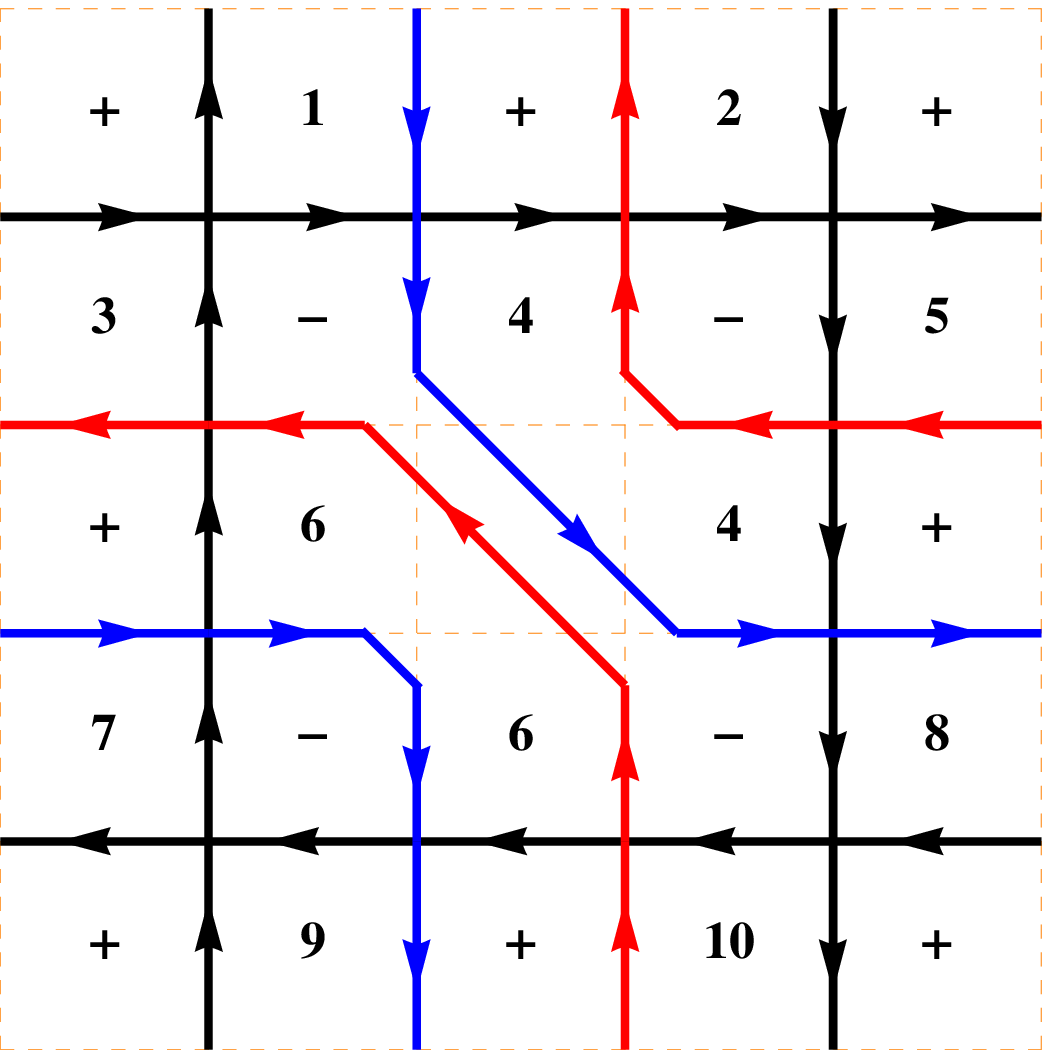}\\
\end{tabular}
\captionof{figure}{\footnotesize{The effect of Operation II, in this case creating a $(-1,1)$ and a $(1,-1)$ path. Two pairs of gauge groups are joined, and five quartic couplings become two cubic and one sextic coupling.}\label{fig:op2families}}
\end{center}
\begin{itemize}
\item[] By the same logic as above, not more than two families can be created here.
If we wish to create $n$ pairs of oppositely oriented paths we may do so according to the prescription in Section~\ref{sec:gulotta}. From Figure~\ref{fig:op2} in that section it is clear that, in creating further zigzag paths the number of fields charged under any given gauge group does not change. Since every gauge group has four matter fields charged under it, there will still be no more than two families.

\item \textbf{Operation III} creates different numbers of $(1,1)$ and $(-1,-1)$ paths, $m$ and $n$, say, where $m>n$. The first step is to create $m-n$ $(1,1)$ paths. This is just Operation I, and creates no more than 3 families. One then removes $n$ pairs of $(1,0)$ and $(-1,0)$ paths, and $n$ pairs of $(0,1)$ and $(0,-1)$ paths. This does not change the number of families. Finally, we create $n$ pairs of adjacent $(1,1)$ and $(-1,-1)$ paths. To avoid additional crossings, these should be created parallel to one of the already existing $(1,1)$ paths from the first step. By creating the  paths in this manner, no extra families are created, and so for Operation III the maximum number of families obtainable is three. 
\end{itemize}
Gulotta's algorithm involves repeated applications of these operations. For example zigzag paths with higher slopes are generated by applying multiple Operations I, II or respectively III. What are the effects of this? The only non-trivial effect appears when such additional operations affect a local region that has already been modified, as this is the only way to modify one gauge group that has been modified already. Typically, joining gauge groups together generates additional crossings which after removal do not lead to extra families.

For example, consider the zigzag diagram on the left of Figure~\ref{fig:familycrossings} which has a single $(1,1)$ path, created by Operation I. If we tried to add a $(-1,-1)$ path to this diagram to form a longer gauge group ``tube'' using the same operation, we would obtain additional crossings in our diagram, as in the right hand side of  Figure~\ref{fig:familycrossings}. Gulotta's algorithm prescribes the removal of these additional crossings in such a way that no further families are introduced.
\begin{center}
\begin{tabular}{c c}
\includegraphics[width=0.38\textwidth]{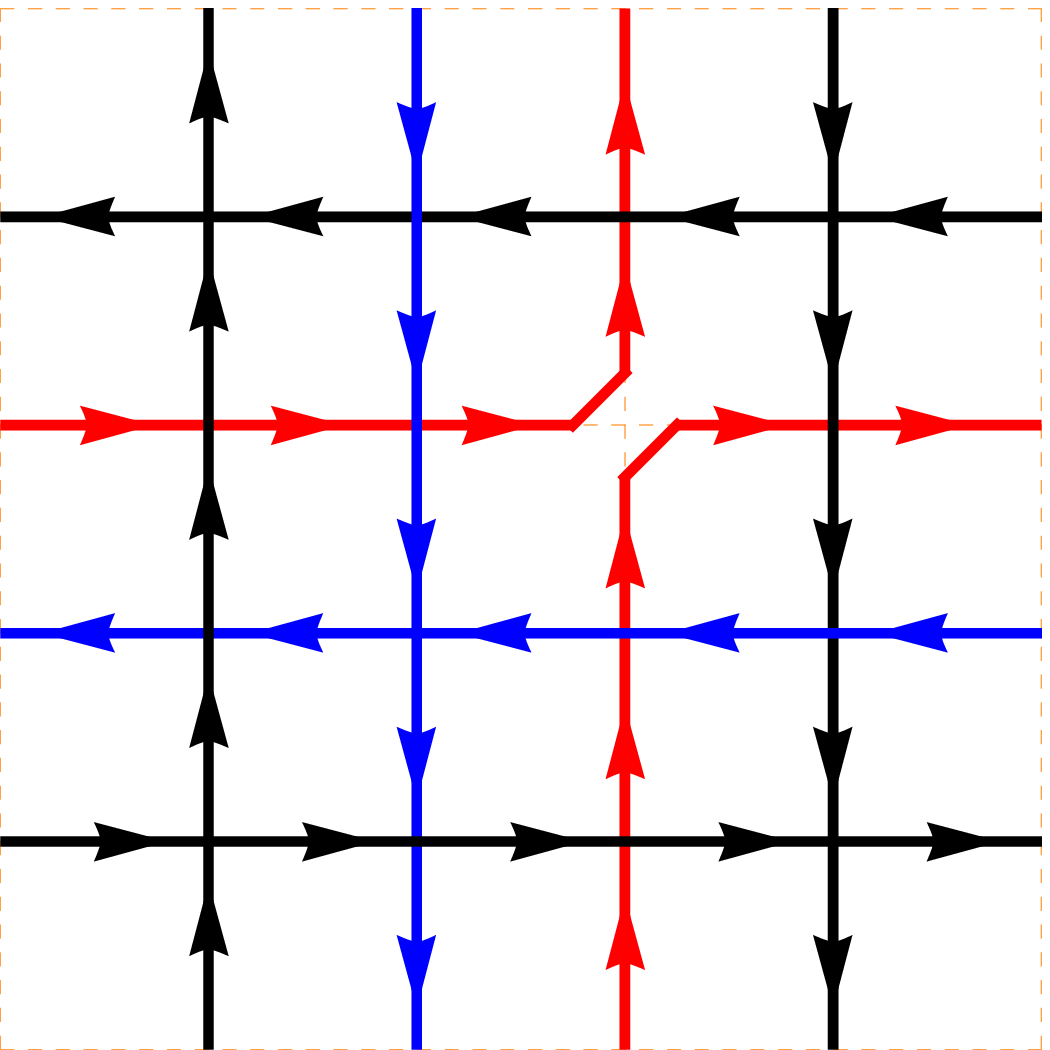} &
\includegraphics[width=0.38\textwidth]{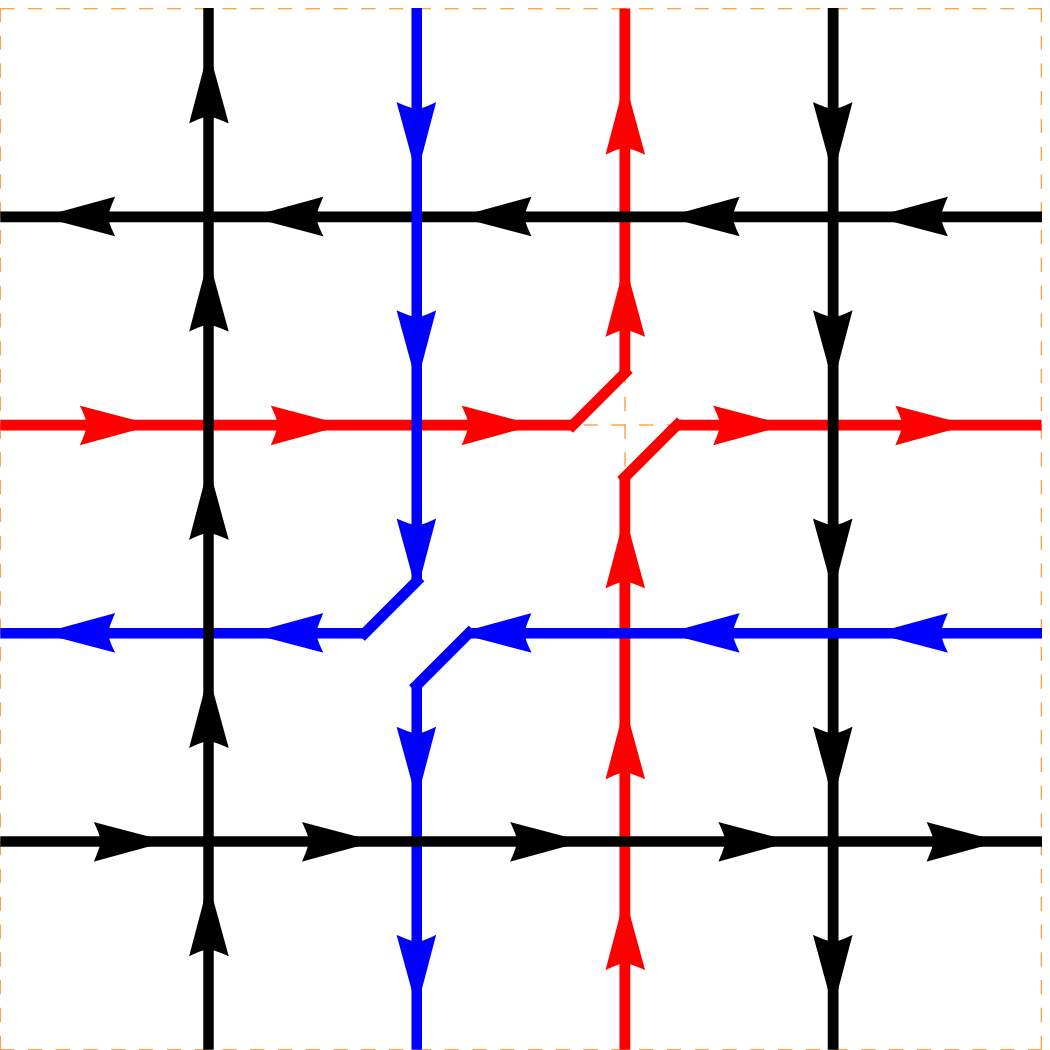} 
\end{tabular}
\captionof{figure}{\footnotesize{Left: A zigzag diagram with a single $(1,1)$ path and two highlighted $(-1,0)$ and $(0,-1)$ paths which will be merged. Right: The same diagram after the paths have been joined. There is one $(1,1)$ and one $(-1,-1)$ path, but with additional crossings.}\label{fig:familycrossings}}
\end{center}

The only case where a non-trivial modification of the gauge group faces takes place without introducing additional crossings is by applying Operation II or III after Operation I. An example of this is in the transition from $dP_2$ to $dP_1$ as shown Figure~\ref{fig:dp1dimer}. Doing this  generates at most 3 families.

\subsection{Additional crossings}

While the dimers obtained with Gulotta's algorithm do not include additional crossings, their presence does not necessarily render the dimer inconsistent.
In this section we therefore consider the possibility that the paths in the zigzag diagram have additional crossings. 
We will show that the maximum number of families is four and that there is only a single toric gauge theory which has four families: the zeroth Hirzebruch surface $\mathbb{F}_0$.

Once we allow the presence of additional crossings, there is no reason for arbitrarily long gauge group tubes not to exist, such as shown in the right hand side of Figure~\ref{fig:familycrossings}. However, we may only create straight gauge group tubes, not ones that 'bend'. Bending gauge group tubes lead to inconsistencies in the dimer, as discussed in \cite{0511063} and  mentioned in Section 2.1. An example of such a path is a $(1,1)$ path made from the combination of a $(2,1)$ and a $(-1,0)$ path. The presence of such paths violates the correspondence between zigzag paths and slopes in the web diagram, and the associated dimer is inconsistent.

Since the gauge group tube cannot have addtional branches it must have only one beginning and one end. There are then two possibilities: that the gauge group tubes in which we are interested are orthogonal to each other or they are parallel. The first case is heuristically shown on the left of Figure~\ref{fig:4fam}, where the orange and purple tubes represent two different gauge groups. The arrows show the  matter fields which are bi-fundamentals of the orange and purple gauge groups. It is clear that a maximum of four families is possible in this case. While the diagram shows a square chess-board, it is possible to generalise this to a rectangular one. 

The right hand side of Figure~\ref{fig:4fam} shows the case of parallel gauge group tubes. 
In this case there are superpotential mass terms for the bi-fundamental matter between the orange and purple gauge groups which must be integrated out. After this there will be at most one family of chiral matter, this occurs if the common length of the gauge group tubes is odd. 

\begin{center}
\begin{tabular}{c c}
\begin{tabular}{c}\includegraphics[width=0.38\textwidth]{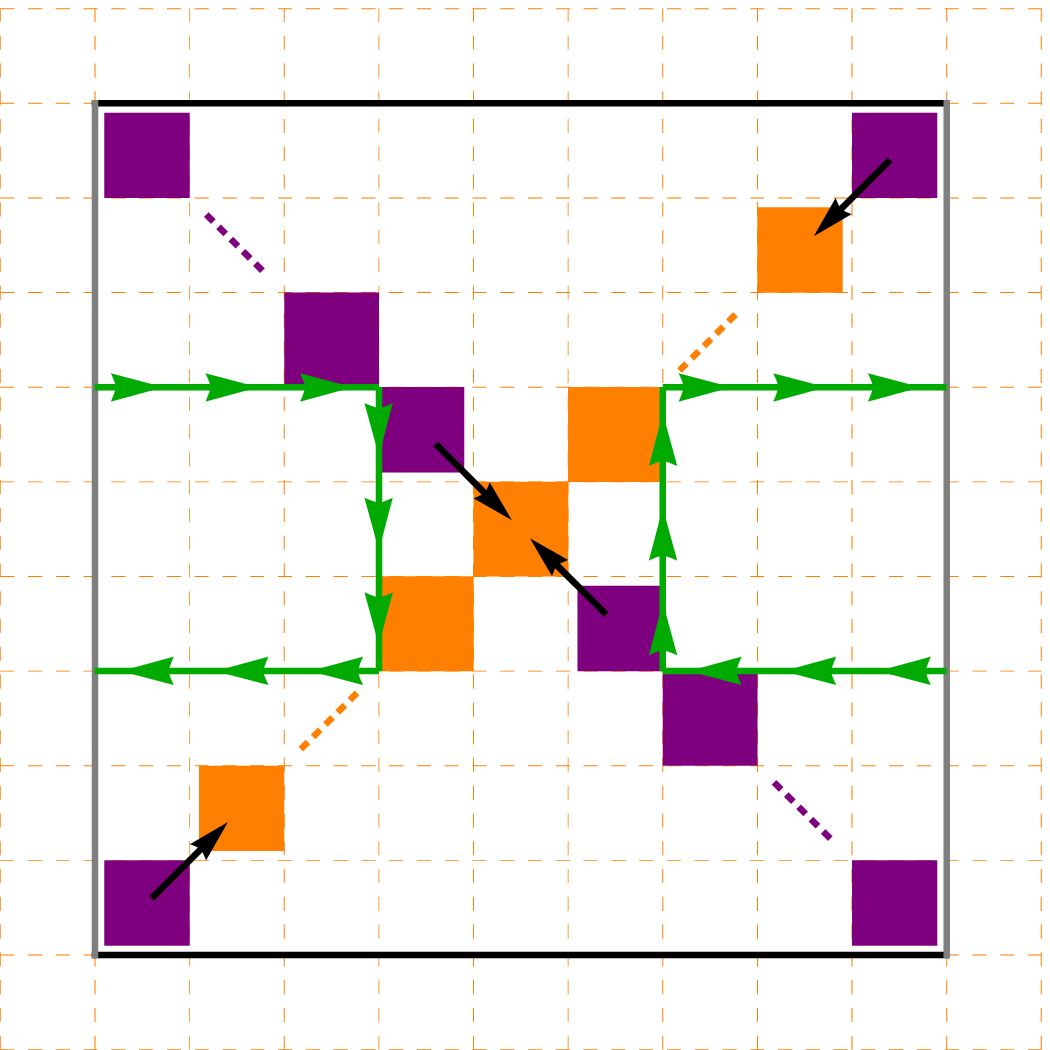} \end{tabular}& \begin{tabular}{c}
\includegraphics[width=0.38\textwidth]{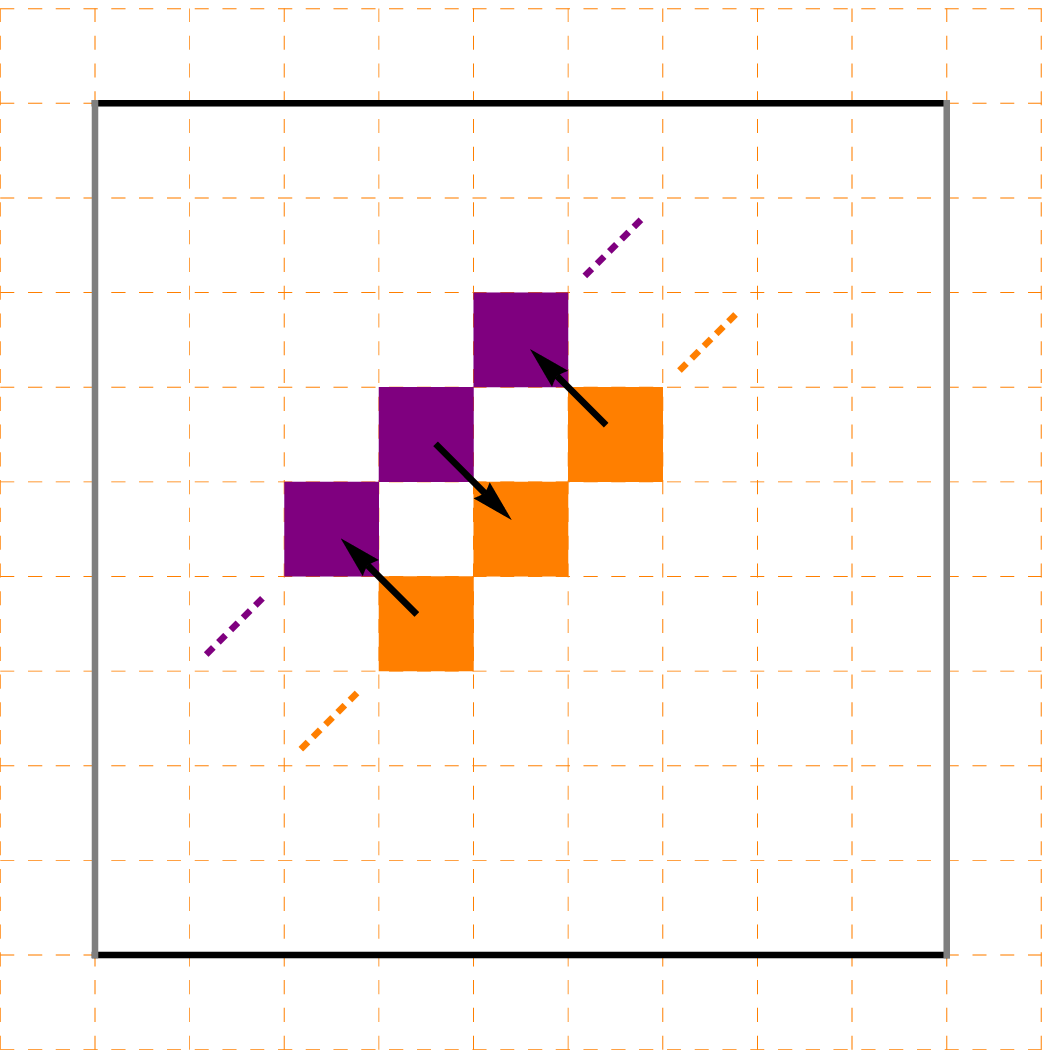} \end{tabular}\\
\end{tabular}
\captionof{figure}{\footnotesize{{\it Left:} A zigzag diagram demonstrating the presence of four families with orthogonal gauge group tubes. The purple and orange tubes are different gauge groups, with arrows highlighting the bi-fundamental matter between purple and orange gauge groups. The green arrows show a zigzag path with an inconsistent winding that would be present in this dimer. {\it Right:} A zigzag diagram with parallel purple and orange gauge group tubes. The matter fields are non-chiral.}\label{fig:4fam}}
\end{center}
The schematic four family case from the left hand side of Figure~\ref{fig:4fam} is not easy to realize since creating orthogonal tubes leads to inconsistent paths, as mentioned in the previous section. The green path in the above figure has winding $(0,0)$, which is inconsistent with the correspondence to the web diagram. 
The only situation that does not contain such inconsistent paths is the most simple one, which is the zeroth Hirzebruch surface 
$\mathbb{F}_{0}$, as illustrated in Figure~\ref{fig:f0toric}.
\begin{center}
\begin{tabular}{c c c c}
\begin{tabular}{c}\includegraphics[width=0.17\textwidth]{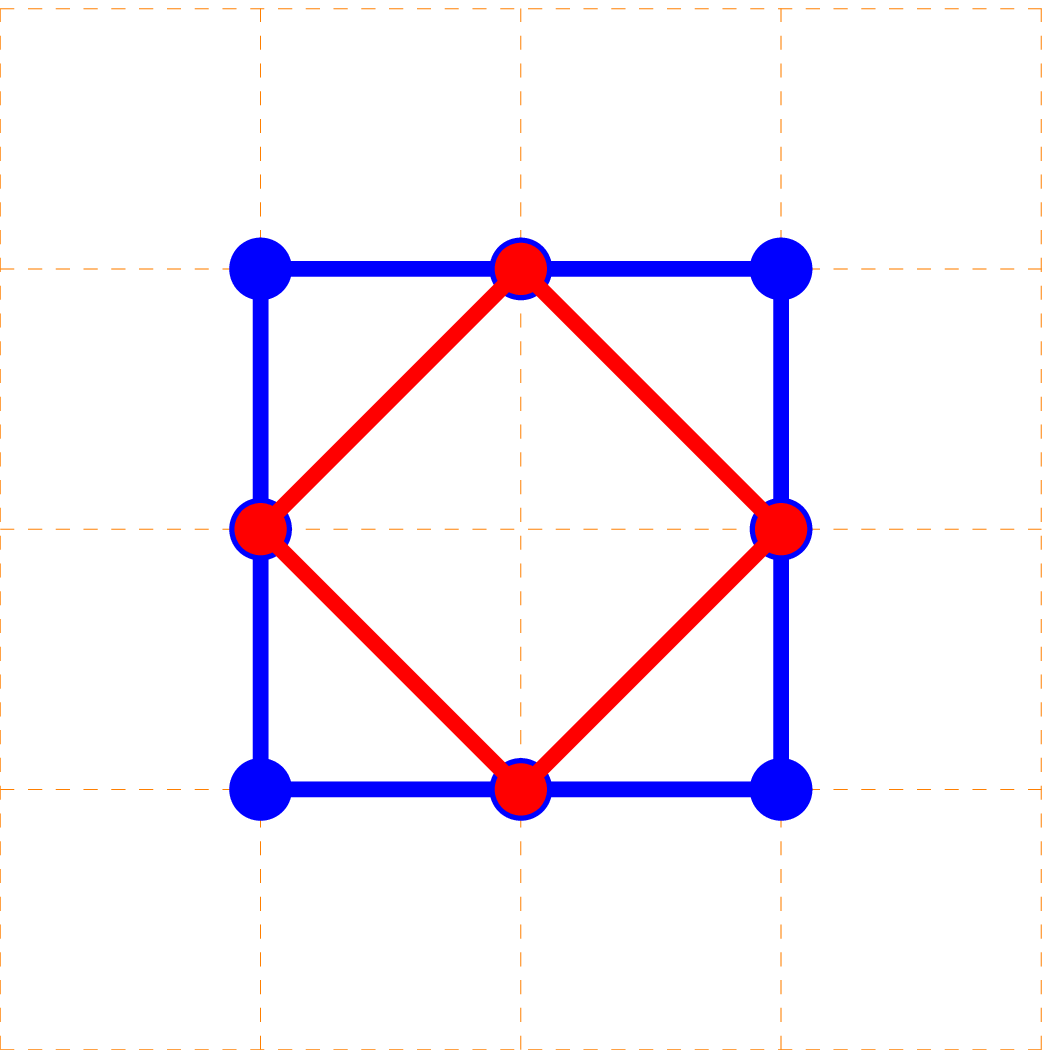}\end{tabular} &\begin{tabular}{c} \includegraphics[width=0.21\textwidth]{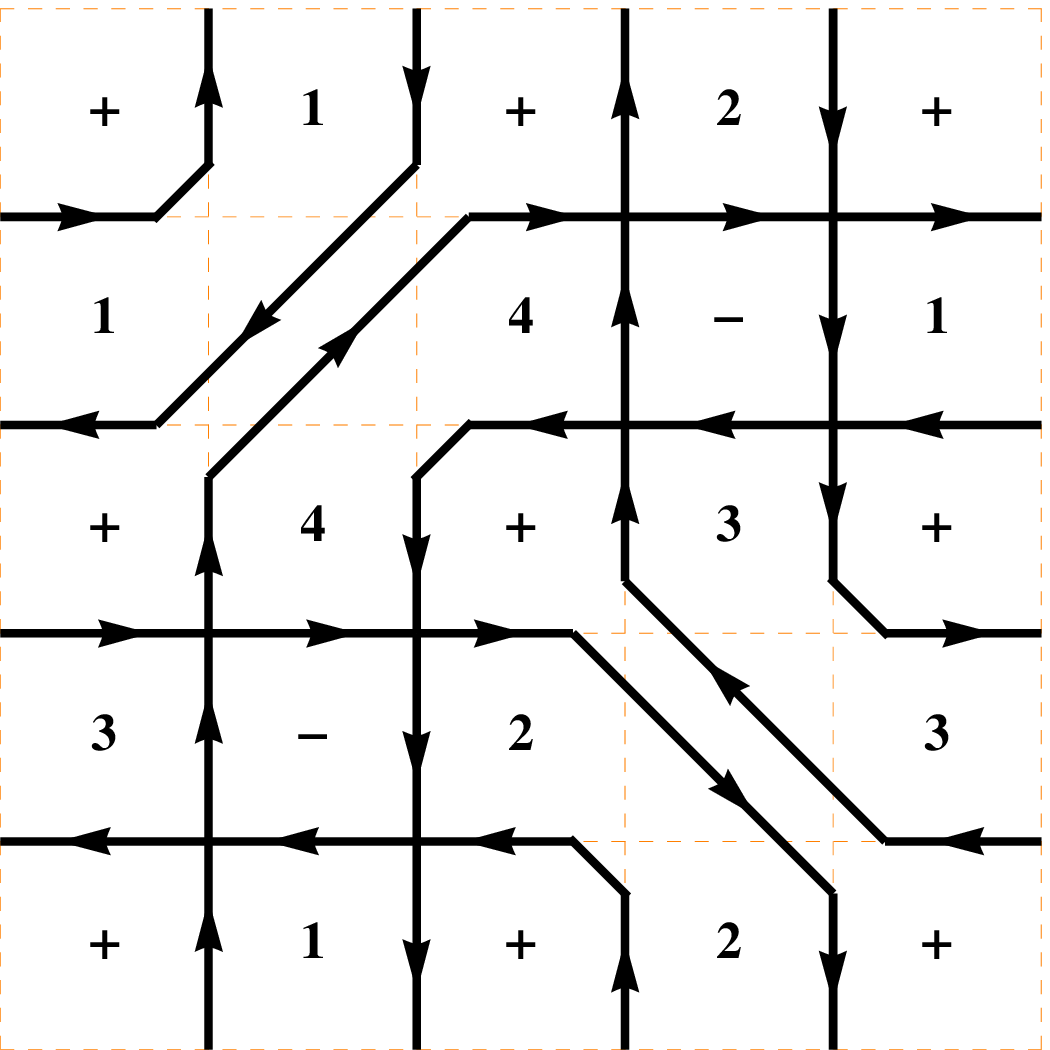} \end{tabular} & \begin{tabular}{c}\includegraphics[width=0.21\textwidth]{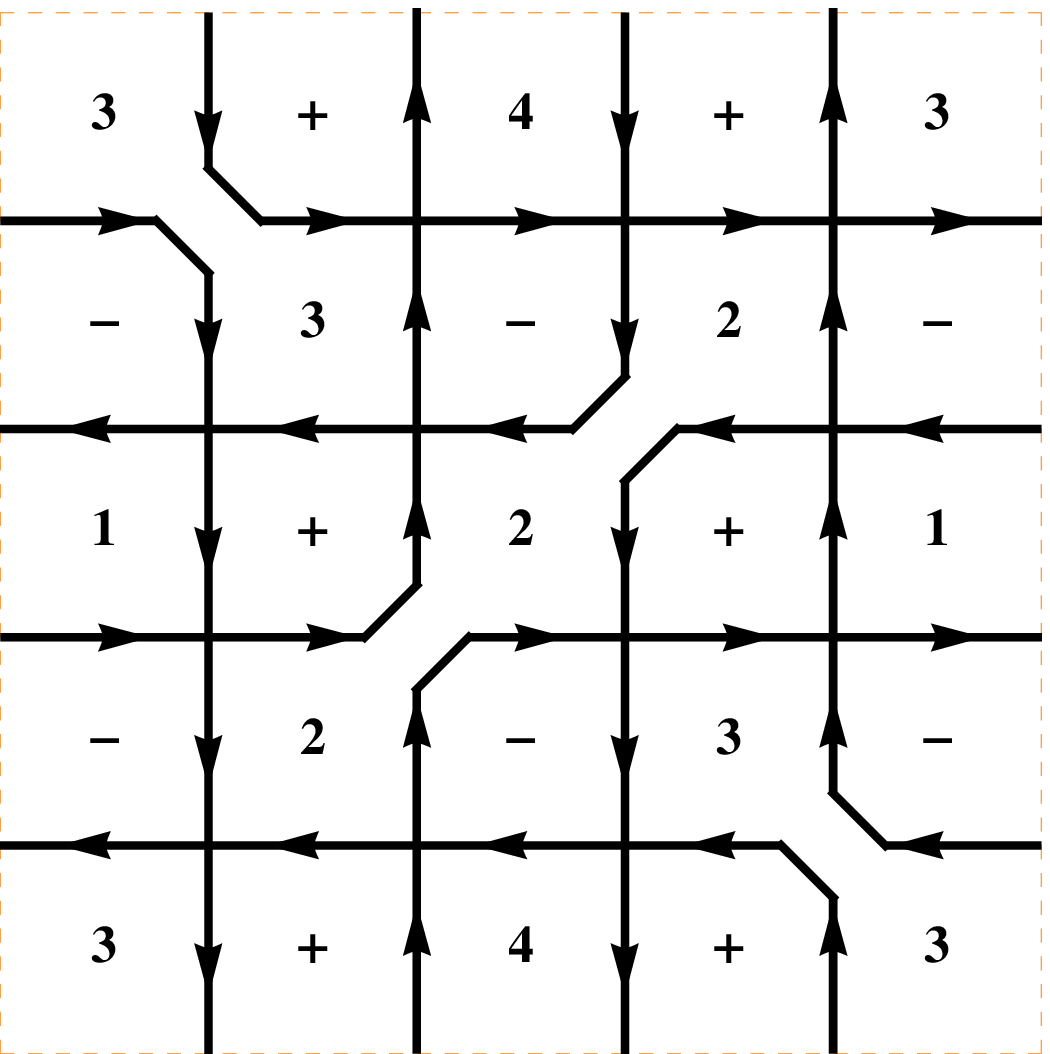} \end{tabular} &\begin{tabular}{c} \includegraphics[width=0.21\textwidth]{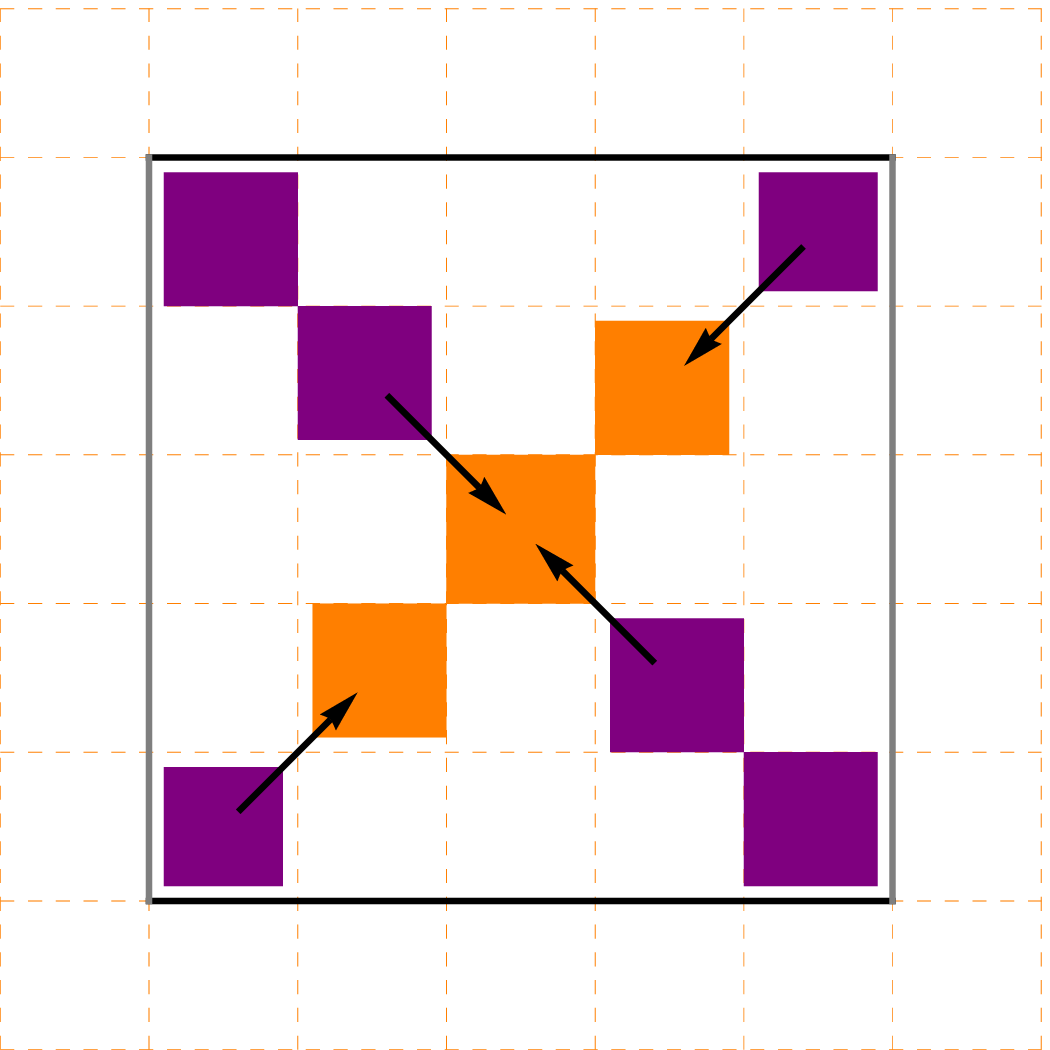} \end{tabular}\\
\end{tabular}
\captionof{figure}{\footnotesize{{\it From Left to Right:} Toric diagram of the zeroth Hirzebruch surface. The dimer obtained with Gulotta's algorithm, containing two families. Four family $\mathbb{F}_0$ dimer, obtained by allowing for additional crossings. A schematic zigzag diagram highlighting how the four families arise.}\label{fig:f0toric}}
\end{center}

We stress again that the 4-family scenario is only possible by allowing for additional crossings in the zigzag diagram. Those additional crossings can be avoided while constructing the dimer according to Gulotta's algorithm. Since we have shown above that Gulotta's algorithm leads to no more than three families, this implies that any singularity which has a toric phase with four families must also have one with less than four. This leads us to some comments on Seiberg duality.

\subsection{Seiberg duality}
There are various gauge theory descriptions for branes probing a toric singularity, all of which are connected by Seiberg duality~\cite{0109063,0109053,0205144}.  Gulotta's algorithm provides us with one gauge theory probing a given singularity in a toric phase. 
There are two ways of obtaining different toric phases:
\begin{enumerate}
\item Start with a gauge theory which corresponds to a toric phase of the singularity. Then Seiberg dualise one of the gauge groups such that the gauge theory remains in a toric phase.
\item Allow for consistent additional crossings in the construction of the gauge theory,  as was done previously for $\mathbb{F}_0.$
\end{enumerate}

We wish to consider Seiberg dualising a gauge group $S$ of a toric gauge theory subject to the requirement that the dual theory also be in a toric phase. Say that the rank of the gauge groups is $N_c$ and that it has $N_f$ flavours. The dualised gauge group has rank $N^{\prime}_c= N_f -N_c$ after the duality.
Since toricity requires $N^{\prime}_c=N_c=N$, the number of flavours must be $2N$ and hence there are two fields transforming in the fundamental and two in the antifundamental representations of $S$.

 In the quiver gauge theory, dualising one particular node replaces all  outgoing fields $Q_{Sa}$, which transform as $(\bar{N}_S,N_a)$, and incoming fields $\tilde{Q}_{bS}$, which transform as $(\bar{N}_b,N_S)$, by their dual fields $\tilde{q}_{aS}$ $(\bar{N}_a,N_S)$ and $q_{Sb}$ $(\bar{N}_S,N_b)$. This implies the arrows corresponding to those fields in the quiver are reversed. In the dual theory one also adds all possible mesons made out of bound states of the original fields, $M_{ij}=\tilde{Q}_{iS}Q_{Sj}$. This can easily be visualised in the quiver, as shown in the left hand side of  Figure \ref{fig:dimerseiberg}. The superpotential is modified to
\begin{equation}
W_{\rm dual}(\tilde{q}q,M,\Phi)=W_{\rm orig}(M,\Phi)+M \tilde{q}q\, ,
\end{equation}
where all indices are suppressed.

In the dimer we can form a picture of Seiberg duality following~\cite{0109063}. The right hand side of Figure~\ref{fig:dimerseiberg} shows a local region of some larger dimer with a gauge group face $S$ with $2N$ flavours and the effects of dualisation. 
Dualising changes the white nodes around gauge group $S$ to black and black to white, which reverses the transformation properties of the bi-fundamental matter. The presence of Seiberg dual mesons is encoded in extra edges between gauge groups 1 to 4 and gauge group $S$ which are shown in red.

\begin{center}
\begin{tabular}{ c c  c c}
\includegraphics[width=0.2\textwidth]{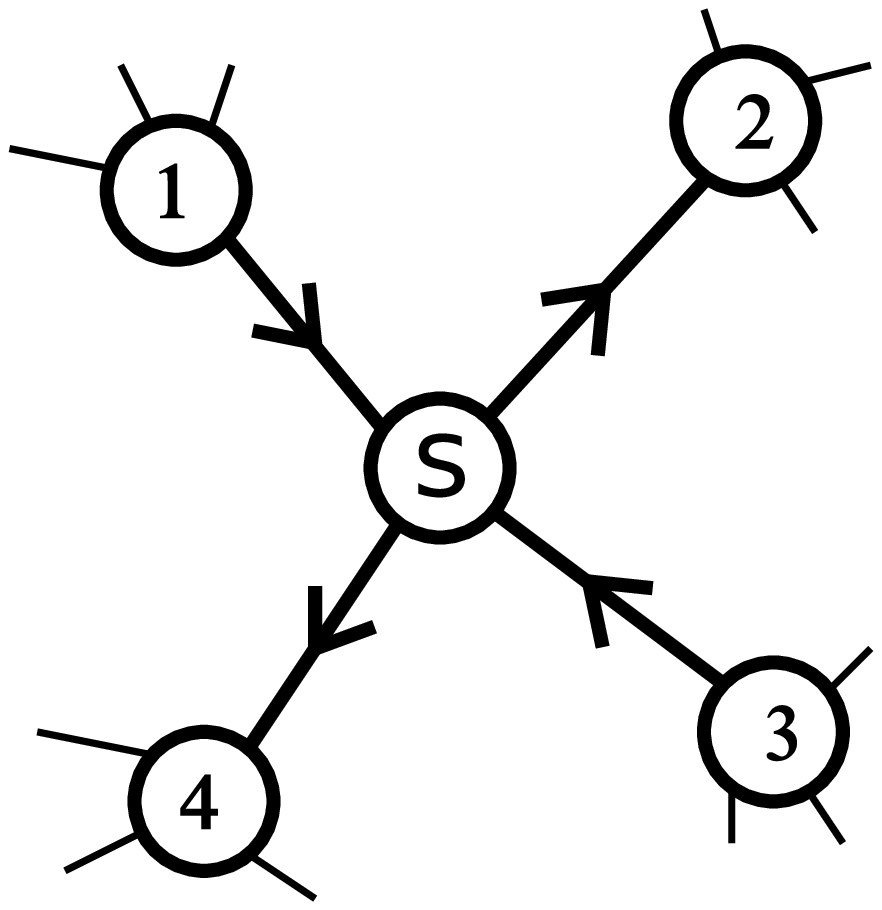}&\includegraphics[width=0.2\textwidth]{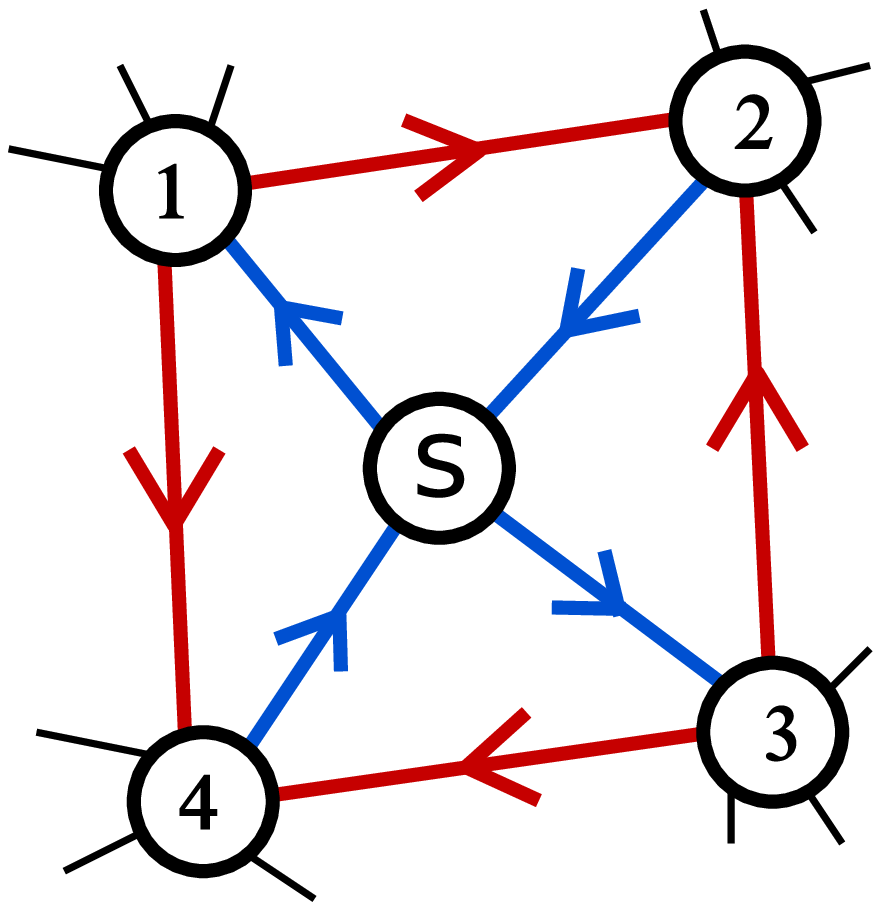} &
\includegraphics[width=0.2\textwidth]{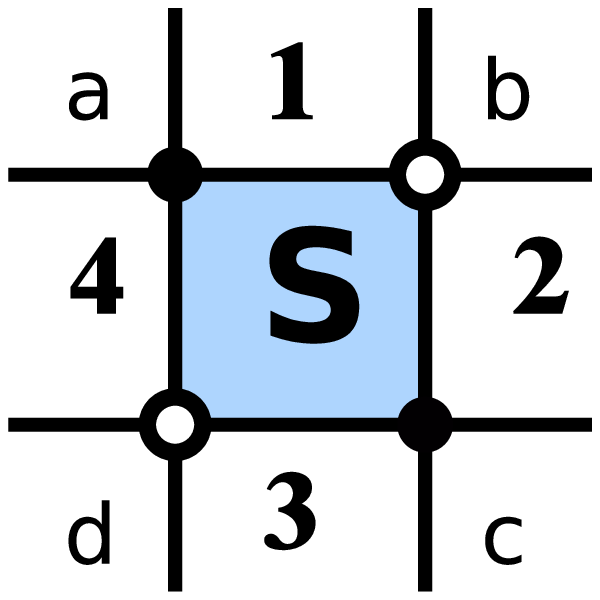} & \includegraphics[width=0.2\textwidth]{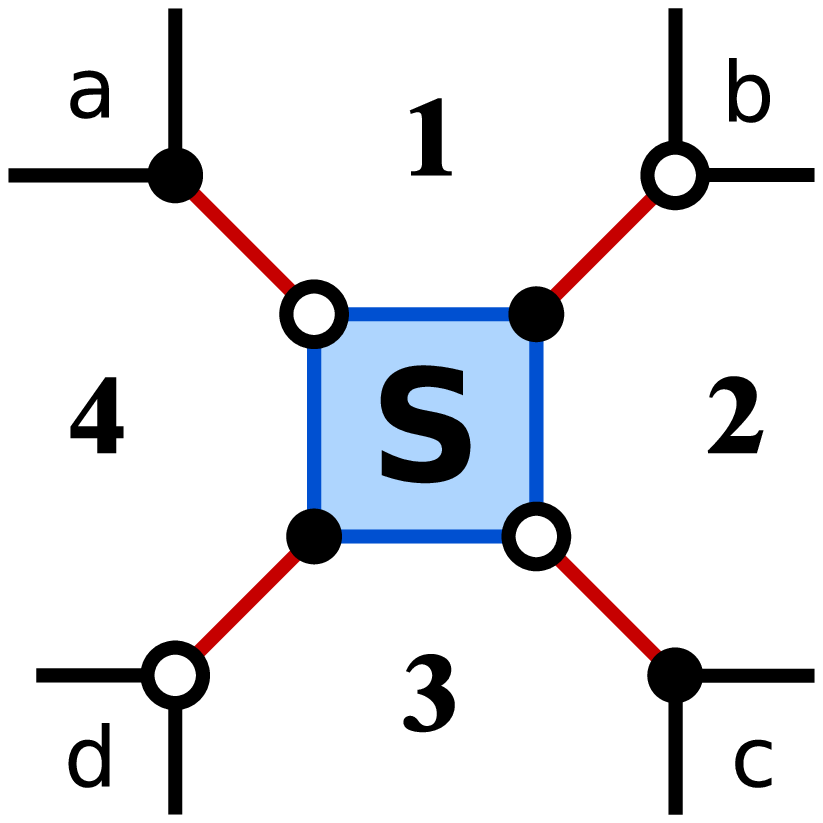}\\
\end{tabular}
\captionof{figure}{\footnotesize{{\bf Seiberg Duality on Quiver.} {\it Left:} A local part of a larger quiver. {\it Right:} The quiver after dualising node S. The fields are reversed (blue) and mesons are added (red).
{\bf Seiberg Duality on Dimer.} \textit{Left}: A local part of a larger dimer. \textit{Right}: The same face after Seiberg dualisation. The white and black nodes are reversed and there are new mesons shown in red. Black and white nodes correspond to opposite signs in the superpotential.}\label{fig:dimerseiberg}}
\end{center}

We can now write down the effect of Seiberg duality on zigzag diagrams. The left hand side of Figure~\ref{fig:seibergzigzag} shows a local section of a zigzag diagram which corresponds to the local section of the dimer shown in Figure~\ref{fig:dimerseiberg}. To dualise this zigzag diagram we must introduce Seiberg dual mesons. In order to reverse the $(N_S,\bar{N}_i)$ fields we also need to change the orientation of the zigzag paths surrounding face $S$. We can do this by twisting the zigzag paths.  The net effect is shown in the right hand figure of Figure~\ref{fig:seibergzigzag} and is local in the zigzag diagram.

\begin{center}
\begin{tabular}{c c}
\includegraphics[width=0.3\textwidth]{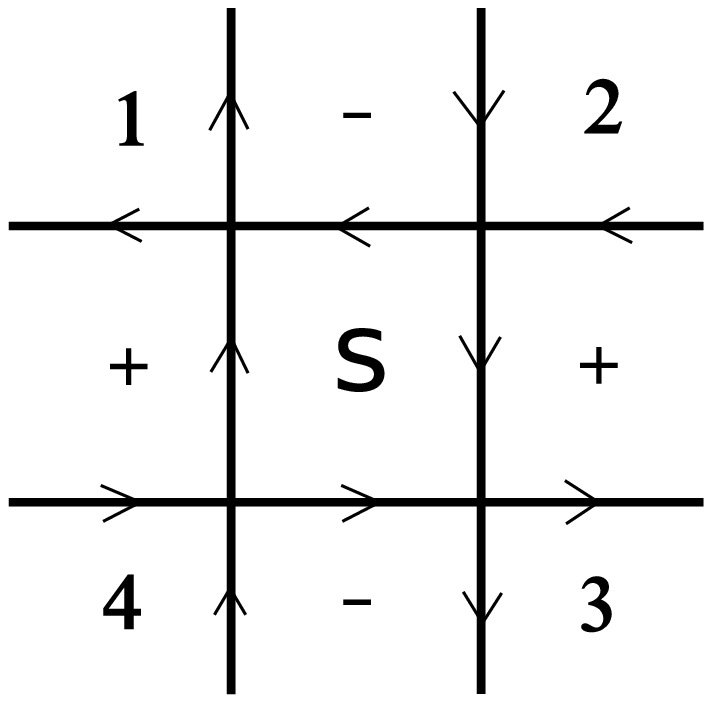} & \includegraphics[width=0.3\textwidth]{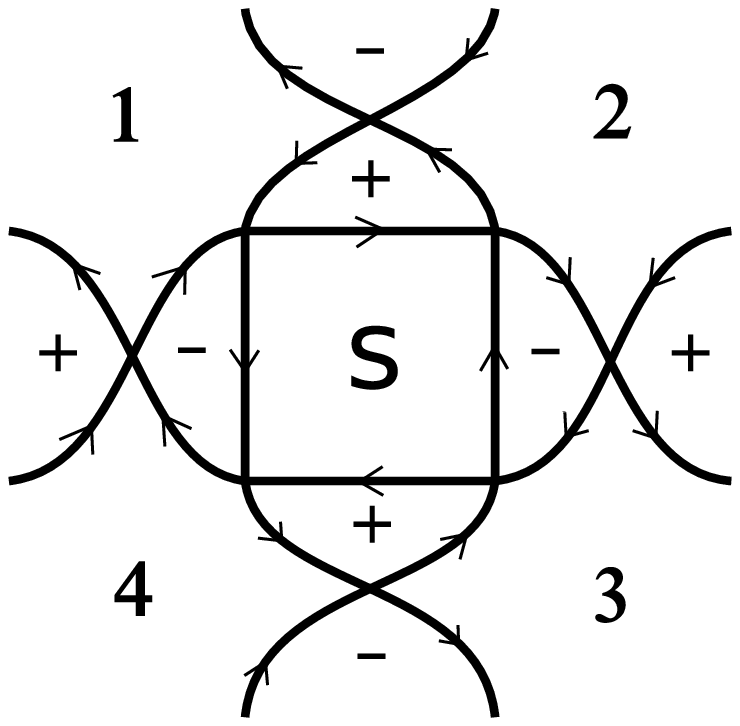}
\end{tabular}
\captionof{figure}{\footnotesize{{\it Left:} A local part of a zigzag diagram  which will be Seiberg dualised. {\it Right}: The effect of dualising the left hand zigzag diagram. The orientation of the paths around face $S$ changes and twists in the zigzag paths are introduced.}\label{fig:seibergzigzag}}
\end{center}

Another way of obtaining toric phases of the gauge theory is by allowing for additional crossings of zigzag paths. 
Specifically, these are additional crossings that are not of mass type and not between paths of same winding number. The canonical example of this is the creation of a pair of paths with opposite winding; unlike in Operation II they are not created at the same time but via Operation I performed twice. The crucial point is that the two paths are created locally distinctly such that they do not generate a mass term unlike in Figure~\ref{fig:addcrossing}.
As an example, in Figure \ref{dp3seibergdual} we show phase 3 of $dP_3$~\cite{0109063}, which is Seiberg dual to the $dP_3$ gauge theory obtained in Section~\ref{sec:toricapp} from Gulotta's algorithm. Phase 3 is obtained if one creates the $(-1,1)$ and $(1,-1)$ paths of $dP_3$ after each other and in a locally distinct fashion.

We should note that using this method only a limited number of toric phases can be found, whereas all toric phases may be obtained by using the twisting procedure described earlier in this section. 

\begin{center}
\begin{tabular}{c c}
\includegraphics[width=0.3\textwidth]{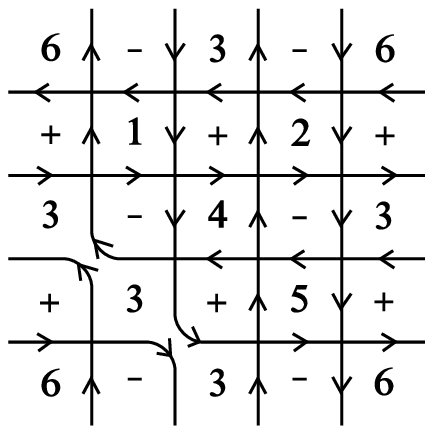} &\includegraphics[width=0.3\textwidth]{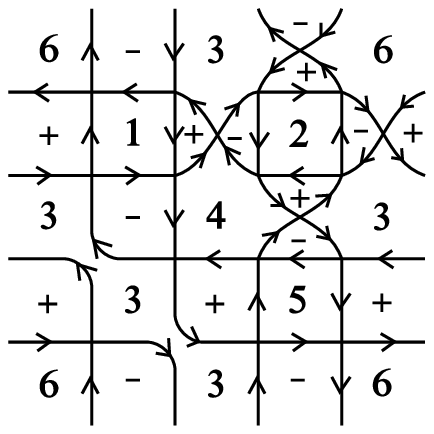}\\
\end{tabular}
\captionof{figure}{\footnotesize{{\it Left:} Dimer for phase 3 of $dP_3$, obtained by allowing for additional crossings. {\it Right:} Dimer of phase 4 of $dP_3$, obtained by dualising gauge group "2."}\label{dp3seibergdual}}
\end{center}

What are the implications of this discussion for the numbers of families? The toric condition limits us to dualising gauge groups with $2N_c$ flavours. This means that there must be two fields transforming in the fundamental and two in the anti-fundamental of the gauge group which will be dualised. Therefore, there can be a maximum of four Seiberg dual mesons and a maximum of four families from Seiberg dualisation is possible.
Given the arguments we have presented this section and the previous one,  $\mathbb{F}_0$ is the only example where Seiberg duality leads to four families.

\subsection{Obtaining further families}
\label{sec:furtherfamilies}

There are a number of ways to obtain further families, thereby evading the bounds we have derived.
Our bound is based on the assumption of toricity. In terms of zigzag paths this implies that there are no self-intersecting or inconsistently constructed zigzag paths.
From a field theory perspective one can Seiberg dualise gauge groups such that one is no longer in a toric phase.
It is known that the non-toric phases often exhibit more than four families~\cite{0212021,0306092}.

There is currently no dimer-based technique available for the investigation of non-toric phases. Using zigzag diagrams one can dualise gauge groups by twisting zigzag paths. 
It may be that such twist operations could be used to understand non-toric phases, possibly leading to more families.
We find the prospect that zigzag diagrams could lead to a greater understanding of the non-toric phases to be an exciting one.

Another way of circumventing the bound is by Higgsing fields in the quiver. 
With Higgsing one can in principle create {\it bendy} tubes which could lead to the proliferation of families.

\section{Yukawa couplings and masses at toric singularities}
\label{sec:masses}

In this section we explore the structure of Yukawa couplings for gauge theories at toric singularities. The models that we shall discuss shall have unequal gauge groups at
the nodes and hence non-conformal. In the absence of supersymmetry
breaking they would flow to strongly coupled superconformal field theories
in the infrared.  We note that at scales below the SUSY breaking scale
the renormalization group flow  would be significantly modified, as
in \cite{0810.5660} we would require to arrange for scales such that
the gauge couplings are weak at the SUSY breaking scale. We show that for models capable of accounting for the multiplicity of families, one quark mass is zero at tree level, whereas the other two are generically non-zero and can differ hierarchically.

Let us start with the quark sector of the MSSM superpotential,
\begin{equation}
W\supset y^u_{ij} Q_L^i H_u u_R^j+y^d_{ij} Q_L^i H_d d_R^j,
\end{equation}
where  $y_u$ and $y_d$ are $3\times 3$ Yukawa coupling matrices. Although the MSSM has only two Higgs fields it is natural in string phenomenology to have more, as for instance in the models of~\cite{0005067,0508089}. The values of the Yukawa couplings are then determined by the values of the Higgs vevs and those of any moduli which may appear in the superpotential, and hence depend on the geometric realisation of the model.
Toric singularities are special in this regard since there are no complex structure moduli fields which appear in the superpotential, and the mass hierarchies and Yukawa couplings depend only on the vevs of the extended Higgs sector. The superpotential for an MSSM-like model at such a singularity is then best written as 
\begin{equation}
W=Q_L^ic_{ij}(H_u^k) u^j+ Q_L^i\tilde{c}_{ij}(H_d^k) d^j,
\end{equation}
where $c_{ij}$ and $\tilde{c}_{ij}$ are functions of the extended Higgs sector.

As an explicit example, consider the {\it Standard-Model} $dP_0$ model of \cite{0005067}, shown in Figure \ref{dp0sm}.
\begin{center}
\includegraphics[width=0.3\textwidth]{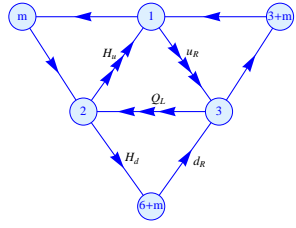}
\captionof{figure}{\footnotesize{The quiver for the Standard Model realised at a $dP_0$ singularity, taken from~\cite{0005067}.}\label{dp0sm}}
\end{center}
The up-quark Yukawa couplings in this model are determined by the $dP_0$ superpotential:
\begin{equation}
W=\left(\begin{array}{c}
Q_L^1\\
Q_L^2\\
Q_L^3
\end{array}\right)
\left(
\begin{array}{ccc}
 0 & Z_{12} & -Y_{12} \\
 -Z_{12} & 0 & X_{12} \\
 Y_{12} & -X_{12} & 0
\end{array}
\right)
\left(\begin{array}{c}
u_R^1\\
u_R^2\\
u_R^3
\end{array}\right),
\end{equation}
where the matrix of Higgs fields is the Yukawa matrix. It is this matrix which determines the flavour structure and masses.
The squares of quark masses associated with the Yukawa matrix are given by the eigenvalues of
\begin{equation}
M=Y.Y^\dagger.
\end{equation}
For $dP_0$ the mass eigenvalues are $0$ which appears once and $|X_{12}|^2+|Y_{12}|^2+|Z_{12}|^2$, which appears twice. Hence the $dP_0$ singularity lacks sufficient structure to provide realistic Yukawa couplings. As a resolution, this model was embedded into $dP_1$ ~\cite{0810.5660} and the eigenvalues were found to be
\begin{equation}
m_i^2 = \left(0,|Y_{12}|^2+ \frac{|\Phi _{61}|^2}{\Lambda^2}\left(|X_{12}|^2+|Z_{12}|^2 \right), |X_{12}|^2+|Y_{12}|^2+|Z_{12}|^2 \right).
\end{equation}
We find that for the higher del Pezzo surfaces the mass eigenvalues have the same structure $(0,m, M)$ where $m\neq M$. 
The zero eigenvalue can be explained as follows.  Firstly, the toric condition implies that every field appears twice in the superpotential, once with a positive sign and once with a negative sign. The determinant of the Yukawa matrix is  the difference of two terms which are equal, each term being the product of all the Higgs fields. Since the matrix $M$ is the product of $Y$ and its Hermitian conjugate, the vanishing determinant of $Y$ guarantees the existence of a zero eigenvalue of $M$. 

In the following section we show that the vanishing determinant is a generic feature of toric singularities, subject to physically motivated constraints regarding the choice of fields labelled as quarks.
We comment further on the possible mass hierarchies in Section~\ref{sec:masshierarchies}.

\subsection{Mass matrix for the rectangular grid}

In order to discuss universal features of the mass matrix in toric singularities, we start with the father theory of all toric singularities, the rectangular grid. The dimer associated to the rectangular grid is a chess-board (cf. Figure~\ref{gridexample}). Our strategy is to convert any statement true for the grid to a corresponding statement in the daughter theories made by partial resolution of the singularity. 
While resolving the singularity we shall allow for generic higgsings, even those which take us to non-toric singularities

For the rectangular grid all interactions are quartic and each field appears exactly twice in the superpotential. Before proceeding we therefore must clarify the meaning of 'quarks' and 'Yukawa couplings'.
A Yukawa coupling in this context is a term in the (quartic) superpotential which involves exactly two quarks, one left-handed and one right-handed. We require that there are no terms in the superpotential which involve more than two quarks\footnote{A set of fields with this property has been discussed in studies of formal aspects of dimers, and is dubbed a perfect matching.} (for both left and right handed).

This implies that the set of the left-handed (and respectively right-handed) quarks is part of a perfect matching of the dimer. 
This leads to an upper bound on the number of fields we may designate as 'quarks'. In particular, for an $n\times m$ chess-board the maximum number of 'quarks' we may choose is $n\cdot m$.

By Higgsing one can break two gauge groups to the diagonal gauge group, leading to family replication from quarks which were originally at different places in the quiver and transforming under different gauge groups. An example of this is given in Appendix~\ref{sec:breakdown}. To take this possibility into account, we adopt a broad definition of quark in this section: the quarks are just some specific choice of fields in the chess-board, free of any constraints of transformation under the same gauge group. They are subject to the constraint above, namely that not more than two fields in any of the quartic superpotential terms are quarks. 

There is one final consistency condition in choosing the quarks. This is that the quarks should be chosen as connected or alternating lines, such as in Figure~\ref{connected}. Choosing the quarks via disconnected lines, as in Figure~\ref{fig:disconnected} makes it impossible to vev the remaining fields in such a way that one ends up with a common gauge symmetry under which all quarks transform. 

In what follows we deal systematically with the three distinct possibilities that arise from the above discussion. The first is that we choose the maximal number of 'quarks', $n\cdot m$.  The second possibility is that there is less than the maximal number of quarks.  Finally and for completeness we discuss the seemingly unphysical case where the quarks are chosen in a disconnected manner.

\subsubsection*{A: Maximal number of quarks}

To illustrate our argument we consider the specific case of the $2\times 2$ chess-board. Our result also holds more generally for the $n\times m$ chess-board. 
Figure~\ref{connected} shows a choice of perfect matchings which corresponds to those chosen for the del Pezzo surfaces in Section~\ref{sec:toricapp}, apart from  the changes in quarks $X_{86}$, $X_{81},$ and $X_{28}.$ In the case of the del Pezzos we had the corresponding quark $X_{58}.$
\begin{center}
\includegraphics[width=0.4\textwidth]{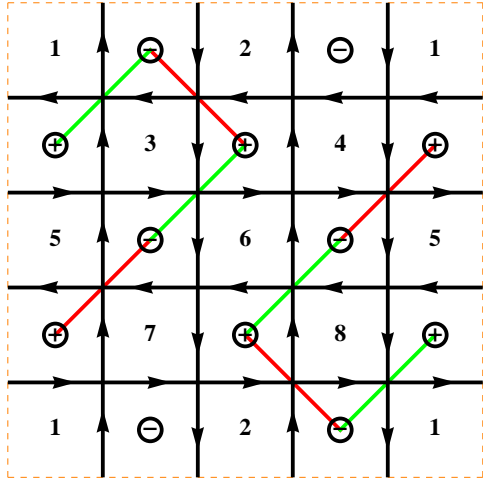}
\captionof{figure}{\footnotesize{Here we display the two perfect matchings which correspond to the choice of quarks in the Yukawa matrix Eq.~\ref{eq:maxyukawa}.}\label{connected}}
\end{center}
The Yukawa matrix in this case is 
\begin{eqnarray}
 W&=&\left(
\begin{array}{c}
 X_{23} \\
 X_{75} \\
 X_{28} \\
 X_{45}
\end{array}
\right)\left(
\begin{array}{cccc}
 -X_{17} X_{72} & X_{42} X_{64} & 0 & 0 \\
 0 & -X_{53} X_{67} & X_{17} X_{58} & 0 \\
 0 & 0 & -X_{14} X_{42} & X_{67} X_{72} \\
 X_{14} X_{53} & 0 & 0 & -X_{58} X_{64}
\end{array}
\right)\left(
\begin{array}{c}
 X_{31} \\
 X_{36} \\
 X_{81} \\
 X_{86}
\end{array}
\right).
\label{eq:maxyukawa}
\end{eqnarray}
The determinant of the Yukawa matrix vanishes and there is one vanishing eigenvalue.

\subsubsection*{B: The case of three quarks}

If we do not choose the maximum number of possible 'quarks' then the situation changes slightly. We consider the case of three quarks in the $2\times 2$ chess-board. 
\begin{center}
\includegraphics[width=0.4\textwidth]{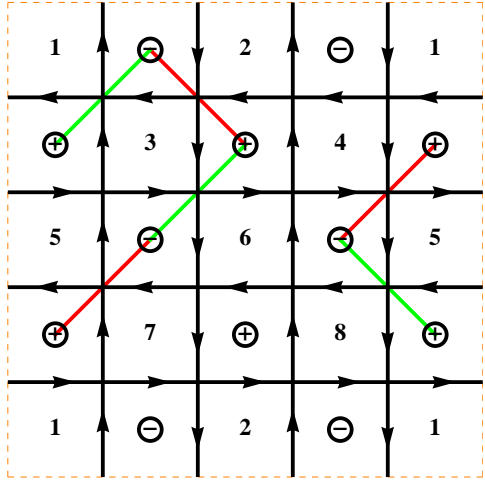}
\captionof{figure}{\footnotesize{Choosing three quarks in the grid.}\label{threequarksingrid}}
\end{center}
We can write the superpotential as follows
\begin{eqnarray}
\nonumber W &=& \left(\begin{array}{c}
X_{23}\\ X_{75}\\ X_{45}
\end{array}\right)
\left(\begin{array}{c c c}
-X_{17}X_{72} & X_{42}X_{64} & 0\\
0 &-X_{53}X_{67}&X_{81}X_{17}\\
X_{14}X_{53}& 0 & -X_{86} X_{64}
\end{array}\right)
 \left(\begin{array}{c}
X_{31}\\ X_{36}\\ X_{58}
\end{array}\right)\\ && +X_{28}X_{86}X_{67}X_{72}-X_{14}X_{42}X_{28}X_{81}~.
\end{eqnarray}
The determinant of the Yukawa matrix is non-zero:
\begin{equation}
\det (Y)=X_{17}X_{53}X_{64}(X_{14} X_{42}X_{81}-X_{67}X_{72}X_{86})
\end{equation}
One can see by vevving fields $\langle X_{72}\rangle=m_1$ and $\langle X_{86}\rangle =m_2$ (joining gauge groups 2 \& 7 and 6 \& 8 as in $dP_3$) the bracket in the determinant simplifies to $X_{14} X_{42}X_{81}-m_1 m_2 X_{62}$.  One finds that the superpotential  has mass terms for the fields $X_{62}$ and $X_{26}$. Integrating them out leads to the vanishing of the determinant.

We observe that although we started with a non-vanishing determinant in the first place, vevving and integrating out lowers the rank of the Yukawa matrix, generating one zero eigenvalue. Again, our argument should generalise to larger numbers of quarks on larger chess-boards.

\subsubsection*{C: Disconnected choice of quarks}

The final possibility is to take a disconnected choice of quarks. It has been mentioned at the beginning of this section that in this case there is no way to Higgs the theory such that the quarks share a common gauge symmetry. We speculate it may be possible to construct realistic theories based upon orientifolding. 
\begin{center}
\includegraphics[width=0.4\textwidth]{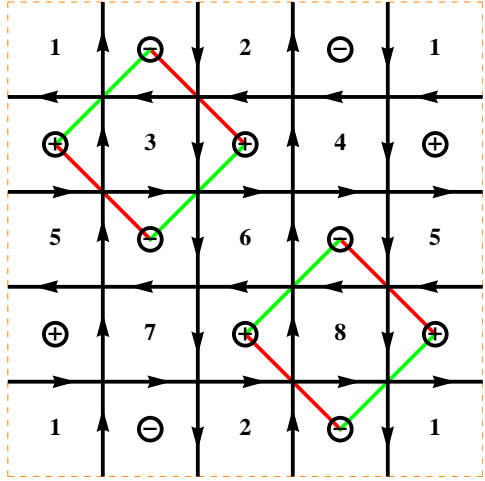}
\captionof{figure}{A disconnected choice of quarks.\label{fig:disconnected}}
\end{center}
For the disconnected choice of quarks in Figure~\ref{fig:disconnected} the Yukawa matrix is
\begin{eqnarray}
\nonumber W&=&X_{14} X_{31} X_{45} X_{53}+X_{23} X_{36} X_{42}
X_{64}-X_{17} X_{23} X_{31} X_{72}-X_{14} X_{28} X_{42} X_{81}\\ \nonumber &&+X_{17} X_{58} X_{75} X_{81}-X_{53} X_{67} X_{75} X_{81}-X_{45} X_{58} X_{64} X_{86}+X_{28}
X_{67} X_{72} X_{86}\\ &=&
\left(
\begin{array}{c}
 X_{23} \\
 X_{53} \\
 X_{58} \\
 X_{28}
\end{array}
\right)\left(
\begin{array}{cccc}
 -X_{17} X_{72} & X_{42} X_{64} & 0 & 0 \\
 X_{14} X_{45} & 0 & -X_{67} X_{75} & 0 \\
 0 & 0 & X_{17} X_{75} & -X_{45} X_{64} \\
 0 & 0 & -X_{14} X_{42} & X_{67} X_{72}
\end{array}
\right)\left(
\begin{array}{c}
 X_{31} \\
 X_{36} \\
 X_{81} \\
 X_{86}
\end{array}
\right).
\end{eqnarray}
In this case the determinant of the Yukawa matrix does not vanish:
\begin{equation}
\det{Y}=X_{14}^2 X_{42}^2 X_{45}^2 X_{64}^2-X_{14} X_{17} X_{42} X_{45} X_{64} X_{67} X_{72} X_{75}\, .
\end{equation}

There are many different ways to embed non-maximal numbers of quarks once the requirement of connectedness is dropped. However, given a non-maximal non-connected choice of quarks, it is always possible to embed them into a maximal embedding, by completing the perfect matching associated with the maximal choice. This makes it clear that whatever the non-maximal choice is, the associated Yukawa matrix can be found by removing rows and column from the maximal Yukawa matrix. Whether the determinant vanishes before Higgsing depends on the choice of quarks, and it is possible to find examples with and without zero eigenvalues. 

\subsection{Mass hierarchies}
\label{sec:masshierarchies}

The hierarchy amongst the quark masses observed in nature is highly puzzling. In string phenomenology these masses are tiny compared to the string scale in most models and it is a very delicate problem to generate the correct quark masses dynamically. Ultimately the size of these masses is related to cycle volumes and the effect of higher order corrections needs to be carefully analysed.  
Thus, zero quark masses for the lightest generation seems to be a natural starting point.
The zero quark mass which we have found to be generic in toric singularities is therefore very attractive.
Let us now explore whether the other two eigenvalues can be hierarchical, neglecting higher order corrections (e.g.~from the K\"ahler potential).

In the case of one zero eigenvalue the characteristic polynomial of the mass matrix simplifies to a quadratic equation.  For $dP_1$ this leads to the non-vanishing eigenvalues:
\begin{eqnarray}
\nonumber\lambda_{1,2}&=&\frac{1}{2} \left(2 |Y_{12}|^2+\left(|Z_{12}|^2+|X_{12}|^2\right)\left(1+\frac{|\Phi_{61}|^2}{\Lambda^2}\right)\pm\left(|X_{12}|^2+|Z_{12}|^2\right)\left(1-\frac{|\Phi_{61}|^2}{\Lambda^2}\right)\right)\\
&=&\begin{cases}
|Y_{12}|^2+|Z_{12}|^2+|X_{12}|^2\, ,\\
|Y_{12}|^2+\frac{|\Phi_{61}|^2}{\Lambda^2}\left(|Z_{12}|^2+|X_{12}|^2\right).
\end{cases}
\end{eqnarray}
For the field theory description to be valid, the vev for $\Phi_{61}$ has to be (much) smaller than the cutoff $\Lambda.$ The physically interesting limit of $\Phi_{61}/\Lambda\ll 1$ leads to a hierarchy between the two non-vanishing eigenvalues, unless $X,Z \ll Y$~\cite{0810.5660}.
The expression for the mass eigenvalues for the higher toric del Pezzo surfaces are complicated. However, it is easy to isolate conditions under which they are hierarchical. In $dP_2$, the discriminant of the solution of the quadratic equation is given by:
\begin{eqnarray}
\nonumber&&\Lambda ^4 |X_{12}|^4 \left(\Lambda ^2-|\Phi _{61}|^2\right){}^2+\left(|Y_{64}|^2 \left(\Lambda ^4-\Lambda ^2 |\Psi _{42}|^2\right)+|Z_{14}|^2 \left(\Lambda
^4-|\Phi _{61}|^2 |\Psi _{42}|^2\right)\right){}^2\\
\nonumber &&-2 \Lambda ^2 |X_{12}|^2 \left(\Lambda ^2-|\Phi _{61}|^2\right) \left(|Y_{64}|^2 \left(\Lambda ^4-\Lambda
^2 |\Psi _{42}|^2\right)+|Z_{14}|^2 \left(-\Lambda ^4+|\Phi _{61}|^2 |\Psi _{42}|^2\right)\right)\\
&\geq& \left(\Lambda ^2 |X_{12}|^2 \left(\Lambda ^2-|\Phi _{61}|^2\right)- \left(|Y_{64}|^2 \left(\Lambda ^4-\Lambda
^2 |\Psi _{42}|^2\right)+|Z_{14}|^2 \left(\Lambda ^4-|\Phi _{61}|^2 |\Psi _{42}|^2\right)\right)\right)^2.\hspace{1cm}
\end{eqnarray}
In the physical scenario of small vevs for $\Phi_{61}$ and $\Psi_{42},$ to obtain a hierarchy we also have to impose the following condition:
 \begin{equation}
\frac{4}{\frac{X_{12}}{Y_{64}}+\frac{Y_{64}}{X_{12}}+\frac{Z_{12}^2}{X_{12}Y_{64}}}\ll 1\, .
\end{equation}
This is satisfied if any of the ratios in the denominator are large. 
Clearly there are more directions in the $dP_2$ moduli space compared to the $dP_1$ moduli space that generate a hierarchy.
In the case that the discriminant equals its lower bound and satisfies the above constraint, the eigenvalues are given by the hierarchichal structure:
\begin{equation}
\lambda_{1,2}=\begin{cases}
|X_{12}|^2+|Y_{64}|^2+|Z_{14}|^2,\\
\frac{|\Phi_{61}|^2}{\Lambda^2}|X_{12}|^2+\frac{|\Psi_{42}|^2}{\Lambda^2}|Y_{64}|^2+\frac{|\Phi_{61}|^2|\Psi_{42}|^2}{\Lambda^4}|Z_{14}|^2.
\end{cases}
\end{equation}
Similarly, the hierarchy in $dP_3$ is at least
\begin{equation}
\lambda_{1,2}=\begin{cases}
|X_{12}|^2+|Y_{64}|^2+\frac{|\rho_{53}|^2}{\Lambda^2}|Z_{14}|^2,\\
\frac{|\Phi_{61}|^2}{\Lambda^2}|X_{12}|^2+\frac{|\Psi_{42}|^2}{\Lambda^2}|Y_{64}|^2+\frac{|\Phi_{61}|^2|\Psi_{42}|^2|\rho_{53}|^2}{\Lambda^6}|Z_{14}|^2.
\end{cases}
\end{equation}

Whether this hierarchy generalises to other singularities depends on the particular Higgs structure and has to be examined on a case by case basis. However, the eigenvalues will generically be different and hence such limits ought to exist.

\section{The CKM matrix}
\label{sec:ckm}

In this section we explore the flavour structure of gauge theories at toric singularities, with the goal of obtaining models which have a realistic CKM matrix. We begin with a short review of flavour mixing in the Standard Model, primarily to set
conventions. When written in terms of quark fields which are gauge eigenstates, the
flavour structure is encoded in  two $3\times 3$ complex matrices, the Yukawa matrices   $Y_{u}$ and $Y_{d}$ of the up and
down generations. The change of variables to mass eigenstates involves rotations by
independent unitary transformations for the left and right handed quarks ($u_{L} \to V_{u} u_{L}$,
$u_{R} \to \tilde{V}_{u} u_{R}$, $d_{L} \to V_{d} d_{L}$ and $d_{R} \to \tilde{V}_{d} d_{R}$). These transformations
diagonalise the Yukawa matrices ($Y_{u} \to V_{u}^{T} Y_{u} \tilde{V}_{u} = D_{u}$, $Y_{d} \to V_{d}^{T} Y_{d} \tilde{V}_{d} = D_{d}$)
and induce flavour changing processes via the weak current. The coupling between various generations is
given by the CKM matrix
\begin{equation}
   V_{\rm CKM} = V_{u}^{\dagger} V_{d}\, .
\end{equation}
A useful method to compute the CKM matrix is to consider the Hermitian matrices
$Y_{u}Y^{\dagger}_{u}$ and $Y_{d}Y^{\dagger}_{d}$.
$V_{u}$ and $V_{d}$  then correspond to the unitary matrices associated
with the similarity transformations which diagonalise $Y_{u}Y^{\dagger}_{u}$ and $Y_{d}Y^{\dagger}_{d}$. 
Experimentally the CKM matrix is~\cite{Amsler:2008zzb}
\begin{equation}
V_{\rm CKM}=\left(\begin{array}{c c c}
0.97419\pm0.00022          & 0.2257\pm0.0010    & 0.00359\pm0.00016   \\
0.2256\pm0.0010            & 0.97334\pm0.00023  & 0.0415^{+0.0010}_{-0.0011} \\
0.00874^{+0.00026}_{-0.00037} & 0.0407\pm0.0010    & 0.999133^{+0.000044}_{-0.000043}\\
\end{array}\right).
\end{equation}
To leading order in each entry of the matrix this can be  parametrised hierarchically as 
\begin{equation}
V_{\rm CKM} = \left(\begin{array}{c c c}
1 & \epsilon & \epsilon^3 \\
\epsilon & 1 & \epsilon^2 \\
\epsilon^3 & \epsilon^2 & 1 \\
\end{array}\right),
\label{ckmscale}
\end{equation}
where $\epsilon \sim 0.2$.
Another useful parametrisation is by three angles and a complex phase,
\begin{eqnarray}
\nonumber
V_{\rm CKM} &=& R_{23}. R_{13}. R_{12} \cr
&=&
\left(\begin{array}{c c c}
1& 0 & 0 \\
0 & c_2 & s_2 \\
0 & -s_2& c_2
\end{array}
\right)
\left(\begin{array}{c c c}
c_3& 0 & s_3 e^{-i \delta} \\
0 & 1 & 0 \\
-s_3 e^{i\delta} & 0& c_3
\end{array}
\right)
\left(\begin{array}{c c c}
c_1& s_1 & 0 \\
-s_1 & c_1 & 0 \\
0 & 0& 1
\end{array}
\right) \label{rot}\\
&=&\left(\begin{array}{c c c}
c_1 c_3& s_1 c_3 & s_3 e^{-i\delta} \\
-s_1c_2-c_1s_2s_3e^{i\delta} & c_1c_2-s_1s_2s_3 e^{i\delta} & s_2 c_3\\
s_1s_2-c_1c_2s_3 e^{i\delta} & -c_1s_2-s_1c_2 s_3 e^{i\delta}& c_2 c_3
\end{array}
\right),
\label{eq:eulerparam}
\end{eqnarray}
using the convention that $c_i=\cos{\theta_i}$ and $s_i=\sin{\theta_i}.$
For now we shall focus only on obtaining the appropriate hierarchies in the CKM matrix, and take its entries to be real which reduces the parametrisation (\ref{rot}) to the Euler angle decomposition of orthogonal matrices. Then, each of the matrices in Eq.~\ref{rot} induces a mixing between two generations.

As discussed in Section~\ref{sec:masses}, models arising from branes at singularities typically involve multiple Higgs fields. We find
that the CKM matrix is very closely related to the ratios of the vevs of these fields. A study of the complete Higgs potential is beyond the scope of this paper, it being a combination of various effects such as D-terms, supersymmetry breaking and radiative corrections. For the present analysis we take a more phenomenological approach, allowing arbitrary vevs of the Higgs fields\footnote{We take care that the vevs of the Higgs fields are always below the cutoff scale as is required for the validity of the effective field theory.} and attempting to identify the regions in field space which give a realistic CKM matrix. Following the philosophy of the  bottom-up approach, our analysis can be thought of as trying to answer the question:  is there enough structure in the singularities to
allow the construction of models with realistic flavour physics? In our analysis we distinguish between two broad classes:
\begin{enumerate}
\item Models with up and down quarks coming from different sources, i.e. D3-D3 and D3-D7 states, and  
\item Models with both up and down quarks coming from D3-D3 states.
\end{enumerate}

\subsection{Up and down quarks from different sources}
\label{sec:d3d7ckm}

We begin with the discussion of the CKM matrix associated to the SM-like model (its $dP_0$ realisation is shown in Figure~\ref{dp0sm}).
The $dP_0$ realisation has the unsatisfying feature that the mass of the two heaviest quarks in the up generation are equal. In addition, as discussed in Section 2, the $SU(3)$ flavour symmetry of the $dP_0$ geometry allows for a generic Yukawa matrix in the down generation. This implies
the from of $V_{d}$ is unrestricted. As a result the form of the CKM matrix is arbitrary and the appropriate
hierarchies can be obtained by tuning the vevs of the Higgs fields. Next, let us discuss the more realistic $dP_1$ model.

\subsubsection*{The $dP_1$ model}

The superpotential for the up quarks in the $dP_1$ model was given in equation~\ref{dp1super}. The  associated unitary
matrix for the up quarks $V_{u}$ is
\begin{equation}
V_{\rm u} =\left(
\begin{array}{ccc}
  a { \Phi_{61} \over \Lambda } X_{12} & - b  Y_{12} X_{12} & - c Z_{12} \\
  a Y_{12}  & b { \Phi _{61} \over \Lambda} ({X_{12}^2 +Z_{12}^2})   & 0 \\
 a { \Phi_{61} \over \Lambda } Z_{12} &  -b Y_{12} Z_{12}  & c X_{12}
\end{array}
\right),
\label{eq:dp1unitary}
\end{equation}
where $a^{-2} = \frac{|\Phi_{61}|^2}{\Lambda^2}  (|X_{12}|^{2} + |Z_{12}|^{2}) + |Y_{12}|^{2} $, $b^{-2} = (|X_{12}|^{2} + |Z_{12}|^{2})\, a^{-2} $ and $c^{-2} = |X_{12}|^{2} + |Z_{12}|^{2}.$
One of the features of the matrix is that one of the entries vanishes. We shall find that such  zero textures are quite common. 
The basic structure of $V_{u}$ is similar to that of $dP_0,$ and $V_u$ for $dP_0$ can be obtained from Eq. \ref{eq:dp1unitary} by setting $\Phi_{61}=\Lambda$, but the form of $V_{d}$ is lot more restricted than in the $dP_0$ model. As discussed in Section 2 the $SU(2)$
flavour symmetry allows for mixing between two generations of quarks and the corresponding matrix
$V_{d}$ induces mixing between only two of the three generations.
Now consider D7 brane wrappings which correspond to mixing between the medium and heavy generations.
Explicitly, we take the down quark Yukawa matrix to be
\begin{equation}
Y_d = \left(\begin{array}{c c c}
m_d & 0 & 0 \\
0 & m_s c_2^2 + m_{b} s_2^2 &  (m_b - m_s) s_2 c_2 \\
0 &  (m_b - m_s) s_2 c_2 & m_b c_2^2 + m_{s} s_2^2   \\
\end{array}
\right)\,.
\label{eq:d3d7yd}
\end{equation}
The corresponding matrix $V_{d}$ is
\begin{equation}
V_d = \left(\begin{array}{c c c}
1& 0 & 0\\
0&c_2 & s_2\\
0& -s_2& c_2
\end{array}\right).
\end{equation}
which is of the form $R_{23}$.

Now, let us examine whether it is possible to obtain a realistic structure of the CKM
matrix. For this purpose we consider  $V_{u}$ in the limit $ {Z_{12} \over X_{12}}  , { {\Lambda Y_{12} \over \Phi_{61} X_{12} } }  \ll 1$, in this limit to leading order
\begin{equation}
V_{u} \approx \left(
\begin{array}{ccc}
 1 & -\frac{\Lambda  Y_{12}}{X_{12} \Phi _{61}} & -\frac{Z_{12}}{ X_{12}} \\
 \frac{\Lambda  Y_{12}}{X_{12} \Phi _{61}} & 1 & 0 \\
 \frac{Z_{12}}{X_{12}} & -\frac{\Lambda  Y_{12} Z_{12}}{\Phi _{61} X_{12}^2 } & 1
\end{array}
\right).
\end{equation}
Note that the Hermitian conjugate of this matrix has precisely the same form  as the product two of the 
three rotation matrices in (\ref{rot})
\begin{equation}
R_{12}.R_{13}=\left(\begin{array}{c c c}
c_1 c_3 & s_1 & c_1 s_3\\
-c_3 s_1 & c_1 & -s_1 s_3\\
-s_3 & 0 & c_3
\end{array}\right)
\end{equation}
with $s_1 \approx -\frac{\Lambda  Y_{12}}{X_{12} \Phi _{61}} $ and $ s_3 \approx \frac{Z_{12}}{X_{12}}$.

Thus the CKM matrix $V_{CKM} = V_{u}^{\dagger}V_{d}$ has the right structure to describe mixing between all the generations. To obtain the correct hierarchical flavour mixing, we require
\begin{equation}
\frac{\Lambda  Y_{12}}{X_{12} \Phi _{61}} \approx \epsilon\, ,\; \frac{Z_{12}}{X_{12}}\approx \epsilon^3\, , \; s_2\approx \epsilon^2\, .
\end{equation}
Note that in this limit the mass eigenvalues are hierarchical.  Therefore $dP_1$ is attractive from the perspective of both mass hierarchies and
flavour mixing.

\subsubsection*{Higher dPs}

One can also consider extending the model to the higher del Pezzo singularities. The absence of
isometries in the geometry implies that at tree-level the mass matrix for the down quarks is diagonal,
and so the matrix $V_{d}$ is equal to the identity matrix, implying $V_{CKM} = V_{u}^{\dagger}$.
The $V_{u}$ matrices are complicated and we do not present their explicit form. One feature is that the texture zero present in $dP_0$ and $dP_1$ is absent. It can also be shown that there are relationships relating the second-third generation mixing to the first-second and first-third which force the second-third generation mixing to be of smaller magnitude than the first-third.\footnote{A realistic value of the second-third generation mixing can arise if some fields take vevs larger than the cutoff.
However, the presence of such vevs  most likely leads to the break down of the effective field theory.} Thus it is not possible to
obtain a fully realistic structure at tree level when the model is extended to the higher del Pezzo surfaces.

However, we note that there is a consistent truncation where the only non-vanishing mixings induced by $V_{u}$ are those between the first and second generations.
This, together with  $V_{d}$ equal to the identity matrix, gives a CKM matrix with  mixings between the first and second generations
and vanishing mixings between the rest. From a pragmatic point of view, it is important to remember that the CKM matrix
is sensitive to kinetic terms and our computations are based on the leading order K\"ahler potential. In general,
the K\"ahler potential receives corrections from various sources and computing extremely small quantities such as the second-third and first-third generation mixings without  incorporating  the corrections to the K\"ahler potential might be unreliable. Thus, while working with the leading order K\"ahler potential a reasonable goal might be to find models where the only non-vanishing
mixing is between first and second generations.

\subsection{Up and down quarks both as D3D3 states}
\label{sec:d3d3}

Models constructed in this class are GUT models with multiple gauge groups such as left-right symmetric, Pati-Salam and trinification~\cite{0810.5660,0005067}. In these models the breaking of the gauge symmetries occurs in two steps: first down to the Standard Model and then electroweak symmetry breaking at a lower scale. In the case of supersymmetric models it is typical that the breaking to the Standard Model preserves supersymmetry. As a result, some fields acquire supersymmetric masses at a very high scale and are non-dynamical in the low energy theory. 

Since both up and down quarks arise from within the singularity, the form of their Yukawa matrices is similar, although characterised by vevs of different fields. In the process of breaking the GUT symmetry, some fields can acquire a vanishing vev thereby giving the Yukawa matrices for the up and down quarks different structures. In order to illustrate this we begin our discussion with a left-right symmetric model constructed on the $dP_1$ quiver.

\subsubsection{Left-right model}

A simple extension of the Standard Model is the left-right model with gauge symmetry $SU(3)_c\times SU(2)_L \times SU(2)_R \times U(1)_{B-L}.$ Its embedding in local models has been studied in \cite{0005067,0810.5660,0001083}. We consider the left-right model from realised on $dP_1$~\cite{0810.5660}, shown in Figure~\ref{fig:leftright}.
Prior to breaking $SU(2)_R$ the Yukawa matrices for the up and down quarks are both of the form of the $dP_1$ Yukawa matrix, as in Eq.~\ref{dp1super}. Each of the Yukawa matrices involves different Higgs fields.
\begin{center}
\includegraphics[width=0.5\textwidth]{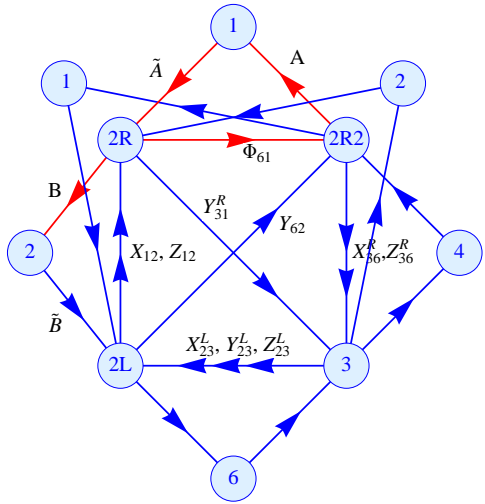}
\captionof{figure}{\footnotesize{The $dP_1$ Left-Right model. We vev the fields $\Phi_{61},$ $A$, $\tilde{A}$ and $B$.}\label{fig:leftright}}
\end{center}
We initially break $U(2)_R^1\times U(2)_R^2$ to $U(1)\times U(1)$ (generated by $T_3^i+Q^i)$ by veving the field $\Phi_{61}$.  
\begin{equation}
\Phi_{61}=\left(
\begin{array}{c c}
0 & 0\\
0 & v\\
\end{array}
\right),
\end{equation}
The Higgs fields $X_{12},$ $Z_{12},$ and $Y_{62},$  are decomposed into
\begin{eqnarray}
(2_L,\bar{2}_R^1) \left(\begin{array}{c}X_{12}\\Z_{12}\end{array}\right)&\to& (2_L,0,0) \left(\begin{array}{c}X_{12}^d\\Z_{12}^d\end{array}\right)+(2_L,-2,0) \left(\begin{array}{c}X_{12}^u\\Z_{12}^u\end{array}\right),\\
(2_L,\bar{2}_R^2) \left(\begin{array}{c}Y_{62}\end{array}\right)&\to& (2_L,0,0) \left(\begin{array}{c}Y_{62}^d\end{array}\right)+(2_L,0,-2) \left(\begin{array}{c}X_{12}^u\\Z_{12}^u\\Y_{62}^u\end{array}\right),
\end{eqnarray}
where the last two entries in the row vectors denote the $U(1)$ quantum numbers. 
The right handed quarks transforming as $(2_R,\bar{3})$ are broken into up and down quarks
\begin{eqnarray}
(2_R^2,\bar{3}) \left(\begin{array}{c} X_{36}\\ Z_{36}\end{array}\right) &\to& (0,0,\bar{3}) \left(\begin{array}{c} X_{36}^d\\ Z_{36}^d\end{array}\right) + (0,2,\bar{3}) \left(\begin{array}{c} X_{36}^u\\ Z_{36}^u\end{array}\right),\\
(2_R^1,\bar{3}) \left(Y_{31}\right) &\to& (0,0,\bar{3}) \left(Y_{31}^d\right) + (2,0,\bar{3}) \left(Y_{31}^u\right),
\end{eqnarray}
where the first two charges indicate again the $U(1)$ charges. 
The superpotential for the quarks is
\begin{eqnarray}
\nonumber W&=&\left(\begin{array}{c}X_{23}^L\\
Y_{23}^L\\
Z_{23}^{L}
\end{array}
\right)
\left(\begin{array}{c c c}
0 & Z_{12}^u& -Y_{62}^u\\
-Z_{12}^u\frac{\varphi}{\Lambda}& 0 & X_{12}^u\frac{\varphi}{\Lambda} \\
Y_{62}^u & -X_{12}^u & 0
\end{array}\right)
\left(\begin{array}{c}
X_{36}^u\\
Y_{31}^u\\
Z_{36}^u
\end{array}\right)\\
&+&\left(\begin{array}{c}X_{23}^L\\
Y_{23}^L\\
Z_{23}^{L}
\end{array}
\right)
\left(\begin{array}{c c c}
0 & Z_{12}^d& -Y_{62}^d\\
-Z_{12}^d\frac{v}{\Lambda}& 0 & X_{12}^d\frac{v}{\Lambda} \\
Y_{62}^d & -X_{12}^d & 0
\end{array}\right)
\left(\begin{array}{c}
X_{36}^d\\
Y_{31}^d\\
Z_{36}^d
\end{array}\right)+W_{\rm D3D7}\, ,
\end{eqnarray}
where $\varphi$ is the remaining field of $\Phi_{61}$ transforming as $(2,-2)$ under the unbroken $U(1)$ factors. As discussed earlier the Yukawa matrix for the up and down quarks have the same structure, although the Higgs fields appearing in each matrix are different.

We must ensure that both the abelian and non-abelian D-term equations are satified. The former can always be satisfied by tuning the FI terms, corresponding to resolving the singularity. The non-abelian D-terms require more fields to acquire vevs. A consistent solution can be obtained by giving the fields $A$, $\tilde{A}$ and  $B$ vevs. These fields are shown in red in Figure~\ref{fig:leftright}.
In Appendix~\ref{sec:dterms} we show explicitly how the non-abelian D-term equations are satisfied. 

The effect of vevving D7D3 states is that we induce large supersymmetric mass terms which then are integrated out, rendering certain couplings absent at low energies. Any terms involving only the high scale vevs may safely be integrated out. However, terms of magnitude $v/\Lambda$ should be kept in the low energy superpotential.
With the vev structure from the Appendix~\ref{sec:dterms}, we find the following term in the superpotential which includes $v$:
\begin{equation}
W\supset v X_{12}^d \tilde{B}\, .
\end{equation}
This generates a large mass for $X_{12}^d,$ and the F-term for $\tilde{B}$ requires that $X_{12}^d=0$. Thus at low energies the superpotential for the down quarks is 
\begin{eqnarray}
 W&=&\left(
\begin{array}{c}
 X_{23}^d \\
 Y_{23}^d \\
 Z_{23}^d
\end{array}
\right)\left(
\begin{array}{ccc}
 0 & Z_{12}^d & -Y_{62}^d \\
 -Z_{12}^d \frac{v_d}{\Lambda} & 0 & 0\\
 Y_{62}^d & 0 & 0
\end{array}
\right)\left(
\begin{array}{c}
 X_{36}^d \\
 Y_{31}^d \\
 Z_{36}^d
\end{array}
\right).
\label{xgoney}
\end{eqnarray}
One can also find consistent vevs for the other Higgs fields which set the fields $Z_{12}^d$ and $Y_{62}^d$ to zero.
The up quark Yukawa matrix is unchanged.

\subsubsection{Flavour mixing}

For the down quark Yukawa matrix in Eq.~\ref{xgoney} the matrix $V_d$ is
\begin{equation}
 V_{d} = \left(
\begin{array}{ccc}
 0 & 0 & 1 \\
 a\frac{\Lambda Y^d_{12}}{Z^d_{12} \Phi _{61}} &  -b \frac{Z^d_{12} \Phi_{61}}{\Lambda  Y^d_{12}} & 0\\
 a & b & 0
\end{array}
\right).
\label{mixone}
\end{equation}
where $a^{-2} = 1 + \left( \frac{\Lambda  Y^d_{12}}{Z^d_{12} \Phi _{61}} \right)^2$ and $b^{-2} = 1+ \left( \frac{Z^d_{12} \Phi _{61}}{\Lambda  Y^d_{12}} \right)^2$. Of course, this matrix could easily be obtained from the matrix in Eq.~\ref{eq:dp1unitary} by simply setting $X^d_{12}=0$.
Similarly, by setting the fields $Z^d_{12}$ and $Y^d_{12}$ to zero one obtains the unitary matrices
\begin{equation}
\left(
\begin{array}{ccc}
c\frac{X^d_{12} \Phi _{61}}{\Lambda  Y^d_{12}} &  -d\frac{\Lambda  Y^d_{12}}{X^d_{12} \Phi _{61}} & 0\\
 c & d & 0 \\
 0 & 0 & 1
\end{array}
\right)\;
\quad {\rm and} \quad
 \left(
\begin{array}{ccc}
 e\frac{X^d_{12}}{Z^d_{12}} & 0& -f\frac{Z^d_{12}}{X^d_{12}}  \\
 0 & 1 & 0 \\
 e & 0 & f
\end{array}
\right),
\label{othermix}
\end{equation}
where $c^{-2} = 1+ \left(\frac{X^d_{12} \Phi _{61}}{\Lambda  Y^d_{12}}\right)^2$, $d^{-2} = 1+ \left( \frac{\Lambda  Y^d_{12}}{X^d_{12} \Phi _{61}} \right)^2$, $e^{-2} = 1+ \left(  \frac{X^d_{12}}{Z^d_{12}} \right)^2 $ and $f^{-2} = 1+ \left( \frac{Z^d_{12}}{X^d_{12}} \right)^2$.

Now let us compute the CKM matrix for $V_d$ in Eq.~\ref{mixone}.
Keeping in mind that the mixing angles in the CKM matrix are small, we take the limit
\begin{equation}
 \frac{\Lambda  Y^d_{12}}{Z^d_{12} \Phi _{61}} \sim  \epsilon^{m}  \ll 1\, .
\end{equation}
We will use the scaling of the CKM matrix in Eq.~\ref{ckmscale} to set the value of $m$ later on.

To obtain a CKM matrix which is close to the identity matrix, at zeroth order in $\epsilon$, $V_{u}$ must be the inverse of $V_{d}^{\dagger}$. 
Motivated by this we take the ratios of fields in the $Y_{u}$ matrix to scale as
\begin{equation}
\frac{X^u_{12}}{Z^u_{12}} \sim \epsilon^{n_{xz}}, \qquad \frac{Y^u_{12}}{Z^u_{12}} \sim \epsilon^{n_{yz}} \quad \mbox{and}\quad
\frac{\Phi_{61}}{\Lambda} \sim \epsilon^{n_{\Phi}}.
\end{equation}
Then to leading order in $\epsilon$
\begin{eqnarray}
V_{\rm u} &=& \left(\begin{array}{c c c}
\frac{X^u_{12}}{Z^u_{12}}&\frac{\Lambda Y^u_{12}X^u_{12}}{\left(Z^u_{12}\right)^2 \Phi_{61}} & 1 \\
\frac{\Lambda Y^u_{12}}{Z^u_{12} \Phi_{61}}&1 & 0\\
1& -\frac{\Lambda Y^u_{12}}{Z^u_{12} \Phi_{61}}& -\frac{X^u_{12}}{Z^u_{12}}
\end{array}
\right) \cr &\sim& \left(\begin{array}{c c c}
\epsilon^{n_{xz}}& \epsilon^{n_{xz}+n_{yz}-n_{\Phi}}& 1\\
\epsilon^{n_{yz}-n_{\Phi}}&1 & 0\\
1& \epsilon^{n_{yz}-n_{\Phi}} & \epsilon^{n_{xz}}\\
\end{array}\right).
\label{sc}
\end{eqnarray}
The resulting CKM matrix is
\begin{eqnarray}
V_{CKM} = V_{u}^{\dagger} V_d  &\approx& \left(\begin{array}{c c c}
\epsilon^{n_{xz}}& \epsilon^{n_{yz}-n_{\Phi}}& 1\\
\epsilon^{n_{xz}+n_{yz}-n_{\Phi}} &1 &  \epsilon^{n_{yz}-n_{\Phi}} \\
1&0 & \epsilon^{n_{xz}}\\
\end{array}\right).\left(\begin{array}{c c c}
0&0 & 1\\
\epsilon^m & 1& 0\\
1& \epsilon^ m& 0
\end{array}
\right) \cr
&=& \left(\begin{array}{c c c}
1& \epsilon^m & \epsilon^{n_{xz}}\\
\epsilon^m &1 & \epsilon^{n_{xz}+n_{yz}-n_{\Phi}} \\
\epsilon^{n_{xz}}&\epsilon^{n_{xz}+m} &1
\end{array}
\right).
\end{eqnarray}
One can see that it is not possible to obtain a realistic second-third generation coupling as the above form of the matrix fixes $m=1$, $n_{xz}=3$ and therefore one of the second-third generation couplings is $\mathcal{O}(\epsilon^4)$. One also can check that the other structures for
$V_{d}$ in (\ref{othermix}) do not give a realistic CKM matrix.

\subsubsection*{Moving to $dP_2$}


This problem can be resolved by considering the model at the $dP_2$ singularity, as described in Appendix~\ref{sec:breakdown}. After integrating the Higgs field $X_{12}^d$ and taking the limit 
\begin{equation}  \frac{\Phi_{61} Z_{14}^d}{\Lambda Y_{64}^d} \sim \epsilon^m \ll 1   \label{eq:d3d3vdscaling} \end{equation}
the matrix $V_d$ takes the form
\begin{equation}
V_d=\left(\begin{array}{c c c}
0&0 & 1\\
1 & \epsilon^m & 0\\
 \epsilon^ m& 1& 0
\end{array}
\right).
\end{equation}
Given this, for the CKM matrix to be close to the identity,  $V_{u}$ at
zeroth order in $\epsilon$ must be of the form
\begin{equation}
V_u \approx \left(\begin{array}{c c c}
0&0 &1 \\
1& 0& 0\\
0& 1&0
\end{array}
\right).
\end{equation}
This form is obtained by taking the scalings
\begin{equation}
\frac{X^u_{12}}{Y^u_{64}},\frac{Z^u_{14}}{Y^u_{64}}\ll 1\; \text{ and } \frac{\Phi_{61}}{\Lambda},\frac{\Psi_{42}}{\Lambda}\ll 1
\label{eq:d3d3vuscaling}
\end{equation}
in $Y_u$, which yield
\begin{equation}
V_u \approx \left(\begin{array}{c c c}
\frac{X^u_{12}\Phi_{61}}{Y^u_{64}\Lambda}&\frac{X^u_{12} Z^u_{14}}{\left(Y^u_{64}\right)^2} & 1\\
1& -\frac{Z^u_{14} \Phi_{61}}{\Lambda Y^u_{64}}& -\frac{X^u_{12}\Phi_{61}}{\Lambda Y^u_{64}}\\
\frac{Z^u_{14} \Phi_{61}}{Y^u_{64} \Lambda}&1 &-\frac{X^u_{12} Z^u_{14}}{\left(Y^u_{64}\right)^2}
\end{array}
\right).
\end{equation}
As introduced earlier, we parametrise the scalings of these ratios by
\begin{equation}
\frac{X^u_{12}}{Y^u_{64}} \sim \epsilon^{n_x}, \qquad \frac{Z^u_{14}}{Y^u_{64}} \sim \epsilon^{n_z} \quad \mbox{and} \quad \frac{\Phi_{61}}{\Lambda} 
\sim \epsilon^{n_{\Phi}}. 
\end{equation}
All contributions which include $\Psi_{42}/\Lambda$ are subleading. The matrix $V_u$ then scales as
\begin{equation}
V_u  \approx \left(\begin{array}{c c c}
\epsilon^{n_x+n_\Phi}&\epsilon^{n_x+n_z} & 1\\
1& \epsilon^{n_z + n_\Phi}&\epsilon^{n_x +n_\Phi} \\
\epsilon^{n_z+n_\Phi}&1 & \epsilon^{n_x+n_z}
\end{array}\right).
\end{equation}
This gives
\begin{eqnarray}
\nonumber V_{\rm CKM} &\approx& \left(\begin{array}{ccc}
\epsilon^{n_x+n_\Phi}& 1 & \epsilon^{n_x+n_\Phi} \\
\epsilon^{n_x+n_z} &  \epsilon^{n_z + n_\Phi}& 1\\
1&\epsilon^{n_x +n_\Phi} & \epsilon^{n_x+n_z}
\end{array}
\right).\left(\begin{array}{c c c}
0&0 & 1\\
1 & \epsilon^m & 0\\
 \epsilon^ m& 1& 0
\end{array}
\right)\\ \cr
&\approx& \left(\begin{array}{ccc}
1 & \epsilon^m & \epsilon^{n_x+n_\Phi}\\
\epsilon^m & 1 & \epsilon^{n_x+n_z}\\
\epsilon^{n_x+n_\Phi} & \epsilon^{n_x+n_z}& 1
\end{array}\right).
\end{eqnarray}
With the following choice of scaling
\begin{equation} m=1,n_x=1,n_z=1,n_\Phi=2 \end{equation} we obtain the correct hierarchical structure for the CKM matrix
\begin{eqnarray}
V_{\rm CKM} \approx 
\left(\begin{array}{ccc}
1& \epsilon& \epsilon^3\\
\epsilon&1 &\epsilon^2 \\
\epsilon^3&\epsilon^2 &1
\end{array}
\right).
\end{eqnarray}

\section{CP Violation}

We have so far neglected the complex phase of the CKM matrix which is the source of CP violation in the Standard Model. A unique measure of the amount of CP violation was introduced by Jarlskog~\cite{Jarlskog:1985ht, Jarlskog:1985cw}.
She introduced the matrix
\begin{equation}
i C=[M_u,M_d]=[Y_uY_u^\dagger,Y_dY_d^\dagger]\, ,
\label{eq:jarlskog}
\end{equation}
and showed that the measure for CP violation is given by $J,$ defined as follows
\begin{equation}
\det{C}=-2 T B J\, ,
\end{equation}
where $T=(m_t^2-m_u^2)(m_t^2-m_c^2)(m_c^2-m_u^2)$ and $B=(m_b^2-m_s^2)(m_b^2-m_d^2)(m_s^2-m_d^2).$ In terms of components of the CKM matrix, $J$ is given as
\begin{equation}
J={\rm Im}{(V_{11}V_{22}V_{12}^*V_{21}^*)}\, .
\end{equation}
In the parametrisation of Equation~\ref{eq:eulerparam} one finds \cite{0605217}
\begin{equation}
J = c_1 c_2 c_3^2 s_1 s_2 s_3 \sin\delta\, . 
\end{equation}
We note that to obtain the scaling in Equation~\ref{ckmscale} requires 
\[ s_1 \approx \epsilon, \quad s_2 \approx \epsilon^2 \quad \mbox{and} \quad s_3 \approx \epsilon^3.  \]
This implies $J\approx \epsilon^6 \sin\delta$. Thus, any model which reproduces the magnitudes of flavour mixing in the CKM matrix automatically provides an $\epsilon^6$ suppression in the Jarlskog invariant. 
Experimentally $J$ is measured as \cite{Amsler:2008zzb}
\begin{equation}
J=3.05^{+0.19}_{-0.20}\times 10^{-5}\approx \epsilon^{6.5} \, .
\end{equation}
Thus in such models an explanation of the strength of CP violation reduces to showing that $\sin\delta \approx 0.5$.
In the following we give expressions for $J$ and $\sin\delta$ in terms of Higgs vevs. This provides a further constraint on our models, but we cannot determine it here.

\subsubsection*{Models with D3D7 and D3D3 states}
For non-zero CP violation Eq.~\ref{eq:jarlskog} requires that the down quark Yukawa matrix, arising from D3D7 states, cannot be proportional to the identity matrix; a rotation between  generations is needed. Recall that in the $dP_1$ model presented in Section~\ref{sec:d3d7ckm} such a rotation between the strange and bottom quarks was necessary to reproduce the CKM matrix. We take the down quark Yukawa matrix to be the same as in Equation~\ref{eq:d3d7yd}, denoting the phase of $Y_{62}Z_{12}\Phi_{61}/\Lambda$ as $\delta$ find 
\begin{equation}
J= \frac{\theta}{2}  \cdot  \frac{|X_{12}|^2}{|X_{12}|^2+|Z_{12}|^2}\cdot \frac{|Y_{62}^*Z_{12}\frac{\Phi_{61}}{\Lambda}|}{|Y_{62}|^2+\frac{|\Phi_{61}|^2}{\Lambda^2}(|X_{12}|^2+|Z_{12}|^2)}\; \sin{\delta} \, .
\end{equation}
Using the scalings of Section 5.1 one can indeed verify that the coefficient of $\sin\delta$ scales as $\epsilon^6$
\begin{equation}
J \approx \epsilon^2\cdot \epsilon^4 \sin{\delta}=\epsilon^6 \sin{\delta}
\end{equation}
in keeping with the general arguments presented earlier in this section. The correct magnitude of the Jarlskog invariant is obtained when the phase of 
$Y_{62}Z_{12}\Phi_{61}/\Lambda$ is approximately $0.5$.

\subsubsection*{Models purely from D3D3 states}

In this section we discuss CP violation for the left-right symmetric model constructed in Section 5.2. On $dP_2$ with $X_d$ integrated out
\begin{equation}
J = \frac{|\Psi_d|^2}{\Lambda^4} \frac{|X_u|^2\; {\rm Im}{(Y_u \bar{Y}_d Z_d \bar{Z}_u \bar{\Phi}_u \Phi_d)}}{(m_s^2-m_d^2)(m_t^2-m_u^2)(m_t^2-m_c^2)} \approx \epsilon^6 \sin\delta
\end{equation}
where we have used the scalings for the ratios of fields in equations \ref{eq:d3d3vdscaling} and \ref{eq:d3d3vuscaling}, and $\delta$ is the phase of $Y_u \bar{Y}_d Z_d \bar{Z}_u\bar{\Phi}_u\Phi_d.$

For both the D3D3 and D3D7 models we find that scalings introduced to achieve the correct hierarchies in the CKM matrix are consistent with the observed levels of CP violation if $\sin\delta \sim \epsilon^{0.5}$.

\section{Conclusion and outlook}
\label{sec:conclusion}
We have explored the phenomenological aspects of a well defined class of string models based on D-branes at toric singularities and substantially extended previous analyses that mostly concentrated on orbifold and del Pezzo singularities to an infinite class of models.
This subject has been widely studied in the past few years from more theoretical perspectives and we have used the recently developed techniques in this field to extract phenomenological information. We have found several encouraging results. The fact that the number of families is bounded by the experimentally known case is remarkable, especially because in other constructions this number 
can be arbitrarily large. Also having a natural hierarchy of quark masses with one vanishing mass eigenvalue is very encouraging. In toric singularities the mass hierarchy at tree level purely arises from the presence of higher dimensional couplings. Instanton effects via E3 branes \cite{0711.1316} can only contribute in very special circumstances. This contrasts with the intersecting brane models \cite{0909.4292, 0909.4701,0905.3044}, where such effects were necessary to create the hierarchy.

The CKM matrix is more model dependent. At tree-level we can obtain the correct result as a unit matrix plus small corrections. These corrections can be affected from yet undetermined contributions from kinetic terms and therefore we cannot be conclusive before exploring them. If they turn out to be subleading, our results can be used to actually select some of the singularities over the rest. 
In this regard, the $dP_1$ singularity stands out by the fact that it allows mixings among all generations for models with quarks from D3D7 and D3D3 states. The $dP_2$ singularity is the simplest example with all mixings for models with quarks purely from D3D3 states. In the limit of correct flavour mixing we automatically find the expected range of CP violation.
It is interesting that both the mass hierarchy and the flavour mixing are achieved at the same time and are explained by demanding only the size of some ratios of Higgs vevs.

This work opens up many new vistas for exploration. As we mentioned, many of the results depend on several higgsings. A proper determination of the vevs of these fields by extremising the corresponding scalar potential should be done. Further phenomenological issues such as neutrino masses and proton stability  are more model dependent and we plan to study them in detail for particular realistic models. 

An important aspect of these models is the fact that the superpotential $W$ (and therefore the Yukawa couplings)  does not depend on the complex structure moduli. It is well known that in general $W$ cannot depend on K\"ahler moduli due to Peccei-Quinn symmetries. But there are no such  symmetries for the complex structure moduli. Nevertheless toric singularities represent a large class of models where the complex structure moduli do not couple directly to the flavour dependant Yukawa couplings. This is relevant for supersymmetry breaking. Even in the cases where  complex structure moduli could contribute significantly 
to supersymmetry breaking \cite{0906.3297}, the soft breaking terms can still be flavour universal.

Clearly the incorporation of these local models into fully fledged Calabi-Yau compactifications with the right singularity structure and all tadpole cancellation constraints satisfied needs to be addressed in order to have a proper string model. 
Also, to date, the techniques for non-toric singularities are not developed to the same extent as for toric singularities. For example, there is no dimer interpretation of non-toric singularities. Extension of these techniques to non-toric cases, such as higher order del Pezzo singularities would be desirable. We hope to come back to some of these issues in future publications.

\section*{Acknowledgments}
We wish to thank Ben Allanach, Cliff Burgess, Sergio Cecotti, Joe Conlon,  Amihay Hanany and Fernando Marchesano for useful discussions. We thank Noppadol a.k.a $\Omega$ Mekareeya for enlightening discussions and for bringing~\cite{gulotta} to our attention. We would especially like to thank Angel Uranga for his invaluable help and guidance throughout this project. Part of this research was performed during stays at  CERN, GGI, ICTP, McMaster University and Perimeter Institute. AM is funded by an STFC rolling grant at DAMTP. MJD thanks St John's College, EPSRC and the CET for financial support.

\newpage
\appendix
\section{More dimer operations}
\label{sec:moredimer}
Here we give some more details on dimer operations.
\begin{itemize}
\item Removing additional crossings when creating $3$ copies of $(1,1)$ paths.
\begin{center}
\begin{tabular}{c c}
\includegraphics[width=0.3\textwidth]{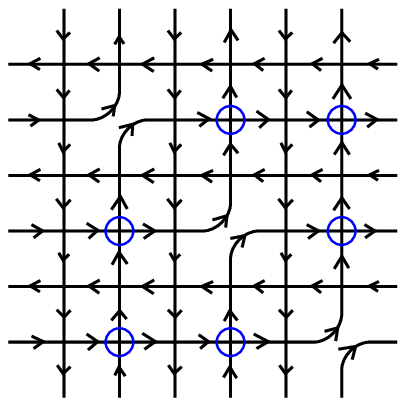} & \includegraphics[width=0.3\textwidth]{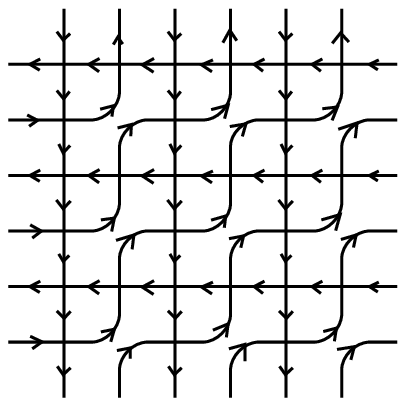}
\end{tabular}
\captionof{figure}{\footnotesize{Creating $3$ copies of $(1,1)$ paths. {\it Left:} Creating them individually generates additional crossings (blue circles highlight additional crossings). {\it Right:} Removing the additional crossings changes the structure of the dimer locally in the same way as creating a single $(1,1)$ path.}\label{fig:appaddcross}}
\end{center}
\end{itemize}

\section{D-term equations}
\label{sec:dterms}

We choose the following VEVs:
\begin{equation}
\Phi_{61}=\left(
\begin{array}{c c}
0 & 0\\
0 & v\\
\end{array}
\right), \; \tilde{A}=\left(\begin{array}{c} 0\\ \sqrt{v^2+\tilde{v}^2} \end{array}\right),\; A=\left(\begin{array}{c} 0\\ v\end{array}\right),\;B=\left(\begin{array}{c c} 0& 0\\  \tilde{v} & 0\end{array}\right).
\end{equation}
The choice for $B$ is so that only $X_{12}^d$ becomes massive in the superpotential. The non-abelian D-term equations are
\begin{equation}
X_{ij} t_{jk}^a X_{ki}^\dagger=Y_{ij} t^a_{jk} Y_{ki}^\dagger\, ,
\end{equation}
where $X_{ij}$ denotes the incoming fields, and $Y_{ij}$ the outgoing ones at a given node of the quiver. The vevs chosen above satisfy these D-term equations.

We do not solve the D-term equations for the D7 gauge groups, which to be satisfied require the inclusion of D7-D7 strings. These D7-D7 strings can also be used to generate mass terms for some unwanted D3D7 states.

\section{Generalising models to higher toric del Pezzos}
\label{sec:breakdown}

All models on lower del Pezzo surfaces can be embedded into models on higher toric del Pezzo surfaces. This can be desirable in the context of flavour physics as discussed in Section 5.

Going from $dP_1$ to $dP_2$ introduces one additional field, $\Psi_{42}$, and one additional gauge group "4". To keep the same matter content as in $dP_1$ the additional gauge group "4" must be at least as large as gauge group "2". One can vev $\Psi_{42}$ to preserve the gauge group which is  a diagonal combination of "4" and "2".

For example, the left-right model on $dP_1$ discussed in Section~\ref{sec:d3d3} can be  generalised to $dP_2$. A vev 
\begin{equation}
\Psi_{42}=\left(\begin{array}{c c}
v & 0\\
0 & v
\end{array}\right)
\end{equation}
breaks the two $U(2)_L$ to the diagonal one.
The quivers of both theories are depicted in Figure~\ref{fig:lrdp1dp2}. The same procedure can be applied to going from $dP_2$ to $dP_3.$
\begin{center}
\begin{tabular}{c c}
\includegraphics[width=0.28\textwidth]{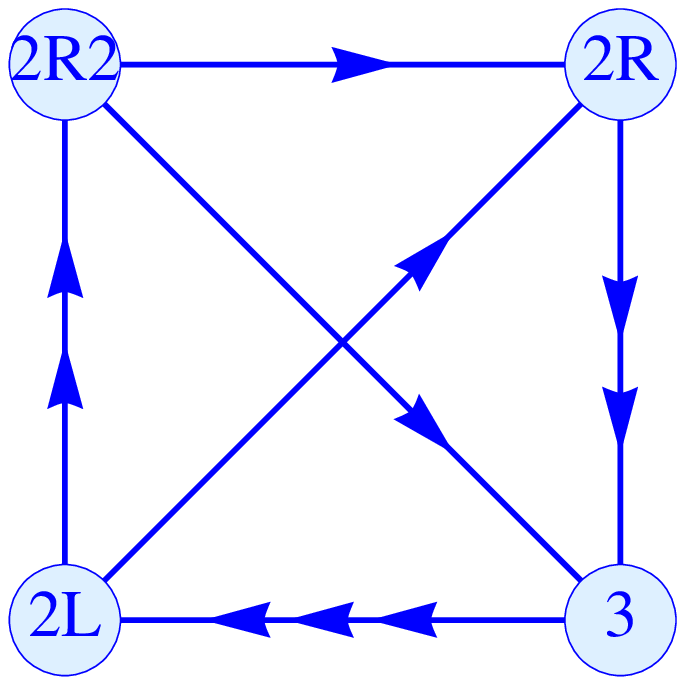} & \includegraphics[width=0.3\textwidth]{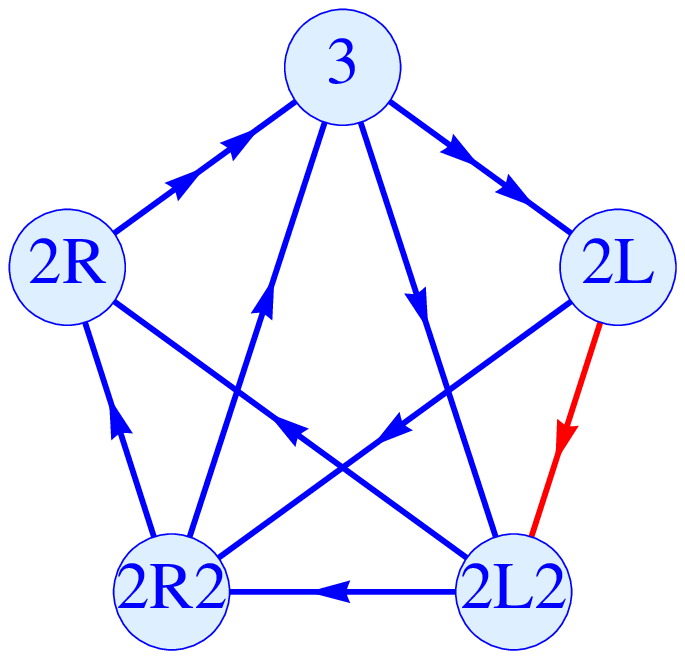}\\
\end{tabular}
\captionof{figure}{\footnotesize{A generalisation of a left-right model from $dP_1$ to $dP_2.$ The additional field $\Psi_{42}$ is indicated in red.}\label{fig:lrdp1dp2}}
\end{center}

\section{Determining all cycles in the quiver diagram}
\label{sec:operators}

In this appendix we present a systematic method to count the number of gauge invariant operators of a given length in a quiver~\cite{stanley}. Gauge invariant operators are given by closed cycles in the quiver.
For a quiver with $n$ nodes, the adjacency matrix is an  $n\times n$ matrix whose entries $a_{ij}$ are given by the number of arrows going from node $i$ to node $j.$ In the case of $dP_1$ the adjacency matrix is
\begin{equation}
a_{ij}=\left(\begin{array}{c c c c}
0&1&0&1\\
0&0&0&2\\
2&1&0&0\\
0&0&3&0
\end{array}\right).
\end{equation}
Raising this matrix to the $l^{th}$ power, the diagonal entries $(a^l)_{ii}$ give the number of gauge invariant operators of length $l$. Note that all gauge invariant operators of length greater than $n$ can be obtained as a product of gauge invariant operators of length less than or equal to $n$.

\newpage
\bibliographystyle{JHEP}
\bibliography{kdmq15}
\end{document}